%% file: draft.tex
\PassOptionsToPackage{usenames,dvipsnames}{xcolor}
\documentclass[a4paper,11pt]{article}
\pdfoutput=1 % if your are submitting a pdflatex (i.e. if you have
\usepackage{amsmath,amssymb}
\usepackage{braket}
\usepackage{fontenc}[T1]
\usepackage{graphicx}
\usepackage[utf8]{inputenc}
\usepackage{jheppub}
\usepackage{multirow}
\usepackage{placeins}
\usepackage{slashed}
\usepackage{xcolor}
\usepackage{xspace}

\allowdisplaybreaks

\title{\boldmath $B\to P$ and $B\to V$ Form Factors
from $B$-Meson Light-Cone Sum Rules beyond Leading Twist}

\author{N.~Gubernari,}
\author{A.~Kokulu,}
\author{D.~van~Dyk}
\affiliation{Physik Department, Technische Universit\"at M\"unchen, James-Franck-Stra\ss{}e 1, D-85748 Garching, Germany}

\note{EOS-2018-02, TUM-HEP-1172/18}

\emailAdd{nico.gubernari@tum.de}
\emailAdd{ahmetkokulu@gmail.com}
\emailAdd{danny.van.dyk@gmail.com}

\abstract{%
  We provide results for the full set of form factors describing semileptonic $B$-meson transitions to
  pseudoscalar mesons $\pi$, $K$, $\bar{D}$ and vector mesons $\rho$, $K^*$, $\bar{D}^*$.
  Our results are obtained within the framework of QCD Light-Cone Sum Rules with $B$-meson distribution amplitudes.
  We recalculate and confirm the results for the leading-twist two-particle contributions
  in the literature. Furthermore, we calculate and provide new expressions for
  the two-particle contributions up to twist four.
  Following new developments for the three-particle
  distribution amplitudes, we calculate and provide new results for the
  complete set of three-particle contributions up to twist four.
  The form factors are computed numerically at several phase space points using
  up-to-date input parameters, including correlations across phase space points and
  form factors.
  We use a model ansatz for all contributing $B$-meson
  distribution amplitudes that is self-consistent up to twist-four accuracy.
  We find that the higher-twist two-particle contributions have a substantial
  impact on the results, and dominate over the three-particle contributions.
  Our numerical results, including correlations, are provided as machine-readable
  ancillary files. We discuss the qualitative phenomenological impact of our results on the
  present $b$ anomalies.
}

\newcommand{\order}[1]{\ensuremath{\mathcal{O}\left(#1\right)}}
\newcommand{\Jint}[1]{\ensuremath{J_\text{int}^{#1}}}
\newcommand{\Jweak}[1]{\ensuremath{J_\text{weak}^{#1}}}
\newcommand{\fp}[1]{\ensuremath{f_+^{B\to #1}}}
\newcommand{\fm}[1]{\ensuremath{f_-^{B\to #1}}}
\newcommand{\fz}[1]{\ensuremath{f_0^{B\to #1}}}
\newcommand{\fT}[1]{\ensuremath{f_T^{B\to #1}}}
\newcommand{\fpm}[1]{\ensuremath{f_{+/-}^{B\to #1}}}
\newcommand{\Azero}[1]{\ensuremath{A_0^{B\to #1}}}
\newcommand{\Aone}[1]{\ensuremath{A_1^{B\to #1}}}
\newcommand{\Atwo}[1]{\ensuremath{A_2^{B\to #1}}}
\newcommand{\Aonetwo}[1]{\ensuremath{A_{12}^{B\to #1}}}
\newcommand{\V}[1]{\ensuremath{V^{B\to #1}}}
\newcommand{\Tone}[1]{\ensuremath{T_1^{B\to #1}}}
\newcommand{\Ttwo}[1]{\ensuremath{T_2^{B\to #1}}}
\newcommand{\Tthree}[1]{\ensuremath{T_3^{B\to #1}}}
\newcommand{\Ttwothree}[1]{\ensuremath{T_{23}^{B\to #1}}}
\renewcommand{\V}[1]{\ensuremath{V^{B\to #1}}}
\newcommand{\Athree}[1]{\ensuremath{A_3^{B\to #1}}}
\newcommand{\Athreezero}[1]{\ensuremath{A_{30}^{B\to #1}}}
\newcommand{\TtwothreeA}[1]{\ensuremath{T_{23A}^{B\to #1}}}
\newcommand{\TtwothreeB}[1]{\ensuremath{T_{23B}^{B\to #1}}}
\newcommand{\GeV}{\ensuremath{\text{GeV}}}
\newcommand{\MeV}{\ensuremath{\text{MeV}}}
\newcommand{\EOS}{\texttt{EOS}\xspace}
\newcommand{\flavio}{\texttt{flavio}\xspace}
\newcommand{\refapp}[1]{appendix~\ref{app:#1}}
\newcommand{\refeq}[1]{eq.~(\ref{eq:#1})}
\newcommand{\reffig}[1]{figure~\ref{fig:#1}}
\newcommand{\refeqs}[2]{eqs.~(\ref{eq:#1})--(\ref{eq:#2})}
\newcommand{\refsec}[1]{section~\ref{sec:#1}}
\newcommand{\reftab}[1]{table~\ref{tab:#1}}
\newcommand{\overbar}[1]{\mkern 1.5mu\overline{\mkern-1.5mu#1\mkern-1.5mu}\mkern 1.5mu}
\begin{document}

\maketitle
\flushbottom

\section{Introduction}
\label{sec:intro}

Form factors for $B$-meson decays to either pseudoscalar ($P$) or vector hadrons ($V$) arise in the
hadronic $B\to P$ and $B\to V$ matrix elements of local flavour-changing currents $\bar{q}_1 \Gamma b$,
where $\Gamma$ denotes some spin structure. The hadronic
form factors are a crucial input for accurate and precise predictions of observables (e.g. branching ratios and
angular coefficients) in various semileptonic $B$-meson decays. In light of the recent $b$ anomalies\footnote{
    See references \cite{Albrecht:2017odf} and \cite{Albrecht:2018vsa} for recent reviews on the topic.
}, the form factors for $B\to K^{(*)}$ and $B\to \bar{D}^{(*)}$ transitions in particular
have moved into the focus of theoretical and phenomenological interest. In this article we study the full set of
form factors that arise in $B\to P,V$ transitions. Results for the tensor form factors in $B\to D^{(*)}$ transitions
are provided at small momentum transfer for the first time.\\

The hadronic matrix elements for the transitions under discussion are genuinely non-perturbative
quantities. Presently, the only ab-initio method for their determination is Lattice QCD (LQCD), which
uses discretized spacetime as a UV regulator \cite{Wilson:1974sk}. In the long run, lattice determinations are expected
to dominantly contribute to our understanding of the hadronic matrix elements discussed here.
At present, lattice results for the form factors are restricted to the phase space region
in which the final state exhibits only small recoil momentum in the $B$ rest frame, corresponding to high
dilepton invariant mass square $q^2$. However,
for the anomalies in $B\to K^{*}\mu^+\mu^-$ decays, the phase space region containing the largest deviations
from the SM predictions corresponds to large recoil of the $K^*$ (or equivalently to low $q^2$).
For semileptonic $B\to \bar{D}^{(*)}$ decays, it has been
demonstrated \cite{Bigi:2017jbd,Jung:2018lfu} that inputs at maximal hadronic recoil can influence the extraction of $|V_{cb}|$
and the SM prediction for the Lepton-Flavour Universality ratios $R_{D^{(*)}}$. The present lattice results for
the relevant form factor therefore require an extrapolation from the phase space region in which they are
obtained to the phase space region in which they are required.\\

An alternative method, which we apply in this work, is to determine the form factors from
Light-Cone QCD Sum Rules (LCSRs).
In the framework of $B$-meson LCSRs
\cite{Khodjamirian:2005ea,Khodjamirian:2006st,Faller:2008tr,Khodjamirian:2010vf,Cheng:2017smj}\footnote{%
    This setup is complementary to more commonly used LCSRs in which the $B$ meson is interpolated and the
    final state meson is taken on-shell; see e.g. \cite{Ball:2004ye,Ball:2004rg,Khodjamirian:2009ys,Khodjamirian:2011ub,Bharucha:2012wy,%
    Imsong:2014oqa,Straub:2015ica,Khodjamirian:2017fxg}.
    Our $B$-LCDA-based setup is similar to the framework of SCET sum rules~\cite{DeFazio:2005dx,DeFazio:2007hw,Wang:2015vgv,Wang:2017jow}.
} the time-ordered product of two local quark currents $\bar{q}_2(x)\Gamma_2 q_1(x)$ and $\bar{q}_1(0) \Gamma_1 b(0)$
is expanded in a series of non-local operators $\bar{q}_2(x) \Gamma h_v(0)$, where $h_v$ denotes a $b$-flavoured
HQET field, $v_\mu$ is the four-velocity of the $B$-meson and $x$ is a light-like separations: $x^2 \simeq 0$.
The operators' hadronic matrix elements between an on-shell $B$ meson state and the hadronic vacuum
can then be expressed as convolutions of hard scattering kernels with Light-Cone Distribution Amplitudes (LCDAs) of the $B$ meson.
The LCDAs are organized in terms of their twist, i.e., the difference between an operator's mass dimension
and the canonical spin of the operator. Here the canonical spin of the $h_{v}$ field
is chosen such that the leading-twist contributions enter at twist two, similar to LCDAs of light mesons \cite{Braun:2017liq}.
While the hard scattering kernels are perburbatively calculable, the LCDAs are genuine but universal non-perturbative input to the sum
rules. The non-local $B$ to vacuum matrix element is then related to the sought-after form factors via dispersion
relations and semi-local quark-hadron duality.
In this process, the relevant duality threshold parameters can be determined from moments of the correlations function. LCSRs therefore
constitute a systematic approach to determine hadronic matrix elements relevant for exclusive $B$ decays.
The light-cone dominance $x^2\simeq 0$ of the expansion of the correlator is only fulfilled for small values of the momentum transfer
$q^2$~\cite{Khodjamirian:2005ea}. As a consequence LQCD and LCSRs provide complementary inputs toward the determination
of the $B$-meson form factors as functions of the momentum transfer $q^2$, albeit with with starkly different levels of
uncertainties.\\

Our work improves upon the results in the literature in several places. First, we include matrix elements
of two-particle operators at twist-four level. Second, we calculate the
contributions arising from the full set of eight independent Lorentz structures \cite{Geyer:2005fb} of three-particle matrix elements.
A set of models for the three-particle LCDAs, consistent with the relevant equations of motion \cite{Kawamura:2001jv} and theoretical constraints,
has been worked out for the first time in ref.~\cite{Braun:2017liq}. Our results therefore go beyond
the three-particle contributions previously discussed \cite{Khodjamirian:2005ea,Khodjamirian:2006st,Faller:2008tr}.
Third, we use a rigorous statistical framework to estimate the theoretical parametric uncertainties of the form factors.
In this, we follow closely a previous work that proved the concept at the hand of $B\to \pi$ form factors in LCSRs with $\pi$ LCDAs.
The $B\to V$ transitions discussed here all involve vector states $V=\rho,K^*,\bar{D}^*$ that are unstable
and decay strongly to pairs of pseudoscalar mesons. Throughout this article we work within the narrow-width
approximation, i.e., we assume the vector mesons to be stable under QCD. To go beyond this approximation
requires further studies. A dedicated programme has emerged in recent years to study $\pi\pi$
\cite{%
Faller:2013dwa,Kang:2013jaa,Hambrock:2015aor,Boer:2016iez,Cheng:2017smj,Cheng:2017sfk,Feldmann:2018kqr%
} and $K \pi$ \cite{%
Meissner:2013pba,Meissner:2013hya,Das:2014sra,Feldmann:2015xsa%
} final states beyond a simple resonance ansatz. LCSR analyses of $B\to P_1 P_2$ form factors require
ongoing updates
\cite{DescotesGenon:2018xxx,DescotesGenon:2018yyy} by other authors to improve our understanding
of the semileptonic processes as well as purely hadronic multi-body $B$ decays \cite{%
Wang:2015uea,Krankl:2015fha,Daub:2015xja,Cheng:2016shb,Wang:2016rlo,Albaladejo:2016mad,Klein:2017xti%
}.\\

The structure of our article is as follows. We provide details on the derivation of the sum rules at next-to-next-to-leading twist
and the master formulas for the analytical results in \refsec{details}. Our numerical results in \refsec{results} are comprised
of the actual LCSR results (in \refsec{results:LCSRs}) and a fit to LCSR and LQCD results (in \refsec{results:fits}).
We continue to discuss some selected phenomenological implications of our results in \refsec{pheno}, before concluding
in \refsec{conclusion}. In a series of appendices we discuss the details of the $B$-meson LCDAs (in \refapp{LCDAs}), provide the
lengthy formulas of our analytical LCSR results in form of scalar coefficient functions (in \refapp{coefficients}), and illustrate
our numerical results at the hand of plots of all form factors considered here (in \refapp{plots}).

\section{Details on the Computation and Analytical Results}
\label{sec:details}

We construct the LCSRs for the full set of form factors in the hadronic matrix elements of
local flavour-changing $b\to q_1$ currents. For $B\to P$ transitions, with $P=\pi,K,\bar{D}$, we discuss
the three independent non-vanishing form factors $\fp{P}$, $\fz{P}$ and $\fT{P}$, which are defined via
\begin{align}
    \label{eq:BtoP:vector}
    \bra{P(k)} \bar{q}_1 \gamma^\mu b \ket{B(p)}
        & = \left[(p + k)^\mu - \frac{m_B^2 - m_P^2}{q^2}q^\mu\right]\, \fp{P}
          + \frac{m_B^2 - m_P^2}{q^2} q^\mu\, \fz{P},\\
    \label{eq:BtoP:tensor}
    \bra{P(k)} \bar{q}_1 \sigma^{\mu\nu}\, q_\nu b \ket{B(p)}
        & = \frac{i \fT{P}}{m_B + m_P}\, \left[q^2\, (p + k)^\mu - (m_B^2 - m_P^2)\, q^\mu\right]\,.
\end{align}
Here and throughout this article the form factors are functions of the momentum transfer $q^2 \equiv (p - k)^2$, and
$p$ and $k$ denote the $B$-meson's and the final-state meson's momentum, respectively.
For $B\to V$ transitions, with $V=\rho,K^*,\bar{D}^*$, we discuss the seven non-vanishing form factors $\Azero{V}$,
$\Aone{V}$, $\Atwo{V}$, $\V{V}$, $\Tone{V}$, $\Ttwo{V}$, and $\Tthree{V}$,
which are defined via:
\begin{align}
    \label{eq:BtoV:vector}
    \bra{V(k, \eta)} \bar{q}_1 \gamma^\mu b \ket{B(p)}
        & = \epsilon^{\mu\nu\rho\sigma} \eta^*_\nu p_\rho k_\sigma \frac{2 \V{V}}{m_B + m_V}\,,\\
    \label{eq:BtoV:axial}
    \bra{V(k, \eta)} \bar{q}_1 \gamma^\mu \gamma_5 b \ket{B(p)}
        & = i \eta^*_\nu\ \bigg[g^{\mu\nu} (m_B + m_V) \Aone{V} - \frac{(p + k)^\mu q^\nu}{m_B + m_V}\, \Atwo{V}\\
    \nonumber
        & \qquad\qquad  - q^\mu q^\nu \frac{2 m_V}{q^2} \left(A_3 - A_0\right)\bigg]\,,\\
    \label{eq:BtoV:tensor}
    \bra{V(k, \eta)} \bar{q}_1  i \sigma^{\mu\nu}\, q_\nu b \ket{B(p)}
        & = \epsilon^{\mu\nu\rho\sigma} \eta^*_\nu p_\rho k_\sigma 2 \Tone{V}\,,\\
    \label{eq:BtoV:tensor5}
    \bra{V(k, \eta)} \bar{q}_1 i \sigma^{\mu\nu}\, q_\nu \gamma_5 b \ket{B(p)}
        & = i \eta^*_\nu \bigg[\left( g^{\mu\nu} (m_B^2 - m_V^2) - (p + k)^\mu q^\nu\right) \Ttwo{V}\\
    \nonumber
        & \qquad\qquad +  q^\nu \left(q^\mu - \frac{q^2}{m_B^2 - m_V^2} (p + k)^\mu\right) \Tthree{V}\bigg]\,,
\end{align}
with $\eta$ representing the polarization of the vector meson, and we use the Bjorken-Drell convention with $\epsilon_{0123} = +1$.
If the final state is either a $\pi^0$ or a $\rho^0$, the l.h.s. of \refeqs{BtoP:vector}{BtoV:tensor5}
have to be multiplied with a factor of $\sqrt{2}$.
Note that $\Athree{V}$ is redundant, since it is a linear combination of the two other axial form factors
\begin{equation}
    \Athree{V} \equiv \frac{m_B + m_V}{2 m_V} \Aone{V} - \frac{m_B - m_V}{2 m_V} \Atwo{V}\,.
\end{equation}
However, the decomposition of the matrix elements including the $\Athree{V}$ form factor is convenient
for the extraction of the form factors within a sum rule approach, as discussed below.\\

The matrix elements defined in \refeq{BtoP:vector} and \refeq{BtoV:axial} exhibit apparently unphysical
singularities at $q^2 = 0$. These are removed by the identities
\begin{align}
    \fp{P}(q^2 = 0)
        & = \fz{P}(q^2 = 0)\,, &
    \Azero{V}(q^2 = 0)
        & = \Athree{V}(q^2 = 0)\,.
\end{align}
In addition, the algebraic relations between $\sigma^{\mu\nu}$ and $\sigma^{\mu\nu}\gamma_5$ give rise to the
identity
\begin{equation}
    \Tone{V}(q^2 = 0)
        = \Ttwo{V}(q^2 = 0)\,.
\end{equation}
It is common to replace the form factors $\Atwo{V}$ and $\Tthree{V}$ with the linear combinations
\begin{align}
    \Aonetwo{V}   & \equiv \frac{(m_B+m_V)^2(m_B^2 - m_V^2 - q^2)A_1 - \lambda (q^2) A_2}{16 m_B m_V^2 (m_B + m_V)}, \\
    \Ttwothree{V} & \equiv \frac{(m_B^2-m_V^2)(m_B^2 + 3m_V^2 - q^2)T_2 - \lambda (q^2) T_3}{8 m_B m_V^2 (m_B - m_V)};
\end{align}
where $\lambda(q^2) \equiv [(m_B+m_V)^2-q^2][(m_B-m_V)^2-q^2)]$ is the K\"all\'en function.
The linear combinations $\Aonetwo{V}$ and $\Ttwothree{V}$ correspond to form factors for the
transition into a longitudinal vector state and therefore simplify the structure of angular
coefficients in the differential decay rate of the semileptonic $B$-meson decays.\\

\begin{table}[t]
    \renewcommand{\arraystretch}{1.25}
    \centering
\begin{tabular}{|c|c|c|c|}
    \hline
      process  & $\Jint{\nu}$ & $\Jweak{\mu}$ & form factor \\\hline\hline
       \multirow{ 2 }{*}{$\bar{B}^0 \to \pi^+$}        &\multirow{ 2 }{*}{$\bar{d} \gamma^\nu \gamma_5 u$}
                                                       & $\bar{u} \gamma^\mu h_v$        & $\fp{\pi},\,\fpm{\pi}$                               \\
                                                      && $\bar{u} \sigma^{\mu\{q\}} h_v$ & $\fT{\pi}$                                          \\\hline
       \multirow{ 2 }{*}{$\bar{B}^0 \to \bar{K}^0$}    &\multirow{ 2 }{*}{$\bar{d} \gamma^\nu \gamma_5 s$}
                                                       & $\bar{s} \gamma^\mu h_v$        & $\fp{K},\,\fpm{K}$                                   \\
                                                      && $\bar{s} \sigma^{\mu\{q\}} h_v$ & $\fT{K}$                                            \\\hline
       \multirow{ 2 }{*}{$\bar{B}^0 \to D^+$}          &\multirow{ 2 }{*}{$\bar{d} \gamma^\nu \gamma_5 c$}
                                                       & $\bar{c} \gamma^\mu h_v$        & $\fp{D},\,\fpm{D}$                                   \\
                                                      && $\bar{c} \sigma^{\mu\{q\}} h_v$ & $\fT{D}$                                            \\\hline
       \multirow{ 4 }{*}{$\bar{B}^0 \to \rho^+$}       &\multirow{ 4 }{*}{$\bar{d} \gamma^\nu u$}
                                                       & $\bar{u} \gamma^\mu h_v$                 & $\V{\rho}$                                 \\
                                                      && $\bar{u} \gamma^\mu \gamma_5 h_v$        & $\Azero{\rho},\,\Aone{\rho},\,\Atwo{\rho}$ \\
                                                      && $\bar{u} \sigma^{\mu\{q\}} h_v$          & $\Tone{\rho}$                              \\
                                                      && $\bar{u} \sigma^{\mu\{q\}} \gamma_5 h_v$ & $\Ttwo{\rho}, \Tthree{\rho}$               \\\hline
       \multirow{ 4 }{*}{$\bar{B}^0 \to \bar{K}^{*0}$} &\multirow{ 4 }{*}{$\bar{d} \gamma^\nu s$}
                                                       & $\bar{s} \gamma^\mu h_v$                 & $\V{K^*}$                                  \\
                                                      && $\bar{s} \gamma^\mu \gamma_5 h_v$        & $\Azero{K^*},\,\Aone{K^*},\,\Atwo{K^*}$    \\
                                                      && $\bar{s} \sigma^{\mu\{q\}} h_v$          & $\Tone{K^*}$                               \\
                                                      && $\bar{s} \sigma^{\mu\{q\}} \gamma_5 h_v$ & $\Ttwo{K^*}, \Tthree{K^*}$                 \\\hline
       \multirow{ 4 }{*}{$\bar{B}^0 \to D^{*+}$}       &\multirow{ 4 }{*}{$\bar{d} \gamma^\nu c$}
                                                       & $\bar{c} \gamma^\mu h_v$                 & $\V{D^*}$                                  \\
                                                      && $\bar{c} \gamma^\mu \gamma_5 h_v$        & $\Azero{D^*},\,\Aone{D^*},\,\Atwo{D^*}$    \\
                                                      && $\bar{c} \sigma^{\mu\{q\}} h_v$          & $\Tone{D^*}$                               \\
                                                      && $\bar{c} \sigma^{\mu\{q\}} \gamma_5 h_v$ & $\Ttwo{D^*}, \Tthree{D^*}$                 \\\hline
\end{tabular}
    \caption{%
    \label{tab:listcurrents}
    Summary of the various combinations of weak and interpolating currents used to extract the form factors. We abbreviate
    $\sigma^{\mu\lbrace q\rbrace} \equiv \sigma^{\mu\nu} q_\nu$.
    }
\end{table}

The starting point for the construction of the $B$-LCSRs is the correlation function
\begin{align}
    \label{eq:correlator}
    \Pi^{\mu\nu}(q, k)
        \equiv i \int \text{d}^4 x\, e^{i k\cdot x}\,
        \bra{0} \mathcal{T}\lbrace \Jint{\nu}(x), \Jweak{\mu}(0)\rbrace \ket{\bar{B}_{q_2}(q + k)}
\end{align}
of two quark currents $\Jint{\nu} \equiv \bar{q}_2(x) \Gamma_2^\nu q_1(x)$ and $\Jweak{\mu}(0) \equiv \bar{q}_1(0) \Gamma_1^\mu h_v(0)$.
The various choices of spin structures $\Gamma_{1,2}$ and quark flavours $q_1$ and $q_2$ for the form factors extracted
in this article are shown in table \ref{tab:listcurrents}.
The correlator (\ref{eq:correlator}) is calculated in the framework of heavy quark effective theory
(HQET), i.e. the $b$-quark field is replaced by the HQET field $h_{v}$.
In the kinematic regime $q^2 \leq m_b^2 + m_b k^2 / \Lambda_\text{had.}$ and $k^2 \ll -\Lambda_\text{had.}^2$, the dominant contributions to the correlator \refeq{correlator} arise at
light-like distances $x^2\simeq 0$~\cite{Khodjamirian:2006st}. This motivates a systematic expansion of the time-ordered product
in terms of bi-local operators with light-like separation $\bar{q}_2(x) \Gamma [x, 0] h_{v}(0)$, where the $[x, 0]$
denotes a gauge link that renders the bi-local operators gauge invariant.
The expansion of the $q_1$ propagator up to next-to-leading power in $x^2$ near the light-cone $x^2 \simeq 0$
gives rise to two-particle and three-particle contributions to the correlator. Four-particle contributions are not taken in account
in this work. The two-particle contributions can be summarized as
\begin{multline}
    \label{eq:correlatorHQET2pt}
    \Pi^{\mu\nu}(q, k)\bigg|_\text{2p}
        \equiv \int \text{d}^4 x\,\int \text{d}^4 p'\, e^{i (k - p')\cdot x}\,
        \left[ \Gamma_2^\nu\, \frac{\slashed{p}'+m_1}{m_1^2-p'^2} \Gamma_1^\mu \right]_{\alpha\beta}\\*
        \bra{0} \bar{q}_{2}^{\alpha}(x) h_{v}^{\beta}(0) \ket{\bar{B}_{q_2}(v)},
\end{multline}
where $\alpha$, $\beta$ are spinor indices. The three-particle contributions involve
a further gluon field:
\begin{multline}
    \label{eq:correlatorHQET3pt}
    \Pi^{\mu\nu}(q, k)\bigg|_\text{3p}
        \equiv \int \text{d}^4 x\,\int \text{d}^4 p'\,\int_0^1 \text{d}u\, e^{i (k - p')\cdot x}\,
        \left[ \Gamma_2^\nu\, S_{\text{3p}}^{\lambda\rho}(u, p') \Gamma_1^\mu \right]_{\alpha\beta}\\
        \bra{0} \bar{q}_{2}^{\alpha}(x) G_{\lambda\rho}(ux) h_{v}^{\beta}(0) \ket{\bar{B}_{q_2}(v)}\,.
\end{multline}
In the above $G_{\lambda\rho}\equiv g_s (\lambda^a/2) G^{a}_{\lambda\rho}(x)$ denotes the gluon field strength tensor,
and
\begin{align}
      S_{\text{3p}}^{\lambda\rho}(u, p') \equiv \frac{ \bar{u} (\slashed{p}'+m_1) \sigma^{\lambda\rho} +  {u} \sigma^{\lambda\rho}
      (\slashed{p}'+m_1)   }{2 (p^{'2}-m_1^2)^2}\,,\qquad \bar{u} \equiv 1 - u\,,
\end{align}
is the momentum-space representation of the next-to-leading-power term in the light-cone expansion of the quark propagator \cite{Balitsky:1988fi}, in which $u$ is the position of the gluon field as a fraction of the light-like distance $\left[ 0, x \right]$ .
The $B$-meson to vacuum matrix elements of the non-local heavy-light currents,
appearing in \refeq{correlatorHQET2pt} and \refeq{correlatorHQET3pt}, are parametrized in terms
of $B$-meson Light-Cone Distribution Amplitudes (LCDAs). The full expressions of these non-local
matrix elements for the $B$-meson LCDAs used in this work are collected in \refapp{LCDAs}.
Previous works~\cite{Khodjamirian:2005ea,Khodjamirian:2006st,Faller:2008tr}
calculate the correlation functions up to twist three for the two-particle LCDAs, and use an incomplete set
of three-particle LCDAs. In our work we go beyond the accuracy of the previous calculations by
including all contributions to the correlator from two- and three-particle Fock states
up to twist four \cite{Braun:2017liq}.\\

In order to construct the sum rule, one has to insert a complete set of hadronic states in \refeq{correlator},
thereby obtaining a hadronic dispersion relation for $\Pi^{\mu\nu}$:
\begin{align}
    \label{eq:hadronicdisp}
    \Pi^{\mu\nu}(q, k) = \frac{\bra{0}  \Jint{\nu}(x) \ket{M(k)} \bra{M(k)} \Jweak{\mu}(0) \ket{\bar{B}_{q_2}(q + k)}}{m_M^2-k^2}
    + \frac{1}{2\pi}\int^{\infty}_{s_0^h} \text{d}s\frac{\rho^{\mu\nu}(s)}{s-k^2}\,,
\end{align}
with $M=P,V$.
The last term involving the spectral density $\rho(s)$ on the r.h.s. of \refeq{hadronicdisp} captures the contributions arising from excited and
continuum states, $s_0^h$ is the corresponding threshold for the lowest-mass excited or continuum state.
The local $M$ to vacuum matrix elements are proportional to decay constants $f_P$ and $f_V$:
\begin{align}
    \bra{0} \bar{q}_2 \gamma^\nu \gamma_5 q_1 \ket{P(k)} &= i k^\nu f_P, \nonumber\\
    \bra{0} \bar{q}_2 \gamma^\nu q_1 \ket{V(k, \eta)} &= i \eta^\nu m_V f_V\,.
\end{align}
The $B\to P$ and $B\to V$ matrix elements have been already introduced in \refeqs{BtoP:vector}{BtoV:tensor5}.
Again, if the initial state of the equation above is a $\pi^0$ or a $\rho^0$, the l.h.s.
receives an additional factor of $\sqrt{2}$.\\

Using the formulas given in \refeq{BLCDAs2pt} and \refeq{BLCDAs3pt}, we can cast the two- and three-particle
terms in \refeq{correlatorHQET2pt} and \refeq{correlatorHQET3pt} into an integral form similar to
\refeq{hadronicdisp}.
Using semi-local quark hadron duality to subtract the continuum contributions we obtain the sum rule.
In the process, we apply a Borel transformation from $k^2$ to $M^2$, which removes surfaces terms in the integrals and improves
the numerical stability of the sum rule. The latter is achieved by accelerating the convergence
of the twist expansion, and by reducing the sensitivity to the duality approximation.
The sum rule can then be written in the following form for all the form factors $F$ and final states $M=P,V$ considered here:
\begin{align}
 F= 
    &\frac{f_B M_B}{K^{(F)}} \sum_{n=1}^{\infty}\Bigg\{(-1)^{n}\int_{0}^{\sigma_0} d \sigma \;e^{(-s(\sigma,q^2)+m^2_{P,V})/M^2} \frac{1}{(n-1)!(M^2)^{n-1}}I_n^{(F)}\nonumber\\
        & - \Bigg[\frac{(-1)^{n-1}}{(n-1)!}e^{(-s(\sigma,q^2)+m^2_{P,V})/M^2}\sum_{j=1}^{n-1}\frac{1}{(M^2)^{n-j-1}}\frac{1}{s'}
        \left(\frac{\text{d}}{\text{d}\sigma}\frac{1}{s'}\right)^{j-1}I_n^{(F)}\Bigg]_{\sigma=\sigma_0}\Bigg\rbrace\,,
        \label{eq:masterformula}
\end{align}
where we use the auxiliary variable $s$ and its derivative
\begin{align}
    s(\sigma,q^2)=\sigma m^2_B +\frac{m_1^2-\sigma q^2}{\bar{\sigma}}\,,
    \qquad
    s'(\sigma,q^2)=\frac{\text{d} s(\sigma,q^2)}{\text{d} \sigma}\,.
\end{align}
In \refeq{masterformula}, the expressions involving powers of differential operators should always be read as
\begin{equation*}
    \left(\frac{\text{d}}{\text{d}\sigma}\frac{1}{s'}\right)^{n} I(\sigma) \to 
    \left(\frac{\text{d}}{\text{d}\sigma}\frac{1}{s'}\left(\frac{\text{d}}{\text{d}\sigma}\frac{1}{s'}\dots I(\sigma)\right)\right)\,.
\end{equation*}
We further abbreviate
$\bar{\sigma}\equiv1-\sigma$ and $\sigma_0\equiv \sigma(s_0,q^2)$, where
$s_0$ is an effective threshold parameter not to be confused with $s_0^h$, from which it differs in general.
The functions $I_n^{(F)}$
can be represented as integrals involving the two-particle and three-particle LCDAs:
\begin{align}
    I_n^{(F,\,\text{2p})}(\sigma,q^2)
     &= \frac{1}{\bar{\sigma}^n} \sum_{\psi_\text{2p}}  C^{(F,\psi_\text{2p})}_n(\sigma,q^2)\, \psi_\text{2p} (\sigma m_B),  \, \hspace{2.3cm}\psi_\text{2p}=\phi_+,\bar{\phi},g_+,\bar{g};
    \label{eq:CoeffFuncs2pt}\\ \nonumber
    I_n^{(F,\,\text{3p})}(\sigma,q^2)
    &= \frac{1}{\bar{\sigma}^n}\int\displaylimits_{0}^{\sigma m_B}\text{d}\omega_1 \int \displaylimits_{\sigma m_B-\omega_1}^{\infty}\frac{\text{d}\omega_2}{\omega_2}
    \sum_{\psi_\text{3p}}  C^{(F,\psi_\text{3p})}_n(\sigma,u,q^2)\, \psi_\text{3p} (\omega_1, \omega_2) \Bigg|_{u=(\sigma m_B-\omega_1)/\omega_2} ,\nonumber\\
    &\hspace{8cm}\psi_\text{3p}=\phi_3,\phi_4,\psi_4,\chi_4;
\label{eq:CoeffFuncs3pt}
\end{align}
with $\sigma=\omega/m_B$ in \refeq{CoeffFuncs2pt} and $\sigma=(\omega_1 + u\omega_2)/m_B$ in \refeq{CoeffFuncs3pt}, respectively.
The coefficients $C^{(F,\psi)}$, as well as the normalization factors $K^{(F)}$ of \refeq{masterformula}
are listed in the \refapp{coefficients}.
In the cases $F = \fp{P}$, $\fT{P}$, $\Aone{V}$, $\V{V}$, and $\Tone{V}$
we can construct the sum rule for the form factor directly, whereas for the remainder of the cases
$F$ denotes one of the following linear combinations of form factors:
\begin{align}
    \fpm{P}
        & \equiv \fp{P} + \fm{P}\,,\\
    \Athreezero{V}
        & \equiv \Athree{V} - \Azero{V}\,,\\
    \label{eq:defTA}
    \TtwothreeA{V}
        & \equiv \Ttwo{V} +\frac{q^2}{m_B^2-m_V^2}\Tthree{V}  \,,\\
    \label{eq:defTB}
    \TtwothreeB{V}
        & \equiv \frac{1}{2}\Ttwo{V} +\frac{1}{2} \left(\frac{q^2}{m_B^2-m_V^2}-1 \right) \Tthree{V} \,.
\end{align}
Here $\fm{P}$ is given by
\begin{align}
    \fz{P} = \fp{P} + \frac{q^2}{m_B^2 - m_P^2} \fm{P}.
\end{align}
Our results for the analytical expressions are always provided for a generic final state meson $P, V$ with valence quark content $(q_1 \bar{q}_2)$. To
the precision we work at, only the mass $m_1$ of the quark field $q_1$ enters the expressions.\\

We fully reproduce the two-particle leading-twist contributions proportional to $\phi_+$ and $\phi_-$ given in ref.~\cite{Khodjamirian:2006st}.
Furthermore, we extend the two-particle results adding the terms containing $g_+$ and $g_-$, that take in account corrections up to twist five.
The results for three-particle contributions in ref.~\cite{Khodjamirian:2006st,Faller:2008tr} are obtained for only a subset
of the three-particle LCDAs: $\psi_{\text{3p}} = \psi_V$, $\psi_A$, $X_A$, and $Y_A$. When artificially restricting the LCDAs
to the same subset, we reproduce the results of refs.~\cite{Khodjamirian:2006st,Faller:2008tr}. Our results for the
coefficients $C^{(F,\psi)}$ provide for the first time the complete results for the two- and three-particle contributions up to and including
twist four.

\section{Numerical Results}
\label{sec:results}

\subsection{LCSR Results}
\label{sec:results:LCSRs}

We implement the sum rules for the full set of $B\to P$ and $B\to V$ form factors as part of the \EOS
software \cite{EOS-v0.2.3}, which is an open source project for the evaluation of flavour observables \cite{EOS}.
Our implementation is agnostic of the concrete parametrization of the various LCDAs entering the sum rules.
This is achieved by computing all contributing integrals numerically. For this work, the LCDAs implemented in \EOS conform to the
exponential model put forward in ref.~\cite{Braun:2017liq}. However, further LCDA models can readily be added to \EOS, in
order to challenge the (implicit) dependency of the sum rules on the LCDA model. Realistically, this
can only be done after measurements of the photo-leptonic decay $B^-\to \gamma\ell^-\bar\nu$; see \cite{Beneke:2018wjp}
for a recent update of the theoretical framework for the extraction of the LCDA model parameters.\\

In order to obtain numerical predictions for the form factors and to estimate
the theory uncertainties due to the input parameters, we follow the statistical
procedure used in ref.~\cite{Imsong:2014oqa}. Within a Bayesian framework we first
define an a-priori Probability Density Function (PDF) for the input parameters.
A summary of the process-specific elements of this PDF is given in \reftab{priors}.
The universal elements of this PDF can be summarized as the following independent
Gaussian PDFs for the $B$-meson to vacuum matrix elements:
\begin{equation}
\begin{aligned}
    f_B               & = (189.4 \pm 1.4)\, \MeV\,,   &
    1 / \lambda_{B,+} & = (2.2 \pm 0.6)\,\GeV^{-1}\,, \\
    \lambda_E^2       & = (0.03 \pm 0.02)\,\GeV^2\,,  &
    \lambda_H^2       & = (0.06 \pm 0.03)\,\GeV^2\,.
\end{aligned}
\end{equation}
We use the $f_B$ value from the most precise LQCD analysis available \cite{Bazavov:2017lyh}, our own
estimate of $1 / \lambda_{B,+}$ and the $\lambda_{E,H}^2$ from ref.~\cite{Nishikawa:2011qk}.
Two classes of the process-specific parameters
deserve a more detailed discussion: the Borel parameters $M^2$ and the duality thresholds $s_0$.\\

\begin{table}[t]
    \centering
    \renewcommand{\arraystretch}{1.25}
    \begin{tabular}{|c|c|c|c|}
        \hline
        meson   & decay constant $f_{P,V}$ $[\MeV]$ & $s_0$ $[\GeV^2]$& $M^2$ $[\GeV^2]$ \\\hline\hline
        $\pi$   &  $130.2 \pm 1.4$        & $0.7 \pm0.014^\times$               & $1.0 \pm 0.5$  \\\hline
        $K$     &  $155.6 \pm 0.4$        & $1.05\pm0.021^\times$               & $1.0 \pm 0.5$  \\\hline
        $D$     &  $212.6 \pm 0.5$        & $[5.8,\, 7.8]^\dagger$              & $4.5 \pm 1.5$  \\\hline
        $\rho$  &  $213 \pm 5    $        & $1.6 \pm0.032^\times$               & $1.0 \pm 0.5$  \\\hline
        $K^*$   &  $204 \pm 7    $        & $[1.4,\, 1.7]^\dagger$              & $1.0 \pm 0.5$  \\\hline
        $D^*$   &  $249 \pm 21   $        & $[6.9,\, 8.0]^\dagger$              & $4.5 \pm 1.5$  \\\hline
    \end{tabular}
    \caption{
    \label{tab:priors}
             Overview of transition and form-factor specific numerical inputs used in our calculations.
             Values marked with $\times$ are taken from refs. \cite{Shifman:1978bx,Colangelo:2000dp,Khodjamirian:2003xk},
             with the unertainties estimated from the uncertainty of the corresponding
             decay constant. Intervals marked with $\dagger$ represent the union of intervals for the individual
             form factors obtained in our analyses.
             For the deacay constants we use values given in
             refs. \cite{Aoki:2016frl,Follana:2007uv,Bazavov:2010hj,Arthur:2012yc,Straub:2015ica,Dowdall:2013rya,
             Carrasco:2014poa,Bazavov:2014wgs,Bazavov:2017lyh,Gelhausen:2013wia}. The Borel parameters are the same
             as the ones used in \cite{Khodjamirian:2006st,Faller:2008tr}.
    }
\end{table}

\textbf{Borel Parameters}~The Borel parameters $M^2_P$ and $M^2_V$ for the pseudoscalar and vector
final states are taken from previous studies
\cite{Khodjamirian:2005ea,Khodjamirian:2006st,Faller:2008tr}. As usual in QCD sum rules,
a window should be chosen for the Borel parameter $M^2$ such that:
\begin{enumerate}
    \item[a)] $M^2$ is not too large, to ensure that excited and continum state contributions to the correlation function
    are exponentially suppressed; and
    \item[b)] $M^2$ is not too small, to ensure that the impact of higher-twist contributions are suppressed by powers of $1/M^2$.
\end{enumerate}
We explicitly confirm that the central values of the form factors for $B\to K^*,\bar{D}^{(*)}$ transitions exhibit a
plateau in their $M^2$ dependence. For the remaining form factors we find no such plateau, which, however, does not preclude
us from applying the sum rules with some increased systematic uncertainties.
Based on the variations of the form factors under change of the Borel parameters, we assign a systematic uncertainty as a percentage
of the central value as follows:
\begin{equation}
\begin{aligned}
    B\to & \pi  : 15\%\,, &
    B\to & \rho : 12\%\,, \\
    B\to & K    : 8\%\,, &
    B\to & K^*  : 5\%\,,  \\
    B\to & \bar{D}^{(*)}
                : 3\%\,.
\end{aligned}
\end{equation}
For $\pi$, $K$ and $\rho$ final states the systematic uncertainties can be further reduced through
a simultaneous analyses of the form factors and the light-meson decay constants within the framework of
QCD sum rules, since both analyses have the Borel parameters and the thresholds in common. This effect
has been previously shown in the case of LCSRs with $\pi$ LCDAs~\cite{Imsong:2014oqa}.
For $K^*$ and $\bar{D}^{(*)}$ final states the uncertainty arising from the variation
of the Borel parameter can be included in the statistical procedure. Given the present
knowledge of the $B$-meson LCDA parameter(s), these uncertainties are presently subleading
to the parametric uncertainties due to thresholds and LCDA parameters.
We leave both of these improvements to future work.\\

\textbf{Power corrections}~Using the full set of LCDAs up to twist-four accuracy, the authors of
ref.~\cite{Braun:2017liq} expect to account for the contributions of HQET operators up to and including $1/m_b$ corrections.
This expectation is based on the observation that an increase by two units of collinear twist corresponds to a suppression by
a factor of $1/m_b$ \cite{Braun:PrivateCommunications}.
Moreover, four-particle LCDAs also start to contribute at the twist-four level and
are presently unknown. Given the small size of the three-particle contributions to the sum rules we do not expect sizeable contributions
from the four-particle terms, which we ignore throughout.
The corrections at order $1/m_b^2$ are presently unknown, and we estimate them based on naive dimensional arguments at $\sim 5\%$.
We add this uncertainty in quadrature to the systematic uncertainty incurred by the Borel parameters.\\

\textbf{Duality Threshold Parameters}~The threshold parameters $s_0^{(F)}$ can in principle be determined by closely following
the procedure carried out in ref. \cite{Imsong:2014oqa}.
First, one defines a prior interval with uniform probability for the threshold parameters. In this step one also varies the
LCDA parameters $1 / \lambda_{B,+}$, $\lambda_E^2$ and $\lambda_H^2$ to determine the correlations between
thresholds and LCDA model parameters.
In a next step, the a-priori PDF is challenged with a theoretical likelihood. The pseudo-observables that
are constrained through the likelihood are the ``first moments'' of the form factors' correlation function.
These moments warrant a more careful definition: for any form factor $F$ we differentiate its scalar-valued
correlator $\Pi^F(q^2; M^2)$ with respect to $-1/M^2$ and normalize it to $\Pi^F$. The resulting ratio
is a pseudo-observable that is expected to yield the final state's mass square $m_P^2$ or $m_V^2$,
respectively, within the accuracy of the light-cone OPE for the correlation function.

We carry out this procedure for the $K^*$ and $\bar{D}^{(*)}$ final states.
Within the likelihood, we impose that the theory prediction for the first moments match the
square of the respective final state hadron mass. We impose relative uncertainties of $5\%$ on these predictions, in order
to account for the impact of $1/m_b^2$ corrections to the correlators. The added uncertainties are considerably
larger than in the $B\to \pi$ analysis \cite{Imsong:2014oqa}. We think our more conservative treatment is warranted
as we expect the ``first moments'' to exhibit a substantial but difficult-to-quantify dependence on the $B$-meson
LCDA model.
For some of the threshold parameters we find a marked non-gaussianity for
the two-dimensional joint posterior PDF of a single threshold parameter and $1 / \lambda_{B,+}$.

For pseudo-Goldstone bosons such as the $\pi$ and $K$, and for the $\rho$ meson with its substantial decay width,
the first moments are not expected to reproduce the meson mass squares. As an exercise, we attempt anyway to apply the
procedure described above and find it to be too unstable to determine the duality threshold for any of these states.
We therefore adopt the thresholds used in ref.~\cite{Khodjamirian:2006st},
which are determined ($\lambda_{B,+}$ independently) from two-point QCD sum rules of the $\pi$, $K$ and $\rho$
decay constants.

In our analysis of form factors to $K^*$ and $\bar{D}^{(*)}$ final states a further complication arises from the fact
that the first moments of the correlation functions exhibit a noticable but mild $q^2$ dependence.
We choose to study this effect as follows: for each form factor, the theory
likelihood includes the form factor's first moment for seven values of $q^2$ in the range $-15\,\GeV^2$
to $0\,\GeV^2$, with increments of $2.5\,\GeV^2$. We make a linear ansatz for the $q^2$ dependence of the
threshold parameters $s_0^{(F)}$:
\begin{equation}
    s_0^{(F)}(q^2) = s_0^{(F)} + q^2 s_0^{\prime,(F)}\,.
\end{equation}
We then determine the two parameters $s_0^{(F)}$ and $s_0^{\prime,(F)}$ for each form factor from the theory likelihood.
Subsequently we repeat the fit while fixing the slope parameters $s_0^{\prime,(F)}$ to zero. For most of the form
factors we find a negligible difference in the constant parts $s_0^{(F)}$. The only exception is
the form factor $\fT{D}$, for which the two parameters $s_0$ and $s^\prime_0$ are very strongly linearly correlated.
We can therefore not reliably obtain the threshold parameter for this form factor, and choose to use
the same threshold as for $\fp{D}$, which is a good approximation for other vector/tensor pairs of form factors
and holds at the $3\%$ level for e.g. the pair $\V{D^*}$, $\Tone{D^*}$. Nevertheless, we increase the systematic
uncertainty on $\fT{D}$ by $5\%$ due to this treatment.
Considering the full set of form factors and final state hadrons,
we find only negligible impact due to our treatment of $q^2$ dependence of the threshold parameters when
comparing to the dominant uncertainties incurred by the $B$-meson LCDA parameters.
We therefore proceed with the assumption of $q^2$-independent duality thresholds. However,
we remark that this problem needs to be revisited once the parametric uncertainties due to the LCDA model-dependence
are under better control.\\

For the $\bar{D}$ and $\bar{D}^*$ final states, we find that increasing $q^2$ to positive values increases the uncertainty in the prediction of the first moments
substantially.
In fact, for $q^2 \simeq 5\,\GeV^2$ we find very broad intervals that include $s_0 = 0$ at $68\%$ probability.
This increase in uncertainty is accompanied by a substantial growth of relative contributions (to $\sim 50\%$ and
beyond) due to the higher-twist two-particle terms.
This clearly poses a problem for the calculation of the $B\to \bar{D}^{(*)}$ form factors at positive $q^2$. It remains to be seen
if this effect is due to the modelling of the LCDAs, or indicates an earlier-than-expected breakdown of the
Light-Cone OPE at positive $q^2$.\\

\begin{table}[tb]
    \renewcommand{\arraystretch}{1.25}
    \centering
    \resizebox{!}{.30\textheight}{%
    \input{tables/tab-ff-details.tex}
    }
    \caption{%
        Detailed budget of the $\phi_\pm$, $g_+$, $g_-^\text{WW}$ and three-particle contributions
        to our LCSR results for the form factors at $q^2 = 0$.
    }
    \label{tab:ff-details}
\end{table}

\textbf{Predictions}~Based on the procedure discussed above, we obtain threshold parameters for the individual
form factors. A summary of these parameters and their uncertainties are listed in \reftab{priors}. We then proceed
to produce posterior-predictive distributions for the form factors at five different $q^2$ points:
$q^2 = \lbrace -15, -10, -5, 0, +5\rbrace\,\GeV^2$. Note that the form factors $\Azero{V}$ and $\Ttwo{V}$ are linearly dependent
on the remaining form factors at $q^2 = 0$, and therefore this particular point is dropped from the predictions for
these two quantities. For heavy final states $M=D,D^*$ we remarked previously that the threshold computation becomes unstable
for $q^2 > 0$. We therefore drop the point $q^2 = +5\,\GeV^2$ for these two final states.
The resulting Probability Density Functions (PDFs) of the form factors at the various $q^2$
points are most readily communicated in form of machine readable files, containing the mean values and covariance matrices
of a multivariate Gaussian density. The results are included in the \EOS software~\cite{EOS-v0.2.3}
as of version v0.2.3 as YAML files, defining the following named constraints:
\begin{center}
\begin{verbatim}
B->pi::FormFactors[f_+,f_0,f_T]@GKvD2018
B->rho::FormFactors[V,A_0,A_1,A_2,T_1,T_2,T_23]@GKvD2018
B->K::FormFactors[f_+,f_0,f_T]@GKvD2018
B->K^*::FormFactors[V,A_0,A_1,A_2,T_1,T_2,T_23]@GKvD2018
B->D^(*)::FormFactors[f_+,f_0,f_T,V,A_0,A_1,A_2,T_1,T_2,T_23]@GKvD2018
\end{verbatim}
\end{center}
We provide a detailed budget of the individual contributions to the form factors at $q^2 = 0$ in
\reftab{ff-details}. We also compare our results and their uncertainties, including all sources
of systematic uncertainties, with results in the literature in \reftab{ff-comparison}.\\

Our numerical results can subsequently be used to fit concrete parametrizations of the respective
form factors. We carry out such fits for the BSZ parametrization \cite{Straub:2015ica} in the next
subsection.

\begin{table}[p]
    \renewcommand{\arraystretch}{1.25}
    \centering
    \resizebox{!}{.44\textheight}{%
    \input{tables/tab-ff-comparison.tex}
    }
    \caption{%
        Comparison of our LCSR results for the form factors at $q^2 = 0$ with previous results in
        the literature. Note that the $B\to D$ form
        factors from ref.~\cite{Faller:2008tr} have been obtained from a different interpolating
        current $\Jint{}$ than our results. Note that the $\fT{P}$, $\Tone{V}$, $\Ttwo{V}$
        and $\Ttwothree{V}$ are scale-dependent quantities, evaluated at $\mu^2 = 1\,\GeV^2$
        for $P=\pi,K$ and $V=\rho,K^*$, and at $\mu^2 = 4.5\,\GeV^2$ for $P=\bar{D}$ and $V=\bar{D}^*$.
    }
    \label{tab:ff-comparison}
\end{table}

\FloatBarrier

\subsection{Parametrization and fits to LCSR and lattice QCD constraints}
\label{sec:results:fits}

With our LCSR results in hand at selected $q^2$ values $\leq 5\,\GeV^2$, we proceed to extrapolate
the form factors to large positive $q^2$ values. This is most readily achieved using a $z$ expansion of
the form factors. We adopt the same parametrization as used in ref.~\cite{Straub:2015ica},
which also facilitates comparisons between the results therein and ours. The parametrization
of any form factor $F$ reads
\begin{equation}
    \label{eq:BSZ-param}
    F(q^2) \equiv \frac{1}{1 - q^2 / m_{R,F}^2}\, \sum_{k=0}^{2} \alpha_k^{(F)} \left[z(q^2) - z(0)\right]^k\,.
\end{equation}
Here $m_{R,F}$ denotes the mass of sub-threshold resonances compatible with the quantum numbers of the
form factor $F$, as listed in \reftab{resonances}. We also apply the conformal map from $q^2$ to $z$:
\begin{equation}
    z(t) \equiv \frac{\sqrt{t_+ - t} - \sqrt{t_+ - t_0}}{\sqrt{t_+ - t} + \sqrt{t_+ - t_0}}\,,
\end{equation}
where $t_\pm = (m_B \pm m_{P,V})^2$ and $t_0$ is a free parameter that governs the size of $z$ in
the semileptonic phase space. As in ref.~\cite{Straub:2015ica} we use
$t_0 \equiv t_+ \left(1 - \sqrt{1 - t_- / t_+}\right)$.

For each final state we perform two fits. The first fit includes only the information at small $q^2$ values,
obtained from the LCSRs, within the likelihood. Within all plots in \refapp{plots}, the results of this fit are displayed as
a dark gray band. For the second fit, we add further information from lattice QCD analyses of the
form factors at large values of $q^2$ as available~\cite{%
Bouchard:2013pna,%
Horgan:2013hoa,%
Bailey:2014tva,%
Horgan:2015vla,%
Na:2015kha,%
Lattice:2015tia,%
Harrison:2017fmw%
}. Due to the absence of
lattice QCD analyses of the $B\to \rho$ transitions there is no combined fit for the respective form factors.
Results arising from the second fit are displayed as blue bands, throughout this work.\\

For the LCSR-only fits we have four data points and three parameters per form factor, equivalent to one degree of freedom
(three data points and two parameters in the case of $\fz{P}$, $\Azero{V}$ and $\Ttwo{V}$). Given the large
uncertainties and small number of degrees of freedom, it is not surprising that we find a p value $\gg 3\%$,
our a-priori threshold, in each of these fits. For the combined fits to LCSR and LQCD inputs, we find p values
very close to one, indicating an excellent fit in each of these analyses.

As for the LCSRs, the posterior PDFs of our fits are most readily provided as machine readable files
containing the mean values and covariance matrices
of a multivariate Gaussian density. The results are included in the \EOS software~\cite{EOS-v0.2.3}
as of version v0.2.3 as YAML files, defining the following named constraints:
\begin{center}
\begin{verbatim}
B->pi::FormFactors[parametric,LCSR]@GKvD2018
B->pi::FormFactors[parametric,LCSRLattice]@GKvD2018
B->rho::FormFactors[parametric,LCSR]@GKvD2018
B->K::FormFactors[parametric,LCSR]@GKvD2018
B->K::FormFactors[parametric,LCSRLattice]@GKvD2018
B->K^*::FormFactors[parametric,LCSR]@GKvD2018
B->K^*::FormFactors[parametric,LCSRLattice]@GKvD2018
B->D^(*)::FormFactors[parametric,LCSR]@GKvD2018
B->D^(*)::FormFactors[parametric,LCSRLattice]@GKvD2018
\end{verbatim}
\end{center}
Moreover, we provide our results also through machine-readable JSON files in the same format as used in ref.~\cite{Straub:2015ica}.
These files are attached to the arXiv preprint of this article as ancillary files. Moreover, our results will be available by default
to users of the \flavio software~\cite{Straub:2018kue} from the next release on.

\begin{table}[t]
    \centering
    \renewcommand{\arraystretch}{1.25}
    \begin{tabular}{|c|c|c|c|c|}
        \hline
              &                                                        & \multicolumn{3}{c|}{resonance masses [$\GeV$]} \\
        $J^P$ & form factors                                           & $B_{u,d}(J^P)$ & $B_{s}(J^P)$ & $B_{c}(J^P)$ \\
        \hline
        $0^-$ & $\Azero{V}$                                            & 5.279          & 5.336        & 6.275 \\\hline
        $0^+$ & $\fz{P}$                                               & 5.540          & 5.630        & 6.420 \\\hline
        $1^-$ & $\fp{P}$, $\fT{P}$, $\V{V}$, $\Tone{V}$                & 5.325          & 5.412        & 6.330 \\\hline
        $1^+$ & $\Aone{V}$, $\Aonetwo{V}$, $\Ttwo{V}$, $\Ttwothree{V}$ & 5.724          & 5.829        & 6.767 \\\hline
    \end{tabular}
    \caption{%
    \label{tab:resonances}
        Overview of the lowest-lying resonances in the individual $b\to \lbrace u, d\rbrace$, $b\to s$
        and $b\to c$ transitions, and the association to the respective form factors. The masses
        above enter the parametrization of the form factors \refeq{BSZ-param} as the resonance
        mass parameter $m_{R,F}$. The $B_{u,d,s}$ masses have been taken from ref.~\cite{Straub:2015ica},
        to ensure interoperability of their and our results. The $B_c$ resonance masses have been
        taken from ref.~\cite{Detmold:2015aaa}.
    }
\end{table}

\section{Selected Phenomenological Implications}
\label{sec:pheno}

We will briefly discuss the impact of our results for the form factors on the present
$b$ anomalies.

\subsection{The $B\to K^*\mu^+\mu^-$ anomaly and $P'_5$}
\label{sec:pheno:btokstarmumu}

Rare semileptonic $b$ decays presently exhibit a number of measurements that
deviate indivdually by about $2\sigma$ from their respective Standard Model (SM)
predictions. These include all exclusive $b\to s\mu^+\mu^-$ branching ratios \cite{Aaij:2014pli,Aaij:2015esa,%
Aaij:2016flj} (with the exception of $\Lambda_b\to\Lambda\mu^+\mu^-$ \cite{Aaij:2015xza}); the full set of
angular observables in $B\to K^*\mu^+\mu^-$ \cite{Aaij:2015oid,Khachatryan:2015isa,Aaboud:2018krd}; and most
notably the Lepton Flavour Universality (LFU) ratios $R_{K}$ \cite{Aaij:2014ora} and $R_{K^*}$
\cite{Aaij:2017vbb}.\\

Several studies \cite{%
Descotes-Genon:2013wba,Altmannshofer:2013foa,Beaujean:2013soa,Hurth:2013ssa,Descotes-Genon:2015uva,Altmannshofer:2017fio,Altmannshofer:2017yso,Capdevila:2017bsm,Geng:2017svp,Ciuchini:2017mik,Mahmoudi:2018qsk%
} come to
the conclusion that a negative shift to the
short-distance coupling $C_9^\mu$, and potentially to some couplings that vanish in the SM, can
explain simultaneously the deviations in all anomalous $b\to s\ell^+\ell^-$ measurements;
see \cite{Albrecht:2018vsa} for a recent review and the definition of the low-energy Lagrangian.
In the case of $e$ vs $\mu$ universality with a lower dilepton mass cut $q^2 \leq 1\,\GeV$,
the SM predictions of the LFU ratios are insensitive to the hadronic form factors \cite{Hiller:2003js,Bobeth:2007dw,%
Bordone:2016gaq}. We will therefore not discuss them here any further. Instead,
we will discuss the qualitative impact of our results on fits of the $b\to s\ell^+\ell^-$
short-distance couplings to the available data on exclusive $B\to K^*\mu^+\mu^-$
decays, which have presently the biggest impact in global $b\to s\mu^+\mu^-$ fits.\\

\begin{figure}
    \centering
    \includegraphics[width=0.8\textwidth]{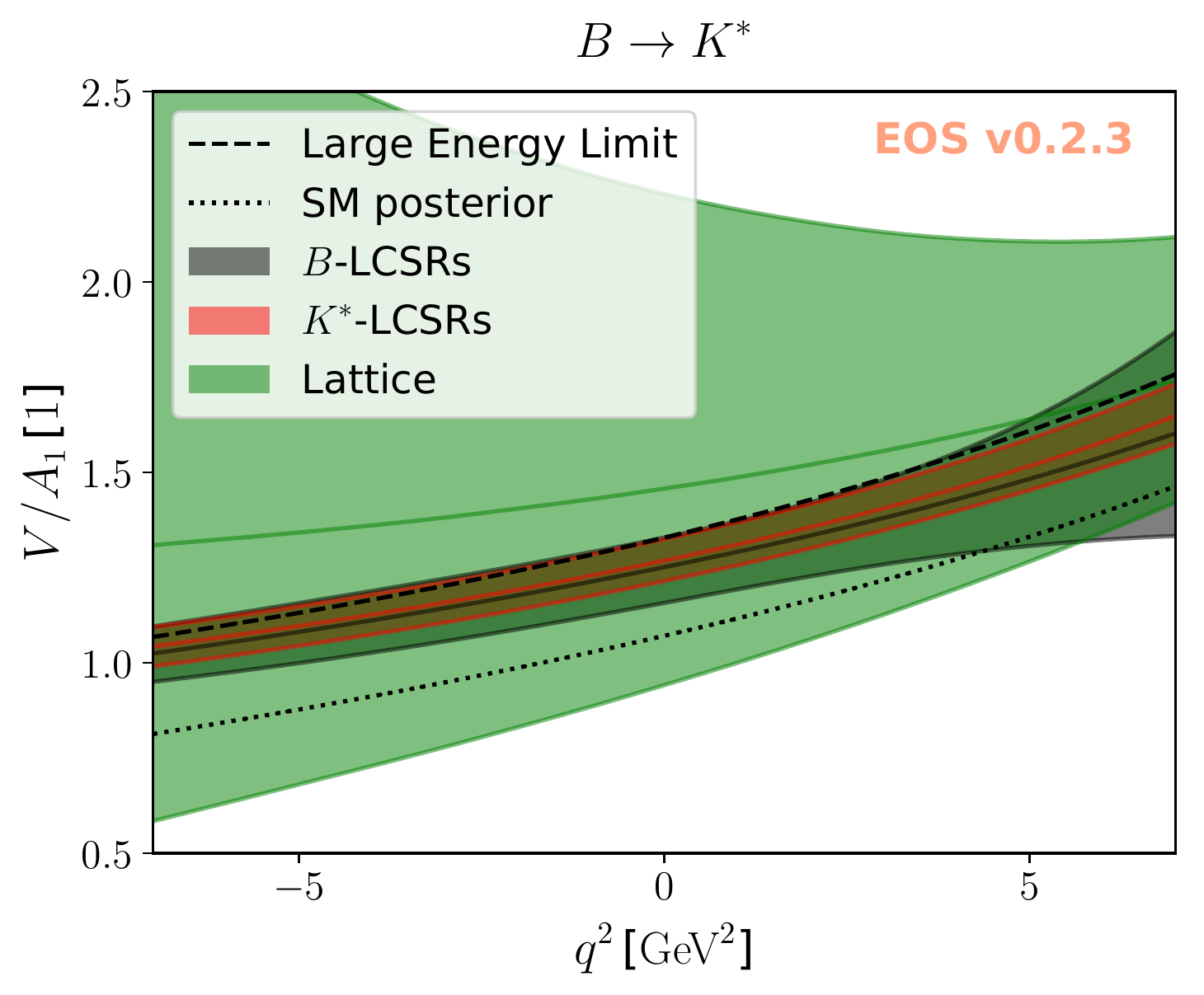}
    \caption{The ratio $V(q^2)/A_1(q^2)$ for $B\to K^*$ transitions.
    The green, red and gray lines and shaded areas correspond to the central
    values and the $68\%$ probability envelope of the form factors obtained
    from fits to only LQCD results, fits to LCSR results from ref.~\cite{Straub:2015ica},
    and fits to our LCSR results, respectively. The dashed line correspond to the large-energy
    limit for this ratio \cite{Charles:1998dr,Beneke:2000wa}. The dotted line corresponds to the
    central value of the SM fit to $B\to K^*\mu^+\mu^-$ data from ref.~\cite{Bobeth:2017vxj}.
    }
    \label{fig:ratio-VoA1}
\end{figure}

Assuming the global fits to correctly account for non-local effects arising from
four-quark operators in the $B\to K^*\mu^+\mu^-$ amplitudes~\cite{Bobeth:2017vxj},
the data leads to two possible
conclusions \cite{Beaujean:2013soa,Descotes-Genon:2015uva}:
\begin{enumerate}
    \item the ratio of form factors $\V{K^*} / \Aone{K^*}$ deviates from the ratio predicted
    by symmetry relations at large kaon energies \cite{Charles:1998dr,Beneke:2000wa} as well as a-priori
    predictions from extrapolations of lattice QCD result \cite{Horgan:2013pva} and light-meson LCSRs
    \cite{Straub:2015ica}, leading to global fits with border-line goodness of fit; or
    \item there is a New Physics (NP) shift to the short-distance coefficient $C_9$ corresponding
    to $\sim 25\%$ of its SM value.
\end{enumerate}
This interpretation has been strengthened recently by a proof-of-concept analysis in which the
non-local matrix elements are further constrained in shape due to their properties following
from analyticity and unitarity \cite{Bobeth:2017vxj}. The particular solution to obtaining a good
fit in the absence of NP effects requiress the ratio $\V{K^*} / \Aone{K^*}$ to not only deviate
in value from the large-energy limit prediction, but also in shape. We show explicitly in \reffig{ratio-VoA1}
that our predictions are compatible with the symmetry limit at large energies; with extrapolations
of lattice QCD results within their large uncertainties; and with the rather precise results obtained
from LCSRs with $K^*$ LCDAs.

\subsection{Standard Model Predicitions for $B\to D^{(*)}\ell\bar\nu$ and $R(D^{(*)})$}
\label{sec:pheno:RD(*)}

The exclusive semileptonic decays $B\to \bar{D}^{(*)}\ell\bar\nu$ are of great phenomenological interest.
One the one hand, they can be used to extract the magnitude of the
Cabibbo-Kobayashi-Maskawa (CKM) matrix element $V_{cb}$. Its determinations from exclusive and
inclusive $B$ decays has been famously in tension with each other for the last decade.
On the other hand, the exclusive decays allow to test the SM through LFU ratios
$R(D)$ and $R(D^*)$
\begin{equation}
    R(D^{(*)})
        \equiv \frac{\mathcal{B}(B\to \bar{D}^{(*)}\tau^-\bar\nu)}{\mathcal{B}(B\to \bar{D}^{(*)}\ell^-\bar\nu)}\,,\qquad \text{with $\ell=e,\mu$.}
\end{equation}
Both the extraction of $|V_{cb}|$ and testing the SM through LFU violation require accurate predictions
of the relevant form factors. The Heavy-Quark-Expansion, in combination with data, can help in this
particular case of heavy-to-heavy flavour-changing quark transitions; see ref.~\cite{Boyd:1995cf,Caprini:1997mu}
and references therein for dispersive bounds, and ref.~\cite{Fajfer:2012vx} for the SM prediction
of $R(D^*)$. It has been recently argued that strict adherence to the so-called CLN parametrization~\cite{Caprini:1997mu}
is, at least partially, responsible for the exclusive-vs-inclusive tension~\cite{%
Bernlochner:2017jka,Bigi:2017jbd,Jung:2018lfu} when determinig $V_{cb}$ from semileptonic $B\to \bar{D}^*$ transitions.
In the case of $B\to \bar{D}$, recent lattice QCD analyses yield $V_{cb}$ values that are compatible
with both the inclusive and the $B\to \bar{D}^*$ determinations.\\

LCSR determinations of $B\to \bar{D}^*$ form factors \cite{Faller:2008tr} play an important role in
some of the phenomenological analyses \cite{Bigi:2017jbd,Jung:2018lfu}, e.g. through form factor ratios at $q^2 = 0$. With our
updated results for the form factors, we are in the position to also update these ratios and also
to provide parametric correlations between them. The ratios under discussion are labelled $R_0$, $R_1$
and $R_2$ (see e.g. ref.~\cite{Caprini:1997mu} for their definitions), which are functions of the recoil parameter
$w$, with $m_B m_D^* w = p\cdot k$. At maximal recoil $w_\text{max}$, corresponding to $q^2 = 0$, one has:
\begin{equation}
\begin{aligned}
    R_0(q^2 \!=\! 0)
        & = \frac{A_0(q^2 = 0)}{A_1(q^2 = 0)}\,, &
    R_1(q^2 \!=\! 0)
        & = \frac{V(q^2 = 0)}{A_1(q^2 = 0)}\,, &
    R_2(q^2 \!=\! 0)
        & = \frac{A_2(q^2 = 0)}{A_1(q^2 = 0)}\,.
\end{aligned}
\end{equation}
Moreover, at $q^2 = 0$ the equation of motion implies that only two of these ratios are linearly independent.
Based on our correlated results for the form factors we obtain
\begin{equation}
\begin{aligned}
    R_0(q^2 \!=\! 0)
        & = 1.117 \pm 0.061\,, &
    R_1(q^2 \!=\! 0)
        & = 1.151 \pm 0.114\,, \\
    R_2(q^2 \!=\! 0)
        & = 0.856 \pm 0.076\,.
\end{aligned}
\end{equation}
The correlation coefficients $\rho$ between $R_1$ and $R_2$ reads
\begin{equation}
    \rho_{12}
        = 0.5154\,.
\end{equation}
Our correlated results are compatible with the previous LCSR determinations of $R_1(q^2 = 0)$ and $R_2(q^2 = 0)$
\cite{Faller:2008tr} at less than one standard deviation.\\

We can also use our results to calculate the values of the LFU observables
$R(D)$ and $R(D^*)$ in the SM and beyond. Using the correlated results for the form factor
parameters obtained in \refsec{results:fits} from the fit to only our LCSR results, we obtain:
\begin{align}
    ~
        & \text{LCSR only} &
    R(D)\bigg|_\text{SM}
        & = 0.269 \pm 0.100\,.  \\
    ~
        & \text{LCSR only} &
    R(D^*)\bigg|_\text{SM}
        & = 0.242 \pm 0.048\,.
\end{align}
Our prediction for $R(D^*)$ is the first theory prediction that does not use either symmetry arguments based on the simultaneous
expansion in $1/m_c$ and $1/m_b$ or experimental data from $\bar{B}\to D^*(e,\mu)\bar\nu$ decays.
Given the substantial uncertainties of our prediction, these values are in good agreement with the predictions obtained from
heavy quark symmetry relations, lattice inputs and $B\to \bar{D}^{(*)}\ell\bar\nu$ data
\cite{Bernlochner:2017jka,Bigi:2017jbd,Jung:2018lfu}.
Using the form factors parameters obtained in \refsec{results:fits} from a fit to both our LCSR results
and two LQCD inputs for the $B\to D^*$ form factor $\Aone{D^*}$ \cite{Bailey:2014tva,Na:2015kha,Harrison:2017fmw}, we obtain:
\begin{align}
    ~
        & \text{LCSR + Lattice} &
    R(D)\bigg|_\text{SM}
        & = 0.296 \pm 0.006\,, \\
    ~
        & \text{LCSR + Lattice} &
    R(D^*)\bigg|_\text{SM}
        & = 0.256 \pm 0.020\,.
\end{align}
Our result for $R(D)$ is dominated by the precise LQCD inputs \cite{Na:2015kha} beyond zero recoil, and the agreement
with the LQCD prediction $R(D) = 0.300\pm 0.008$ is therefore not surprising. Our result for $R(D^*)$, on
the other hand, is supported by two LQCD inputs for the $\Aone{D^*}$ form factor at only the zero recoil point.
We find excellent agreement with the values obtained using heavy quark symmetry relations, i.e.:
\begin{align}
    R(D^*)\bigg|_\text{SM,\cite{Bernlochner:2017jka}}
        & = 0.257 \pm 0.003\,, \\
    R(D^*)\bigg|_\text{SM,\cite{Bigi:2017jbd}}
        & = 0.260 \pm 0.008\,.
\end{align}
As a closing remark, we wish to emphasize that the predictions in the framework of heavy quark symmetry relations
are complicated by the proliferation of matrix elements associated with $1/m_c$ corrections that are not present
in our LCSR-derived results.

\section{Summary and Outlook}
\label{sec:conclusion}

We have presented a comprehensive update of Light-Cone Sum Rules (LCSRs) results for the full set of form factors relevant to semileptonic $B$ decays.
Our update includes, for the first time, a consistent treatment of \emph{all} two-particle and three-particle Light-Cone
Distribution Amplitudes (LCDAs) up to twist four. Moreover, our work also updates the numerical inputs across the board.\\

We have implemented our analytical results agnostic of the concrete expressions for the distribution amplitudes, thereby ensuring
that our analysis can be readily repeated once our knowledge of either the properties of the amplitudes or their parameters
improves in the future. The relevant computer code is publicly available \cite{EOS-v0.2.3} under an open source license as part of
the \EOS software \cite{EOS}.
Moreover, all of our numerical results are available as machine-readable files. For form factors that are common to our
and a previous LCSR analysis using light-meson LCDAs \cite{Straub:2015ica} we have ensured interoperability of the ancillary
data files attached to this preprint.\\

Within our analyses we find sizable contributions from two-particle states at the twist-four level, which exceed the
twist-three and twist-four three-particle contributions by one order of magnitude. Our analysis has been carried out strictly
in the framework of Heavy-Quark Effective Theory, which enables us to be agnostic of the final state quark flavour, thereby
facilitating the analysis. However, it also precludes us from using the $\order{\alpha_s}$ corrections to the leading-twist
two-particle results obtained in the framework of SCET Sum Rules for massless \cite{Wang:2015vgv} and massive \cite{Wang:2017jow}
pseudoscalar final states. The logical next step is
therefore to extend our present framework with these radiative corrections, and to check if the combined twist and $\alpha_s$
expansion of the non-local operators in the light-cone OPE is well behaved. Particularly, we wonder if the instability of
inferring the duality thresholds for $\pi$, $K$ and $\rho$ final states can be overcome by including the radiative corrections, or
by including contributions at the twist-five and twist-six levels; see ref.~\cite{Lu:2018cfc} for recent efforts in the latter
direction.\\

Finally, we have selected two phenomenological applications connected to the present $b$ anomalies to highlight the usefulness
of our results. Our finding weaken the interpretation of the $B\to K^*\mu^+\mu^-$ angular anomalies as effects of our lack of
knowledge of hadronic form factors. Furthermore, we have updated the form factor ratios $R_1$ and $R_2$, relevant for
$V_{cb}$ extractions and predictions of the LFU ratio $R(D^*)$. Our results permit for the first time to account for
correlations among the relevant $B\to \bar{D}^*$ form factors, and are in agreement with previous results at less than one
standard deviation.

\acknowledgments

This work is supported by the DFG within the Emmy Noether Programme under grant DY-130/1-1
and the DFG Collaborative Research Center 110 ``Symmetries and the Emergence of Structure in QCD''.
We are very grateful to S\'ebastien Descotes-Genon, Paolo Gambino and Javier Virto for helpful discussions,
and to Martin Jung, David Straub and Javier Virto for comments on the manuscript.
We are also grateful to the authors of \cite{DescotesGenon:2018xxx,DescotesGenon:2018yyy} for discussing
their results prior to publication.

\appendix

\section{$B$-meson Distribution Amplitudes}
\label{app:LCDAs}

In this appendix we collect formulas relevant to the parametrization in terms of momentum-space of $B$-LCDAs
of the non-local matrix elements in eqs.~(\ref{eq:correlatorHQET2pt}) and (\ref{eq:correlatorHQET3pt}).
The two-particle $B$-LCDAs are defined via
\begin{multline}
\bra{0} \bar{q}_{2}^{\alpha}(x) h_{v}^{\beta}(0) \ket{\bar{B}_{q_2}(v)} =\\
    -\frac{i f_B m_B}{4} \int^\infty_0 \text{d}\omega e^{-i\omega v\cdot x} \bigg\{
        (1 + \slashed{v}) \bigg[
            \phi_+(\omega) -g_+(\omega) \partial_\sigma \partial^\sigma
            \\+\left(\frac{\bar{\phi}(\omega)}{2}
            -\frac{\bar{g}(\omega)}{2} \partial_\sigma \partial^\sigma\right) 
        \gamma^\mu \partial_\mu
        \bigg] \gamma_5
    \bigg\}^{\beta\alpha},
    \label{eq:BLCDAs2pt}
\end{multline}
while, for the three-particle $B$-LCDAs, we have
\begin{align}
    \bra{0} &\bar{q}_2^\alpha(x) G_{\mu\nu}(u x) h_{v}^\beta(0) \ket{\bar{B}_{q_2}(v)}= \nonumber \\
    = & \frac{f_B m_B}{4} \int_{0}^{\infty}\text{d}\omega_1 \int_{0}^{\infty}\text{d}\omega_2 e^{-i(\omega_1 + u \omega_2)\, v\cdot x} \bigg\{(1+\slashed{v})\bigg[
        (v_\mu\gamma_\nu-v_\nu\gamma_\mu)[\psi_A-\psi_V] 
        -i\sigma_{\mu\nu}\psi_V  \nonumber\\
    & + (\partial_\mu v_\nu-\partial_\nu v_\mu) \overbar{X}_A
      - (\partial_\mu\gamma_\nu-\partial_\nu\gamma_\mu) [\overbar{W} + \overbar{Y}_A]
      + i\epsilon_{\mu\nu\alpha\beta}\partial^\alpha v^\beta\gamma_5 \overbar{\tilde{X}}_A \nonumber\\
    & - i\epsilon_{\mu\nu\alpha\beta}\partial^\alpha \gamma^\beta\gamma_5 \overbar{\tilde{Y}}_A
      - u (\partial_\mu v_\nu-\partial_\nu v_\mu)\slashed{\partial} \overbar{\overbar{W}}
      + u (\partial_\mu \gamma_\nu-\partial_\nu \gamma_\mu)\slashed{\partial} \overbar{\overbar{Z}}
    \bigg]\gamma_5\bigg\}^{\beta\alpha}(\omega_1,\omega_2)\,,
    \label{eq:BLCDAs3pt}
\end{align}
where a gauge link is implied in the above, and the derivatives are abbreviated as $\partial_\mu \equiv \partial / \partial l^\mu$,is the momentum-space representation
where $l^\mu=\omega v^\mu$ in the two-particle case and $l^\mu=(\omega_1 + u \omega_2)v^\mu$ in the three-particle case.
Throughout, these derivatives are understood to act on the hard-scattering kernel.
In addition, we define the following shorthand notation:
\begin{equation}
\label{eq:def:barred-LCDAs}
\begin{aligned}
    \bar{\phi}(\omega) & \equiv \int_0^{\omega} \text{d}\eta\, \left(\phi_+(\eta) - \phi_-(\eta)\right)\,,\\
    \bar{g}(\omega) & \equiv \int_0^{\omega} \text{d}\eta\, \left(g_+(\eta) - g_-(\eta)\right)\,,\\
    \overbar{\psi}_{\text{3p}}(\omega_1, \omega_2)
        & \equiv \int_0^{\omega_1} \mathrm{d}\eta_1 \, \psi_{\text{3p}}(\eta_1, \omega_2)\,,\\
    \overbar{\overbar{\psi}}_{\text{3p}}(\omega_1, \omega_2)
        & \equiv \int_0^{\omega_1} \mathrm{d}\eta_1 \, \int_0^{\omega_2} \mathrm{d}\eta_2 \, \psi_{\text{3p}}(\eta_1, \eta_2)\,,
\end{aligned}
\end{equation}
where $\psi_\text{3p}$ represents any of the three-particle LCDAs.
We use $\epsilon_{0123} = +1$ in both the definition of the form factors and
in \refeq{BLCDAs3pt}, which matches the conventions of refs.~\cite{Ball:2004rg,Straub:2015ica}.
Accounting for the different convention, we reproduce the results of refs.~\cite{Khodjamirian:2006st,Faller:2008tr}.
The ``traditional'' basis of three-particle LCDAs can related to a basis of LCDAs with definite twist as follows \cite{Braun:2017liq}:
\begin{equation}
\begin{aligned}
    \phi_3(\omega_1, \omega_2)         & =[ \psi_A - \psi_V                                     ](\omega_1, \omega_2) \,,\\
    \phi_4(\omega_1, \omega_2)         & =[ \psi_A + \psi_V                                     ](\omega_1, \omega_2) \,,\\
    \psi_4(\omega_1, \omega_2)         & =[ \psi_A + X_A                                        ](\omega_1, \omega_2) \,,\\
    \chi_4(\omega_1, \omega_2)         & =[ \psi_V -\tilde{X}_A                                 ](\omega_1, \omega_2) \,, \\
    \tilde{\phi}_5(\omega_1, \omega_2) & =[ \psi_A + \psi_V  + 2 Y_A - 2 \tilde{Y}_A + 2 W      ](\omega_1, \omega_2) \,, \\
    \psi_5(\omega_1, \omega_2)         & =[ -\psi_A + X_A  - 2 Y_A                              ](\omega_1, \omega_2) \,, \\
    \chi_5(\omega_1, \omega_2)         & =[ -\psi_V - \tilde{X}_A  +2  \tilde{Y}_A              ](\omega_1, \omega_2)  \,, \\
    \phi_6(\omega_1, \omega_2)         & =[ \psi_A - \psi_V  + 2 Y_A + 2 \tilde{Y}_A + 2 W - 4 Z](\omega_1, \omega_2) \,.
\end{aligned}
\end{equation}
Note that we adopt the same nomenclature for the LCDAs as in ref.~\cite{Braun:2017liq}, except for renaming $\tilde{\psi}_{4,5} \to
\chi_{4,5}$ such that our notation involving barred LCDAs (see \refeq{def:barred-LCDAs}) becomes more legible.
It is possible to invert these relation. We obtain:
\begin{equation}
\label{eq:basis-Braun}
\begin{aligned}
    \psi_A      & = \frac{1}{2}\, [ \phi_3 + \phi_4                                             ](\omega_1, \omega_2) \,, \\
    \psi_V      & = \frac{1}{2}\, [-\phi_3 + \phi_4                                             ](\omega_1, \omega_2) \,, \\
    X_A         & = \frac{1}{2}\, [-\phi_3 - \phi_4 + 2\psi_4                                   ](\omega_1, \omega_2) \,, \\
    Y_A         & = \frac{1}{2}\, [-\phi_3 - \phi_4 +  \psi_4 - \psi_5                          ](\omega_1, \omega_2) \,, \\
    \tilde{X}_A & = \frac{1}{2}\, [-\phi_3 + \phi_4 - 2\chi_4                                   ](\omega_1, \omega_2) \,, \\
    \tilde{Y}_A & = \frac{1}{2}\, [-\phi_3 + \phi_4 -  \chi_4 + \chi_5                          ](\omega_1, \omega_2) \,, \\
    W           & = \frac{1}{2}\, [ \phi_4 - \psi_4 -  \chi_4 + \tilde{\phi}_5 + \psi_5 + \chi_5](\omega_1, \omega_2) \,, \\
    Z           & = \frac{1}{4}\, [-\phi_3 + \phi_4 - 2\chi_4 + \tilde{\phi}_5 + 2 \chi_5 - \phi_6](\omega_1, \omega_2) \,.
\end{aligned}
\end{equation}
A parametrization of the set of three-particle LCDAs at the twist-five and twist-six level has been recently suggested \cite{Lu:2018cfc}.
This set includes three twist-five LCDAs and one twist-six LCDA. However, to obtain the full set of three-particle LCDAs one has to expand
the position-space non-local matrix elements around the light-cone $x^2 \simeq 0$ in a consistent manner. Including the terms $\propto
x^2$ for the structures multiplied by $\phi_3$, $\phi_4$, $\psi_4$ and $\chi_4$, the full set of momentum-space matrix elements at the twist-six
level can be obtained from \refeq{BLCDAs3pt} by using \refeq{basis-Braun} in combination with the replacements
\begin{equation}
\begin{aligned}
    \phi_3 & \mapsto \phi_3 - g_{\phi_3}\,\partial_\sigma\,\partial^\sigma\,, \\
    \phi_4 & \mapsto \phi_4 - g_{\phi_4}\,\partial_\sigma\,\partial^\sigma\,, \\
    \psi_4 & \mapsto \psi_4 - g_{\psi_4}\,\partial_\sigma\,\partial^\sigma\,, \\
    \chi_4 & \mapsto \chi_4 - g_{\chi_4}\,\partial_\sigma\,\partial^\sigma\,.
\end{aligned}
\end{equation}
The twist of the new $g_{\psi_\text{3p}}$ functions corresponds
to the twist of their partner $\psi_\text{3p}$ plus two units of twist. Up to the twist-six level we therefore find twelve independent three-particle
LCDAs: one at twist three, three at twist four, four at twist-five, and further four at twist-six; in variance with the ansatz
of ref.~\cite{Lu:2018cfc}. Our argument here is in full analogy to the approach to the off-the-light-cone contributions for
two-particle LCDAs in form of $g_+$ and $g_-$ introduced in ref.~\cite{Braun:2017liq}.\\

In order to evaluate numerically the form factors, we use the exponential models proposed in ref.~\cite{Grozin:1996pq} and adapted in ref.~\cite{Braun:2017liq} to the LCDAs $\phi_+$, $\phi_-$, $g_+$, $\phi_3$, $\phi_4$, $\psi_4$ and $\chi_4$. Since $g_-$ receives contributions
from the three-particle DA $\psi_5$, for which no model is given in ref.~\cite{Braun:2017liq}, we approximate $g_-$
 in the Wandzura-Wilczek (WW) limit. We use
\begin{align}
\label{eq:gpmWW}g_-^{WW}(\omega)
        & =  \frac{1}{4} \int_0^{\omega} \mathrm{d}\eta_2 \, \int_0^{\eta_2} \mathrm{d}\eta_1 \, \left[ \phi_+ (\eta_1) - \phi_-^{WW} (\eta_1) \right]
        - \frac{1}{2} \int_0^{\omega} \mathrm{d}\eta_1 \, (\eta_1 - \bar{\Lambda}) \phi_-^{WW} (\eta_1) \\*
        & =  \frac{3 \omega}{4}\,  e^{-\omega/\lambda_{B,+}} \,,
\end{align}
where in the second line we use the Grozin-Neubert relation $2 \lambda_{B,+} = 4 \bar{\Lambda} /3$.

%%%%%%%%%%%%%%%%%%%%
\section{Coefficients of the LCSR formula}
\label{app:coefficients}

Here we list all the coefficients of \refeq{masterformula}. The normalization factors are:
\begin{equation}
\begin{aligned}
    &K^{(\fp{P})}=K^{(\fpm{P})}=f_P,&
    &K^{(\fT{P})}=\frac{f_P(m^2_B-m^2_P-q^2)}{m_B(m_B+m_P)},&\\
    &K^{( \V{V})}=\frac{2f_V m_V}{m_B(m_B+m_V)},&
    &K^{( \Aone{V})}=\frac{2f_V m_V (m_B+m_V)}{m_B^2},&        \\
    &K^{(\Atwo{V} )}=\frac{2f_V m_V}{m_B+m_V},&
    &K^{(\Athreezero{V} )}=\frac{4f_V m_V^2}{q^2},&              \\
    &K^{(\Tone{V} )}=K^{(\TtwothreeA{V} )}=K^{(\TtwothreeB{V} )}=\frac{2f_V m_V}{m_B}.&&&
\end{aligned}
\end{equation}
In the next subsections we give the $C^{(F,\psi)}_n$ coefficients of eqs. (\ref{eq:CoeffFuncs2pt}) and (\ref{eq:CoeffFuncs3pt}).
For all the form factors, the following relations hold among the three-particle contributions:
\begin{align}
&C^{(F,\overbar{\overbar{\psi}}_4)}_n     = -C^{(F,\overbar{\overbar{\phi}}_3)}_n -C^{(F,\overbar{\overbar{\phi}}_4)}_n \,, &&
&C^{(F,\overbar{\overbar{\chi}}_4)}_n     =  C^{(F,\overbar{\overbar{\phi}}_3)}_n -C^{(F,\overbar{\overbar{\phi}}_4)}_n \,.
\end{align}

%%%%%%
\subsection{$B\to P$}
\subsubsection{Two-particle Contributions}

The coefficients of \refeq{CoeffFuncs2pt}, for the two-particle DAs, are listed in the following.
For $\fp{P}$ we find the non-vanishing coefficients:
\begin{equation}
\begin{aligned}
    C^{(\fp{P},\phi_+)}_1     & = -\bar{\sigma}\,, &  \\
    C^{(\fp{P},\bar{\phi})}_2 & = -m_B\bar{\sigma}^2\,, \\
    C^{(\fp{P},g_+)}_2        & = -4\bar{\sigma}, &
    C^{(\fp{P},g_+)}_3        & = 8 m_1^2 \bar{\sigma}\,, \\
    C^{(\fp{P},\bar{g})}_3    & = -8 m_B \bar{\sigma}^2\,, &
    C^{(\fp{P},\bar{g})}_4    & = 24 m_1^2 m_B\bar{\sigma}^2\,.
\end{aligned}
\end{equation}
For $\fpm{P}$ we find:
\begin{equation}
    \begin{aligned}
    C^{(\fpm{P},\phi_+)}_1     & = 2\sigma-1\,,                             & \\
    C^{(\fpm{P},\bar{\phi})}_2 & = 2 m_B\sigma\bar{\sigma}-m_1\,,           & \\
    C^{(\fpm{P},g_+)}_2        & = 4(2\sigma-1)\,,                           &
    C^{(\fpm{P},g_+)}_3        & = -8 m_1^2 (2\sigma-1)\,,                     \\
    C^{(\fpm{P},\bar{g})}_3    & = 16 m_B \sigma\bar{\sigma}\,,             &
    C^{(\fpm{P},\bar{g})}_4    & = 24 m_1^2(m_1-2 m_B \sigma\bar{\sigma})\,.
\end{aligned}
\end{equation}
For $\fT{P}$ we find:
\begin{equation}
    \begin{aligned}
    &C^{(\fT{P},\bar{\phi})}_1= \frac{1}{m_B},&
    &C^{(\fT{P},\bar{\phi})}_2= \frac{-(m_B^2 \bar{\sigma}^2-m_1^2+2q^2\sigma-q^2)}{m_B},& \\
    &C^{(\fT{P},\bar{g})}_2= \frac{8}{m_B},&
    &C^{(\fT{P},\bar{g})}_3= \frac{-8(m_B^2 \bar{\sigma}^2+2 m_1^2+2q^2\sigma-q^2)}{m_B},&\\
    &C^{(\fT{P},\bar{g})}_4= \frac{24 m_1^2(m_B^2 \bar{\sigma}^2-m_1^2+2q^2\sigma-q^2)}{m_B}.&
    \end{aligned}
\end{equation}

%%%%%%%%%%%
\subsubsection{Three-particle Contributions}

The coefficients of \refeq{CoeffFuncs3pt}, for the three-particle DAs, for $\fp{P}$ follow. For $\phi_3$:
\begin{equation}
\begin{aligned}
C^{(\fp{P},\phi_3)}_2                        & =  -\frac{2m_1}{m_B}- u \bar{\sigma}                                       \,, & \\
C^{(\fp{P},\overbar{\phi}_3)}_2              & = \frac{u}{m_B}                                                            \,, &
C^{(\fp{P},\overbar{\phi}_3)}_3              & =  -\frac{2}{m_B}(u(m_B^2\bar{\sigma}^2+q^2)+4m_B m_1 \bar{\sigma}+um_1^2) \,,   \\
                                               &                                                                                &
C^{(\fp{P},\overbar{\overbar{\phi}}_3)}_4    & = -6m_1 \bar{\sigma}(2m_B\bar{\sigma}+m_1(2u-1))                           \,.
\end{aligned}
\end{equation}
For $\phi_4$:
\begin{equation}
\begin{aligned}
C^{(\fp{P},\phi_4)}_2                        & =  \bar{\sigma}(1-u)                                                      \,, & \\
C^{(\fp{P},\overbar{\phi}_4)}_2              & =  \frac{u-1}{m_B}                                                        \,, & \\
C^{(\fp{P},\overbar{\phi}_4)}_3              & =  2 um_B\bar{\sigma}^2+4 m_1\bar{\sigma}+2\frac{(1-u)(m_1^2+q^2)}{m_B}   \,,   \\
C^{(\fp{P},\overbar{\overbar{\phi}}_4)}_3    & = \frac{2}{m_B}(m_B \bar{\sigma}(2u-1)+2m_1)                              \,, &\\
C^{(\fp{P},\overbar{\overbar{\phi}}_4)}_4    & = \frac{6}{m_B}(m_B^2\bar{\sigma}^2-q^2)(m_B\bar{\sigma}(2u-1)+2m_1)      &\,.
\end{aligned}
\end{equation}
For $\psi_4$:
\begin{equation}
\begin{aligned}
&C^{(\fp{P},\overbar{\psi}_4)}_2     =     \frac{1-2u}{m_B}                                    \,, &\;\;\;&
C^{(\fp{P},\overbar{\psi}_4)}_3     =     \frac{2}{m_B}(2u-1)(m_1^2-m_B^2\bar{\sigma}^2+q^2)  \,.&
\end{aligned}
\end{equation}
For $\chi_4$:
\begin{equation}
\begin{aligned}
C^{(\fp{P},\overbar{\chi}_4)}_2    & =  \frac{1}{m_B}                                                              \,, \\
C^{(\fp{P},\overbar{\chi}_4)}_3    & = -\frac{2}{m_B}{(m_B^2\bar{\sigma}^2(2u-1)+4m_Bm_1 \bar{\sigma}+m_1^2+q^2)}  \,.
\end{aligned}
\end{equation}
The coefficients of \refeq{CoeffFuncs3pt}, for the three-particle DAs, for $\fpm{P}$ follow. For $\phi_3$:
\begin{equation}
\begin{aligned}
C^{(\fpm{P},\phi_3)}_2                        = &  (3-2\bar{\sigma})u-\frac{4m_1}{m_B}                                   \,, & \\
C^{(\fpm{P},\overbar{\phi}_3)}_2              = &  2u\frac{(\bar{\sigma}-1)}{m_B\bar{\sigma}}                             \,, \\
C^{(\fpm{P},\overbar{\phi}_3)}_3              = & -\frac{2}{m_B\bar{\sigma}}(m_B^2\bar{\sigma}^2(2\bar{\sigma}-3)u  \\
& +(2\bar{\sigma}-1)(4m_Bm_1\bar{\sigma}+uq^2)+um_1^2(2\bar{\sigma}+1))        \,,   \\
C^{(\fpm{P},\overbar{\overbar{\phi}}_3)}_4     = &   -6m_1(4m_B(\bar{\sigma}-1)\bar{\sigma}+m_1(2\bar{\sigma}+1)(2u-1))     \,.
\end{aligned}
\end{equation}
For $\phi_4$:
\begin{equation}
\begin{aligned}
C^{(\fpm{P},\phi_4)}_2                        = &  (1-u)(2 \bar{\sigma}+1)    \,, & \\
C^{(\fpm{P},\overbar{\phi}_4)}_2              = & \frac{2}{m_B\bar{\sigma}}(\bar{\sigma}-1)(u-1)         \,, \\
C^{(\fpm{P},\overbar{\phi}_4)}_3              = &\frac{2}{m_B\bar{\sigma}}(m_B^2\bar{\sigma}^2(2\bar{\sigma}u-u-1)+m_Bm_1\bar{\sigma}(4\bar{\sigma}-1) \\
& +m_1^2(2\bar{\sigma}+1)(1-u)+q^2(2\bar{\sigma}-2\bar{\sigma}u+u-1))    \,,   \\
C^{(\fpm{P},\overbar{\overbar{\phi}}_4)}_3    = &    \frac{2}{m_B\bar{\sigma}}(2m_B(\bar{\sigma}-2)\bar{\sigma}(2u-1)+m_1(4\bar{\sigma}-3))         \,, \\
C^{(\fpm{P},\overbar{\overbar{\phi}}_4)}_4    = &        \frac{6}{m_B\bar{\sigma}}(m_1(4\bar{\sigma}-1)(m_B^2\bar{\sigma}^2-q^2)+2m_B(\bar{\sigma}-1)\bar{\sigma}(2u-1)(m_B^2\bar{\sigma}^2-q^2))  \\
&+\frac{6}{m_B\bar{\sigma}}(m_B m_1^2\bar{\sigma}(2u-1)-m_1^3)    \,.
\end{aligned}
\end{equation}
For $\psi_4$:
\begin{equation}
\begin{aligned}
C^{(\fpm{P},\overbar{\psi}_4)}_2    & =   \frac{2}{m_B\bar{\sigma}}(\bar{\sigma}-1)(1-2u)           \,, \\
C^{(\fpm{P},\overbar{\psi}_4)}_3    & =   \frac{2}{m_B\bar{\sigma}}((2\bar{\sigma}-1)(2u-1)(q^2-m_B^2\bar{\sigma}^2)-2m_Bm_1\bar{\sigma}+m_1^2(2\bar{\sigma}+1)(2u-1))    \,.
\end{aligned}
\end{equation}
For $\chi_4$:
\begin{equation}
\begin{aligned}
C^{(\fpm{P},\overbar{\chi}_4)}_2    = &  \frac{2}{m_B\bar{\sigma}}(\bar{\sigma}-1)               \,, \\
C^{(\fpm{P},\overbar{\chi}_4)}_3    = & -\frac{2}{m_B\bar{\sigma}}(m_B^2\bar{\sigma}^2(4u(\bar{\sigma}-1)-2\bar{\sigma}+1) \\
& +4m_Bm_1\bar{\sigma}(2\bar{\sigma}-1)+m_1^2(2\bar{\sigma}+1)+q^2(2\bar{\sigma}-1))            \,.
\end{aligned}
\end{equation}
The coefficients of \refeq{CoeffFuncs3pt}, for the three-particle DAs, for $\fT{P}$ follow. For $\phi_3$:
\begin{equation}
\begin{aligned}
C^{(\fT{P},\phi_3)}_1                        & =  \frac{2u}{m_B^2\bar{\sigma}} \,,  \\
C^{(\fT{P},\phi_3)}_2                        & =  -\frac{2u}{m_B^2\bar{\sigma}}(m_B^2\bar{\sigma}^2-m_1^2-2q^2\bar{\sigma}+q^2) \,,  \\
C^{(\fT{P},\overbar{\phi}_3)}_2              & = \frac{4}{m_B^2\bar{\sigma}}(m_B\bar{\sigma}u+m_1)   \,, \\
C^{(\fT{P},\overbar{\phi}_3)}_3              & = -\frac{4}{m_B^2\bar{\sigma}}(m_B^2\bar{\sigma}^2-m_1^2-2q^2\bar{\sigma}+q^2)(m_B\bar{\sigma}u+m_1)     \,,   \\
C^{(\fT{P},\overbar{\overbar{\phi}}_3)}_3    & = 12\frac{m_1}{m_B}     \,,\\
C^{(\fT{P},\overbar{\overbar{\phi}}_3)}_4    & = -12\frac{m_1}{m_B}(m_B^2 \bar{\sigma}^2-m_1^2-2q^2\bar{\sigma}+q^2)    \,.
\end{aligned}
\end{equation}
For $\phi_4$:
\begin{equation}
\begin{aligned}
C^{(\fT{P},\overbar{\phi}_4)}_2              & =   -\frac{2}{m_B}   \,, \\
C^{(\fT{P},\overbar{\phi}_4)}_3              & = \frac{2}{m_B}(m_B^2\bar{\sigma}^2-m_1^2-2q^2\bar{\sigma}+q^2)  \,,   \\
C^{(\fT{P},\overbar{\overbar{\phi}}_4)}_2    & =   -\frac{4}{m_B^2\bar{\sigma}}(2u-1)       \,, \\
C^{(\fT{P},\overbar{\overbar{\phi}}_4)}_3    & =   \frac{2}{m_B^2\bar{\sigma}}(2u-1)(m_1^2-m_B^2\bar{\sigma}^2+q^2(5-4\bar{\sigma}))       \,, \\
C^{(\fT{P},\overbar{\overbar{\phi}}_4)}_4    & =  \frac{6}{m_B^2\bar{\sigma}}(2u-1)(m_1^2-m_B^2\bar{\sigma}^2+q^2)(m_1^2-m_B^2\bar{\sigma}^2+q^2(2\bar{\sigma}-1))    \,.
\end{aligned}
\end{equation}
For $\psi_4$:
\begin{equation}
\begin{aligned}
C^{(\fT{P},\overbar{\psi}_4)}_2    & =  -\frac{4m_1}{m_B^2\bar{\sigma}} \,, \\
C^{(\fT{P},\overbar{\psi}_4)}_3    & =  \frac{4m_1}{m_B^2\bar{\sigma}} (m_B^2\bar{\sigma}^2-m_1^2-2q^2\bar{\sigma}+q^2)   \,.
\end{aligned}
\end{equation}
For $\chi_4$:
\begin{equation}
\begin{aligned}
C^{(\fT{P},\overbar{\chi}_4)}_2    & =  \frac{4}{m_B^2\bar{\sigma}}(m_B\bar{\sigma}u+m_1)     \,, \\
C^{(\fT{P},\overbar{\chi}_4)}_3    & =  -\frac{4}{m_B^2\bar{\sigma}}(m_B\bar{\sigma}u+m_1)(m_B^2\bar{\sigma}^2-m_1^2-2q^2\bar{\sigma}+q^2)   \,.
\end{aligned}
\end{equation}

%%%%%%
\subsection{$B\to V$}
\subsubsection{Two-particle Contributions}
The coefficients of \refeq{CoeffFuncs2pt}, for the two-particle DAs, are listed in the following. For $ \V{V}$ we find:
\begin{equation}
\begin{aligned}
C^{( \V{V},\phi_+)}_1     & =  -\frac{1}{m_B}  \,, &  \\
C^{( \V{V},\bar{\phi})}_2 & =   -\frac{m_1}{m_B} \,, \\
C^{( \V{V},g_+)}_2        & = -\frac{4}{m_B}  , &
C^{( \V{V},g_+)}_3        & =  \frac{8m_1^2}{m_B} \,, \\
C^{( \V{V},\bar{g})}_4    & =  \frac{24m_1^3}{m_B}\,.
\end{aligned}
\end{equation}
For $\Aone{V}$ we find:
\begin{equation}
\begin{aligned}
C^{(\Aone{V},\phi_+)}_1     & =  \frac{q^2-(m_B\bar{\sigma}+m_1)^2}{m_B^2\bar{\sigma}}  \,, &  \\
C^{(\Aone{V},\bar{\phi})}_1 & = -\frac{m_1}{m_B^2\bar{\sigma}}    ,&
C^{(\Aone{V},\bar{\phi})}_2& =\frac{m_1(q^2-(m_B\bar{\sigma}+m_1)^2)}{m_B^2\bar{\sigma}}  \,, \\
C^{(\Aone{V},g_+)}_1        & = -\frac{4}{m_B^2\bar{\sigma}}   , &
C^{(\Aone{V},g_+)}_2       & = \frac{4(q^2-m_B^2\bar{\sigma}^2+m_1^2)}{m_B^2\bar{\sigma}}    \,, \\
C^{(\Aone{V},g_+)}_3      & = \frac{8m_1^2((m_B\bar{\sigma}+m_1)^2-q^2)}{m_B^2\bar{\sigma}}  \,, \\
C^{(\Aone{V},\bar{g})}_2    & = -\frac{8}{m_B}  \,, &
C^{(\Aone{V},\bar{g})}_3    & = \frac{8m_1^2(2m_B\bar{\sigma}+3m_1)}{m_B^2\bar{\sigma}}  \,, \\
C^{(\Aone{V},\bar{g})}_4    & =\frac{24m_1^3((m_B\bar{\sigma}+m_1)^2-q^2)}{m_B^2\bar{\sigma}}   \,.
\end{aligned}
\end{equation}
For $\Atwo{V}$ we find:
\begin{equation}
\begin{aligned}
C^{(\Atwo{V},\phi_+)}_1     & =  2\sigma-1   \,, &  \\
C^{(\Atwo{V},\bar{\phi})}_2 & =  2m_B \sigma\bar{\sigma}-m_1  \,,\\
C^{(\Atwo{V},g_+)}_2        & =  4( 2\sigma-1) \,, &
C^{(\Atwo{V},g_+)}_3        & =  -8m_1^2( 2\sigma-1)   \,, \\
C^{(\Atwo{V},\bar{g})}_3    & =  16m_B \sigma\bar{\sigma} \,, &
C^{(\Atwo{V},\bar{g})}_4    & =  24m_1^2(m_1-2m_B\sigma\bar{\sigma})\,.
\end{aligned}
\end{equation}
For $\Athreezero{V}$ we find:
\begin{equation}
\begin{aligned}
C^{(\Athreezero{V},\phi_+)}_1     & =  2\sigma+1   \,, &  \\
C^{(\Athreezero{V},\bar{\phi})}_2 & =  m_1-2m_B \sigma(\sigma+1)  \,,\\
C^{(\Athreezero{V},g_+)}_2        & =  4( 2\sigma+1) \,, &
C^{(\Athreezero{V},g_+)}_3        & =  -8m_1^2( 2\sigma+1)   \,, \\
C^{(\Athreezero{V},\bar{g})}_3    & =  -16m_B \sigma(\sigma+1) \,, &
C^{(\Athreezero{V},\bar{g})}_4    & =  24m_1^2(2m_B\sigma(\sigma+1)-m_1)\,.
\end{aligned}
\end{equation}
For $\Tone{V}$ we find:
\begin{equation}
\begin{aligned}
C^{(\Tone{V},\phi_+)}_1     & = -\frac{(m_B \bar{\sigma}+m_1)}{m_B}    \,, &  \\
C^{(\Tone{V},\bar{\phi})}_2 & =   -m_1\frac{(m_B \bar{\sigma}+m_1)}{m_B}  \,,\\
C^{(\Tone{V},g_+)}_2        & = -4\bar{\sigma}   \,, &
C^{(\Tone{V},g_+)}_3        & =  8 m_1^2\frac{(m_B \bar{\sigma}+m_1)}{m_B}   \,, \\
C^{(\Tone{V},\bar{g})}_2    & = -\frac{4}{m_B}    \,, &
C^{(\Tone{V},\bar{g})}_3    & =  \frac{8m_1^2}{m_B}   \,,\\
C^{(\Tone{V},\bar{g})}_4    & =  24 m_1^3\frac{(m_B \bar{\sigma}+m_1)}{m_B}  \,. &
\end{aligned}
\end{equation}
For $\TtwothreeA{V}$ we find:
\begin{equation}
\begin{aligned}
C^{(\TtwothreeA{V},\phi_+)}_1     & = -\frac{(m_B \bar{\sigma}+m_1)}{m_B}    \,, &  \\
C^{(\TtwothreeA{V},\bar{\phi})}_2 & =  -\frac{(m_1(m_B \bar{\sigma}+m_1)-2q^2\sigma)}{m_B}     \,,\\
C^{(\TtwothreeA{V},g_+)}_2        & = -4\bar{\sigma }  \,, &
C^{(\TtwothreeA{V},g_+)}_3        & =   8 m_1^2\frac{(m_B \bar{\sigma}+m_1)}{m_B}   \,, \\
C^{(\TtwothreeA{V},\bar{g})}_2    & =   -\frac{4}{m_B}     \,, &
C^{(\TtwothreeA{V},\bar{g})}_3    & =     \frac{8(m_1^2+2q^2\sigma)}{m_B}  \,,\\
C^{(\TtwothreeA{V},\bar{g})}_4    & = 24 m_1^2\frac{(m_1(m_B \bar{\sigma}+m_1)-2q^2\sigma)}{m_B}   \,. &
\end{aligned}
\end{equation}
For $\TtwothreeB{V}$ we find:
\begin{equation}
\begin{aligned}
C^{(\TtwothreeB{V},\phi_+)}_1     & = \frac{(m_B {\sigma}-m_1)}{m_B}    \,, &  \\
C^{(\TtwothreeB{V},\bar{\phi})}_1 & = \frac{\sigma}{m_B \bar{\sigma}}    \,, &
C^{(\TtwothreeB{V},\bar{\phi})}_2 & =  \frac{m_B m_1 \sigma\bar{\sigma}-m_B^2\sigma\bar{\sigma}^2+(2\sigma-1)(m_1^2-q^2\sigma)}{m_B\bar{\sigma}}     \,,\\
C^{(\TtwothreeB{V},g_+)}_2        & = 4{\sigma }  \,, &
C^{(\TtwothreeB{V},g_+)}_3        & =   8 m_1^2\frac{(m_1-m_B \sigma)}{m_B}   \,, \\
C^{(\TtwothreeB{V},\bar{g})}_2    & =   4\frac{(3\sigma-1)}{m_B\bar{\sigma}}     \,, &
C^{(\TtwothreeB{V},\bar{g})}_3    & =    -8 \frac{\sigma(m_B^2 \bar{\sigma}^2+3m_1^2+q^2(2\sigma-1))-m_1^2}{m_B\bar{\sigma}}  \,,\\
\end{aligned}
\end{equation}
\begin{equation*}
\begin{aligned}
C^{(\TtwothreeB{V},\bar{g})}_4    & = \frac{24 m_1^2}{m_B\bar{\sigma}} {( m_B^2\sigma\bar{\sigma}^2-m_B m_1\sigma\bar{\sigma}+(2\sigma-1)(q^2\sigma-m_1^2 ))} \,.&\hspace{3.45cm}
\end{aligned}
\end{equation*}
Where $\TtwothreeA{V}$ and $\TtwothreeB{V}$ are defined in eqs. (\ref{eq:defTA}) and (\ref{eq:defTB}).

%%%
\subsubsection{Three-particle Contributions}
The coefficients of \refeq{CoeffFuncs3pt}, for the three-particle DAs, for $\V{V}$ follow. For $\phi_3$:
\begin{equation}
\begin{aligned}
C^{(\V{V},\phi_3)}_2                        & = \frac{u}{m_B}  \,,  \\
C^{(\V{V},\overbar{\phi}_3)}_2              & =  \frac{2u}{m_B^2\bar{\sigma}}  \,, \\
C^{(\V{V},\overbar{\phi}_3)}_3              & = \frac{2u}{m_B^2\bar{\sigma}}(m_B^2\bar{\sigma}^2+m_1^2-q^2)    \,,   \\
C^{(\V{V},\overbar{\overbar{\phi}}_3)}_4    & =   \frac{6m_1^2}{m_B}(2u-1)   \,.
\end{aligned}
\end{equation}
For $\phi_4$:
\begin{equation}
\begin{aligned}
C^{(\V{V},\phi_4)}_2                        & =  \frac{u-1}{m_B}   \,,  \\
C^{(\V{V},\overbar{\phi}_4)}_2              & =   2\frac{(u-1)}{m_B^2\bar{\sigma}}    \,, \\
C^{(\V{V},\overbar{\phi}_4)}_3              & = \frac{2}{m_B^2\bar{\sigma}}(u-1)(m_1^2-q^2)- \frac{2m_1}{m_B}+2\bar{\sigma}(u-1)  \,,   \\
C^{(\V{V},\overbar{\overbar{\phi}}_4)}_3    & = -\frac{6m_1}{m_B^2\bar{\sigma}}         \,, \\
C^{(\V{V},\overbar{\overbar{\phi}}_4)}_4    & =-\frac{6m_1}{m_B^2\bar{\sigma}} (m_B^2\bar{\sigma}^2+m_Bm_1\bar{\sigma}(1-2u)+m_1^2-q^2)        \,.
\end{aligned}
\end{equation}
For $\psi_4$:
\begin{equation}
\begin{aligned}
C^{(\V{V},\overbar{\psi}_4)}_2    & = \frac{2-4u}{m_B^2\bar{\sigma}}  \,, \\
C^{(\V{V},\overbar{\psi}_4)}_3    & =  \frac{2}{m_B^2\bar{\sigma}}((2u-1)(q^2-m_B^2\bar{\sigma}^2)+2m_Bm_1\bar{\sigma}+m_1^2(1-2u))    \,.
\end{aligned}
\end{equation}
For $\chi_4$:
\begin{equation}
\begin{aligned}
C^{(\V{V},\overbar{\chi}_4)}_2    & =  \frac{2}{m_B^2\bar{\sigma}}     \,, \\
C^{(\V{V},\overbar{\chi}_4)}_3    & =   \frac{2}{m_B^2\bar{\sigma}}(m_1^2-q^2)+2\bar{\sigma}  \,.
\end{aligned}
\end{equation}
The coefficients of \refeq{CoeffFuncs3pt}, for the three-particle DAs, for $\Aone{V}$ follow. For $\phi_3$:
\begin{equation}
\begin{aligned}
C^{(\Aone{V},\phi_3)}_1                        & = \frac{u}{m_B^2\bar{\sigma}}   \,,  \\
C^{(\Aone{V},\phi_3)}_2                        & = u \frac{(m_1^2-q^2)}{m_B^2\bar{\sigma}}+\frac{2m_1}{m_B}+u\bar{\sigma} \,,  \\
C^{(\Aone{V},\overbar{\phi}_3)}_1              & =   \frac{2u}{m_B^3\bar{\sigma}^2}   \,, \\
C^{(\Aone{V},\overbar{\phi}_3)}_2              & =\frac{1}{m_B^3\bar{\sigma}^2}(4m_Bm_1\bar{\sigma}-2 u m_B^2\bar{\sigma}^2+4m_1^2u-4q^2u)     \,, \\
C^{(\Aone{V},\overbar{\phi}_3)}_3              & =\frac{2}{m_B^3\bar{\sigma}^2}(m_B^2\bar{\sigma}^2+m_1^2-q^2)(m_B^2\bar{\sigma}^2u+2m_Bm_1\bar{\sigma}+m_1^2u-q^2u)      \,,   \\
C^{(\Aone{V},\overbar{\overbar{\phi}}_3)}_3    & =  \frac{6m_1}{m_B^2\bar{\sigma}}(2m_B\bar{\sigma}+m_1(2u-1))     \,.\\
C^{(\Aone{V},\overbar{\overbar{\phi}}_3)}_4    & = \frac{6m_1^2}{m_B^2\bar{\sigma}}(m_B^2\bar{\sigma}^2(2u-1)+2m_Bm_1\bar{\sigma}+(2u-1)(m_1^2-q^2))     \,.
\end{aligned}
\end{equation}
For $\phi_4$:
\begin{equation}
\begin{aligned}
C^{(\Aone{V},\phi_4)}_1                        = & \frac{u-1}{m_B^2\bar{\sigma}}     \,,  \\
C^{(\Aone{V},\phi_4)}_2                        = & \frac{u-1}{m_B^2\bar{\sigma}}(m_B^2\bar{\sigma}^2+m_1^2-q^2)     \,, \\
C^{(\Aone{V},\overbar{\phi}_4)}_1              = &    2\frac{(u-1)}{m_B^3\bar{\sigma}^2}  \,, \\
C^{(\Aone{V},\overbar{\phi}_4)}_2              = &   \frac{1}{m_B^3\bar{\sigma}^2}(2m_B^2\bar{\sigma}^2u-2m_Bm_1\bar{\sigma}+4(u-1)(m_1^2-q^2)) \,, \\
C^{(\Aone{V},\overbar{\phi}_4)}_3              = &  \frac{2}{m_B^3\bar{\sigma}^2}((m_B\bar{\sigma}+m_1)^2-q^2)(m_B^2\bar{\sigma}^2(u-1) \\
&+m_Bm_1\bar{\sigma}(1-2u)+(u-1)(m_1^2-q^2))  \,,   \\
C^{(\Aone{V},\overbar{\overbar{\phi}}_4)}_2    = & \frac{2}{m_B^3\bar{\sigma}^2}(2m_B \bar{\sigma}(2u-1)-3m_1)        \,, \\
C^{(\Aone{V},\overbar{\overbar{\phi}}_4)}_3    = &   \frac{12m_1}{m_B^3\bar{\sigma}^2}(q^2-m_1^2)- \frac{2}{m_B^2\bar{\sigma}} (2u-1)(m_1^2+2q^2)-\frac{4m_1}{m_B}+4  \bar{\sigma}(2u-1)   \,, \\
C^{(\Aone{V},\overbar{\overbar{\phi}}_4)}_4    = & -\frac{6m_1}{m_B^3\bar{\sigma}^2}(m_B m_1 \bar{\sigma} (2u - 1) (m_B^2 \bar{\sigma}^2 - q^2) + (q^2 - m_B^2\bar{\sigma}^2)^2  \\
&+ m_1^3 m_B \bar{\sigma} (2u - 1) + m_1^4 - 2 m_1^2 q^2)  \,.
\end{aligned}
\end{equation}
For $\psi_4$:
\begin{equation}
\begin{aligned}
C^{(\Aone{V},\overbar{\psi}_4)}_1    & = -\frac{2}{m_B^3\bar{\sigma}^2}(2u-1)  \,, \\
C^{(\Aone{V},\overbar{\psi}_4)}_2    & = -\frac{2}{m_B^3\bar{\sigma}^2}(2u-1)(m_B^2\bar{\sigma}^2+2m_1^2-2q^2)  \,, \\
C^{(\Aone{V},\overbar{\psi}_4)}_3    & = -\frac{2}{m_B^3\bar{\sigma}^2}(2u-1)((q^2-m_B^2\bar{\sigma}^2)^2-2m_1^2(m_B^2\bar{\sigma}^2+q^2)+m_1^4)     \,.
\end{aligned}
\end{equation}
For $\chi_4$:
\begin{equation}
\begin{aligned}
C^{(\Aone{V},\overbar{\chi}_4)}_1    = &  \frac{2}{m_B^3\bar{\sigma}^2}    \,, \\
C^{(\Aone{V},\overbar{\chi}_4)}_2    = &  \frac{2}{m_B^3\bar{\sigma}^2}(m_B^2\bar{\sigma}^2(1-2u)+2m_Bm_1\bar{\sigma}+2m_1^2-2q^2)     \,, \\
C^{(\Aone{V},\overbar{\chi}_4)}_3    = & \frac{2}{m_B^3\bar{\sigma}^2}(m_B^4\bar{\sigma}^4+2m_B^3m_1\bar{\sigma}^3-2m_B^2\bar{\sigma}^2(m_1^2(1-2u)+q^2)  \\
&+2m_Bm_1\bar{\sigma}(m_1^2-q^2)+(m_1^2-q^2)^2)   \,.
\end{aligned}
\end{equation}
The coefficients of \refeq{CoeffFuncs3pt}, for the three-particle DAs, for $\Atwo{V}$ follow. For $\phi_3$:
\begin{equation}
\begin{aligned}
C^{(\Atwo{V},\phi_3)}_2                        = & \frac{4m_1}{m_B}-(2\bar{\sigma}+1)u  \,,  \\
C^{(\Atwo{V},\overbar{\phi}_3)}_2              = & \frac{2u}{m_B\bar{\sigma}} (\bar{\sigma}-1)   \,, \\
C^{(\Atwo{V},\overbar{\phi}_3)}_3              = &   \frac{1}{m_B\bar{\sigma}}(2u(m_B^2(3-2\bar{\sigma})\bar{\sigma}^2-2q^2\bar{\sigma}+q^2) \\
&+8m_Bm_1\bar{\sigma}(2\bar{\sigma}-1)-2m_1^2(2\bar{\sigma}u+u))   \,,   \\
C^{(\Atwo{V},\overbar{\overbar{\phi}}_3)}_4    = &  6m_1(4m_B(\bar{\sigma}-1)\bar{\sigma}+m_1(2\bar{\sigma}+(6-4\bar{\sigma})u-3))    \,.
\end{aligned}
\end{equation}
For $\phi_4$:
\begin{equation}
\begin{aligned}
C^{(\Atwo{V},\phi_4)}_2                        = &   (1-u)(2\bar{\sigma}-3) \,,  \\
C^{(\Atwo{V},\overbar{\phi}_4)}_2              = &     \frac{2}{m_B\bar{\sigma}}(u-1) (\bar{\sigma}-1) \,, \\
C^{(\Atwo{V},\overbar{\phi}_4)}_3              = & \frac{2}{m_B\bar{\sigma}}(m_B^2\bar{\sigma}^2(2\bar{\sigma}u-u-1)+m_Bm_1(3-4\bar{\sigma})\bar{\sigma}+m_1^2(2\bar{\sigma}+1)(1-u) \\
&+q^2(2\bar{\sigma}-2\bar{\sigma}u+u-1))  \,,   \\
C^{(\Atwo{V},\overbar{\overbar{\phi}}_4)}_3    = &  \frac{2}{m_B\bar{\sigma}}(2 m_B(\bar{\sigma}-2)\bar{\sigma}(2u-1)+m_1(9-4\bar{\sigma}))       \,, \\
C^{(\Atwo{V},\overbar{\overbar{\phi}}_4)}_4    = &  \frac{6}{m_B\bar{\sigma}}(2m_B(\bar{\sigma}-1)\bar{\sigma}(2u-1)(m_B^2\bar{\sigma}^2-q^2)  \\
&+3m_Bm_1^2\bar{\sigma}(1-2u)+3m_1^3+m_1(4\bar{\sigma}-3)(q^2-m_B^2\bar{\sigma}^2))       \,.
\end{aligned}
\end{equation}
For $\psi_4$:
\begin{equation}
\begin{aligned}
C^{(\Atwo{V},\overbar{\psi}_4)}_2    = & \frac{2}{m_B\bar{\sigma}}(1-2u)(\bar{\sigma} - 1)   \,, \\
C^{(\Atwo{V},\overbar{\psi}_4)}_3    = &  \frac{1}{m_B\bar{\sigma}}(2(2\bar{\sigma}-1)(2u-1)(q^2-m_B^2\bar{\sigma}^2)  \\
&-4m_Bm_1\bar{\sigma}+2m_1^2(2\bar{\sigma}+1)(2u-1))    \,.
\end{aligned}
\end{equation}
For $\chi_4$:
\begin{equation}
\begin{aligned}
C^{(\Atwo{V},\overbar{\chi}_4)}_2    = &  \frac{2}{m_B\bar{\sigma}}(\bar{\sigma}-1)     \,, \\
C^{(\Atwo{V},\overbar{\chi}_4)}_3    = &  -\frac{2}{m_B\bar{\sigma}}(m_B^2\bar{\sigma}^2(-2\bar{\sigma}+4(\bar{\sigma}-1)u+1)  \\
&+4m_Bm_1(1-2\bar{\sigma})\bar{\sigma}+m_1^2(2\bar{\sigma}+1)+q^2(2\bar{\sigma}-1))  \,.
\end{aligned}
\end{equation}
The coefficients of \refeq{CoeffFuncs3pt}, for the three-particle DAs, for $\Athreezero{V}$ follow. For $\phi_3$:
\begin{equation}
\begin{aligned}
C^{(\Athreezero{V},\phi_3)}_2                        = &  \frac{4m_1}{m_B}+(5-2\bar{\sigma})u   \,,  \\
C^{(\Athreezero{V},\overbar{\phi}_3)}_2              = &   \frac{2u}{m_B\bar{\sigma}}(\bar{\sigma}-3) \,, \\
C^{(\Athreezero{V},\overbar{\phi}_3)}_3              = &  -\frac{2}{m_B\bar{\sigma}}(u(m_B^2\bar{\sigma}(\bar{\sigma}(2\bar{\sigma}-9)+8)+q^2(2\bar{\sigma}-3)) \\
&+4m_Bm_1(3-2\bar{\sigma})\bar{\sigma}+m_1^2(2\bar{\sigma}+3)u)    \,,   \\
C^{(\Athreezero{V},\overbar{\overbar{\phi}}_3)}_4    = & 6m_1(4m_B(\bar{\sigma}-2)(\bar{\sigma}-1)+m_1(2\bar{\sigma}+(2-4\bar{\sigma})u-1))     \,.
\end{aligned}
\end{equation}
For $\phi_4$:
\begin{equation}
\begin{aligned}
C^{(\Athreezero{V},\phi_4)}_2                        = &  2\bar{\sigma}-2\bar{\sigma}u+u-1  \,,  \\
C^{(\Athreezero{V},\overbar{\phi}_4)}_2              = &   \frac{2}{m_B\bar{\sigma}}(u-1)(\bar{\sigma}-3)   \,, \\
C^{(\Athreezero{V},\overbar{\phi}_4)}_3              = & \frac{1}{m_B\bar{\sigma}}(2m_B^2\bar{\sigma}(2\bar{\sigma}^2u-3\bar{\sigma}(u+1)+4)+2m_Bm_1(5-4\bar{\sigma})\bar{\sigma}  \\
&-2m_1^2(2\bar{\sigma}+3)(u-1) -2q^2(2\bar{\sigma}-3)(u-1))  \,,   \\
C^{(\Athreezero{V},\overbar{\overbar{\phi}}_4)}_3    = & \frac{2}{m_B\bar{\sigma}}(2m_B(( \bar{\sigma}-6)\bar{\sigma}+6)(2u-1)+m_1(15-4\bar{\sigma}))        \,, \\
C^{(\Athreezero{V},\overbar{\overbar{\phi}}_4)}_4    = & \frac{6}{m_B\bar{\sigma}}(2m_B^3(\bar{\sigma}-2)(\bar{\sigma}-1)\bar{\sigma}^2(2u-1)+m_B^2m_1(5-4\bar{\sigma})\bar{\sigma}^2  \\
&-m_B(2u-1)(m_1^2(\bar{\sigma}-4) +2q^2(\bar{\sigma}-2)(\bar{\sigma}-1))\\
&+m_1(5m_1^2+q^2(4\bar{\sigma}-5)))      \,.
\end{aligned}
\end{equation}
For $\psi_4$:
\begin{equation}
\begin{aligned}
C^{(\Athreezero{V},\overbar{\psi}_4)}_2    & = \frac{2}{m_B\bar{\sigma}}(1-2u)(\bar{\sigma}-3) \,, \\
C^{(\Athreezero{V},\overbar{\psi}_4)}_3    & = \frac{2}{m_B\bar{\sigma}}((2\bar{\sigma}-3)(2u-1)(q^2-m_B^2\bar{\sigma}^2)+2m_Bm_1\bar{\sigma}+m_1^2(2\bar{\sigma}+3)(2u-1))    \,.
\end{aligned}
\end{equation}
For $\chi_4$:
\begin{equation}
\begin{aligned}
C^{(\Athreezero{V},\overbar{\chi}_4)}_2    = & \frac{2}{m_B\bar{\sigma}}(\bar{\sigma}-3)      \,, \\
C^{(\Athreezero{V},\overbar{\chi}_4)}_3    = & -\frac{2}{m_B\bar{\sigma}}(m_B^2\bar{\sigma}((3-2\bar{\sigma})\bar{\sigma}+4(\bar{\sigma}-2)(\bar{\sigma}-1)u)\\
&+4m_Bm_1(3-2\bar{\sigma})\bar{\sigma}+m_1^2(2\bar{\sigma}+3)+q^2(2\bar{\sigma}-3))   \,.
\end{aligned}
\end{equation}
The coefficients of \refeq{CoeffFuncs3pt}, for the three-particle DAs, for $\Tone{V}$ follow. For $\phi_3$:
\begin{equation}
    \begin{aligned}
        C^{(\Tone{V},\phi_3)}_2                        & = \frac{m_1}{m_B}+ u\bar{\sigma} \,,  \\
        C^{(\Tone{V},\overbar{\phi}_3)}_2              & = \frac{1}{m_B^2\bar{\sigma}}(2m_1-m_B\bar{\sigma}u)  \,, \\
        C^{(\Tone{V},\overbar{\phi}_3)}_3              & = \frac{2}{m_B^2\bar{\sigma}}(m_B^2\bar{\sigma}^2 +m_1^2-q^2)(m_B\bar{\sigma}u+m_1)     \,,   \\
        C^{(\Tone{V},\overbar{\overbar{\phi}}_3)}_3    & = 6\frac{m_1}{m_B}     \,.\\
        C^{(\Tone{V},\overbar{\overbar{\phi}}_3)}_4    & =  6\frac{m_1^2}{m_B}(m_B\bar{\sigma}(2u-1)+m_1)   \,.
    \end{aligned}
\end{equation}
For $\phi_4$:
\begin{equation}
    \begin{aligned}
        C^{(\Tone{V},\phi_4)}_2                        & =   (u-1)\bar{\sigma}  \,,  \\
        C^{(\Tone{V},\overbar{\phi}_4)}_2              & =     \frac{u}{m_B} \,, \\
        C^{(\Tone{V},\overbar{\phi}_4)}_3              & = \frac{2}{m_B}(-(m_B\bar{\sigma}+m_1)(m_1 u-m_B\bar{\sigma}(u-1))-q^2u+q^2)  \,,   \\
        C^{(\Tone{V},\overbar{\overbar{\phi}}_4)}_2    & = \frac{2}{m_B^2\bar{\sigma}} (2u-1)        \,, \\
        C^{(\Tone{V},\overbar{\overbar{\phi}}_4)}_3    & =   \frac{1}{m_B^2\bar{\sigma}} (-2(2u-1)(q^2-m_B^2\bar{\sigma}^2)-2m_B m_1\bar{\sigma}+m_1^2(4-8u))      \,, \\
        C^{(\Tone{V},\overbar{\overbar{\phi}}_4)}_4    & = \frac{6 m_1}{m_B^2\bar{\sigma}} (-m_B^3\bar{\sigma}^3+m_B q^2 \bar{\sigma}-m_1(2u-1)(m_1^2-q^2))      \,.
    \end{aligned}
\end{equation}
For $\psi_4$:
\begin{equation}
    \begin{aligned}
        C^{(\Tone{V},\overbar{\psi}_4)}_2    & = \frac{1}{m_B^2\bar{\sigma}}(m_B \bar{\sigma}(1-2u)-2m_1) \,, \\
        C^{(\Tone{V},\overbar{\psi}_4)}_3    & = \frac{2}{m_B^2\bar{\sigma}}(m_B^3 \bar{\sigma}^3(1-2u)+m_B^2 m_1 \bar{\sigma}^2+m_B\bar{\sigma}(2u-1)(m_1^2+q^2)-m_1^3+m_1 q^2)     \,.
    \end{aligned}
\end{equation}
For $\chi_4$:
\begin{equation}
    \begin{aligned}
        C^{(\Tone{V},\overbar{\chi}_4)}_2    & =  \frac{1}{m_B^2\bar{\sigma}}(m_B \bar{\sigma}(1-2u)+2m_1)     \,, \\
        C^{(\Tone{V},\overbar{\chi}_4)}_3    & =  2\left(\frac{m_1^3-m_1q^2}{m_B^2\bar{\sigma}} -\frac{-2m_1^2 u +m_1^2+q^2}{m_B}+m_B\bar{\sigma}^2+m_1\bar{\sigma} \right)  \,.
    \end{aligned}
\end{equation}
The coefficients of \refeq{CoeffFuncs3pt}, for the three-particle DAs, for $\TtwothreeA{V}$ follow. For $\phi_3$:
\begin{equation}
    \begin{aligned}
        C^{(\TtwothreeA{V},\phi_3)}_2                        = & \frac{m_B m_1-4q^2 u}{m_B^2}+\bar{\sigma}u \,,  \\
        C^{(\TtwothreeA{V},\overbar{\phi}_3)}_2              = & \frac{1}{m_B^2\bar{\sigma}}(2m_1-m_B\bar{\sigma}u)   \,,  \\
        C^{(\TtwothreeA{V},\overbar{\phi}_3)}_3              = & \frac{2}{m_B^2\bar{\sigma}}(m_B^3\bar{\sigma}^3u+\bar{\sigma}(m_B u(m_1^2+3q^2)+4m_1q^2) \\
        & +m_B\bar{\sigma}^2(m_Bm_1-4q^2u)+m_1^3-m_1q^2)  \,,  \\
        C^{(\TtwothreeA{V},\overbar{\overbar{\phi}}_3)}_3    = &  6\frac{m_1}{m_B}    \,.\\
        C^{(\TtwothreeA{V},\overbar{\overbar{\phi}}_3)}_4    = &  6\frac{m_1}{m_B} (m_Bm_1\bar{\sigma}(2u-1)+m_1^2+4q^2(\bar{\sigma}-1) )     \,.
    \end{aligned}
\end{equation}
For $\phi_4$:
\begin{equation}
    \begin{aligned}
        C^{(\TtwothreeA{V},\phi_4)}_2                        = & \bar{\sigma}(u-1)   \,,  \\
        C^{(\TtwothreeA{V},\overbar{\phi}_4)}_2              = &  \frac{u}{m_B}    \,, \\
        C^{(\TtwothreeA{V},\overbar{\phi}_4)}_3              = & -\frac{2}{m_B}((m_B\bar{\sigma}+m_1)(m_1u- m_B\bar{\sigma}(u-1))+q^2(-2\bar{\sigma}+u+1))  \,,   \\
        C^{(\TtwothreeA{V},\overbar{\overbar{\phi}}_4)}_2    = & \frac{2}{m_B^2\bar{\sigma}}(2u-1)         \,, \\
        C^{(\TtwothreeA{V},\overbar{\overbar{\phi}}_4)}_3    = &   \frac{2}{m_B^2\bar{\sigma}}((2u-1)(m_B^2\bar{\sigma}^2+q^2(4\bar{\sigma}-7))-m_Bm_1\bar{\sigma}+m_1^2(2-4u))       \,, \\
        C^{(\TtwothreeA{V},\overbar{\overbar{\phi}}_4)}_4    = & -\frac{6}{m_B^2\bar{\sigma}} (m_1(m_B^3\bar{\sigma}^3-m_Bq^2\bar{\sigma})+2q^2(\bar{\sigma}-1)(2u-1)(q^2-m_B^2\bar{\sigma}^2)\\
        &+m_1^4(2u-1)+m_1^2q^2(2\bar{\sigma}+1)(2u-1))      \,.
    \end{aligned}
\end{equation}
For $\psi_4$:
\begin{equation}
    \begin{aligned}
        C^{(\TtwothreeA{V},\overbar{\psi}_4)}_2    = & \frac{1}{m_B^2\bar{\sigma}}(m_B\bar{\sigma}(1-2u)-2m_1)  \,, \\
        C^{(\TtwothreeA{V},\overbar{\psi}_4)}_3    = &  \frac{2}{m_B^2\bar{\sigma}} (m_B^3\bar{\sigma}^3(1-2u)+m_B^2m_1\bar{\sigma}^2   \\
        &+\bar{\sigma}(m_B(2u-1)(m_1^2+q^2)-4m_1q^2)-m_1^3+m_1q^2)   \,.
    \end{aligned}
\end{equation}
For $\chi_4$:
\begin{equation}
    \begin{aligned}
        C^{(\TtwothreeA{V},\overbar{\chi}_4)}_2    = &  \frac{1}{m_B^2\bar{\sigma}}(m_B\bar{\sigma}-2m_B\bar{\sigma}u+2m_1)     \,, \\
        C^{(\TtwothreeA{V},\overbar{\chi}_4)}_3    = &\frac{2}{m_B^2\bar{\sigma}}(m_B^3\bar{\sigma}^3+\bar{\sigma}(2m_Bu(m_1^2+2q^2)-m_B(m_1^2+q^2)+4m_1q^2)\\
        &+m_B\bar{\sigma}^2(m_Bm_1-4q^2u)+m_1^3-m_1q^2)    \,.
    \end{aligned}
\end{equation}
The coefficients of \refeq{CoeffFuncs3pt}, for the three-particle DAs, for $\TtwothreeB{V}$ follow. For $\phi_3$:
\begin{equation}
    \begin{aligned}
        C^{(\TtwothreeB{V},\phi_3)}_1                        = &  -2\frac{u}{m_B^2\bar{\sigma}} \,,  \\
        C^{(\TtwothreeB{V},\phi_3)}_2                        = &  \frac{1}{m_B^2\bar{\sigma}}(u(m_B^2\bar{\sigma}(3\bar{\sigma}-1)+q^2(2-4\bar{\sigma}))+m_Bm_1\bar{\sigma}-2m_1^2u) \,,  \\
        C^{(\TtwothreeB{V},\overbar{\phi}_3)}_2              = &  \frac{1}{m_B^2\bar{\sigma}}(m_B(2-5\bar{\sigma})u+6m_1)  \,, \\
        C^{(\TtwothreeB{V},\overbar{\phi}_3)}_3              = &\frac{2}{m_B^2\bar{\sigma}}(m_1(q^2(4\bar{\sigma}-3)-m_B^2\bar{\sigma}^2)+ m_B(\bar{\sigma}-1)u(3m_B^2\bar{\sigma}^2-4q^2\bar{\sigma}+q^2) \\
        & -m_Bm_1^2(\bar{\sigma}-1)u+3m_1^3)     \,,   \\
        C^{(\TtwothreeB{V},\overbar{\overbar{\phi}}_3)}_3    = &  6\frac{m_1}{m_B\bar{\sigma}}(3\bar{\sigma}-2)    \,.\\
        C^{(\TtwothreeB{V},\overbar{\overbar{\phi}}_3)}_4    = &  6\frac{m_1}{m_B\bar{\sigma}}((3\bar{\sigma}-2)m_1^2+m_B(2u-1)(\bar{\sigma}-1)\bar{\sigma}m_1\\
        & -2(\bar{\sigma}-1)(m_B^2\bar{\sigma}^2-2q^2\bar{\sigma}+q^2))    \,.
    \end{aligned}
\end{equation}
For $\phi_4$:
\begin{equation}
    \begin{aligned}
        C^{(\TtwothreeB{V},\phi_4)}_2                        = &(\bar{\sigma}-1)(u-1)    \,,  \\
        C^{(\TtwothreeB{V},\overbar{\phi}_4)}_2              = &   \frac{1}{m_B\bar{\sigma}}(2\bar{\sigma}+(\bar{\sigma}-2)u)   \,, \\
        C^{(\TtwothreeB{V},\overbar{\phi}_4)}_3              = & \frac{2}{m_B\bar{\sigma}}((\bar{\sigma}-1)(m_B^2\bar{\sigma}^2(u-2)+2q^2\bar{\sigma}-q^2u)-m_Bm_1(\bar{\sigma}-1)\bar{\sigma}  \\
        &-m_1^2(\bar{\sigma}(u-1)+u))  \,,   \\
        C^{(\TtwothreeB{V},\overbar{\overbar{\phi}}_4)}_2    = &   \frac{6}{m_B^2\bar{\sigma}^2}(\bar{\sigma}-1)(2u-1)       \,, \\
        C^{(\TtwothreeB{V},\overbar{\overbar{\phi}}_4)}_3    = &  \frac{1}{m_B^2\bar{\sigma}^2} (4(2u-1)(m_B^2\bar{\sigma}^3+q^2(2(\bar{\sigma}-3)\bar{\sigma}+3))  \\
        &-2m_Bm_1(\bar{\sigma}-3)\bar{\sigma}-6m_1^2(\bar{\sigma}+2)(2u-1))       \,, \\
        C^{(\TtwothreeB{V},\overbar{\overbar{\phi}}_4)}_4    = & -\frac{6}{m_B^2\bar{\sigma}^2}((2u-1)(2\bar{\sigma}+1)m_1^4-m_B\bar{\sigma}m_1^3+(2u-1)(m_B^2(1-2\bar{\sigma})\bar{\sigma}^2\\
        &+q^2(2\bar{\sigma}^2+\bar{\sigma}-2))m_1^2+m_B(\bar{\sigma}-1)\bar{\sigma}(m_B^2\bar{\sigma}^2-q^2)m_1\\
        &+(2u-1)(\bar{\sigma}-1)(m_B^2\bar{\sigma}^2-q^2)(m_B^2\bar{\sigma}^2-2q^2\bar{\sigma}+q^2))       \,.
    \end{aligned}
\end{equation}
For $\psi_4$:
\begin{equation}
    \begin{aligned}
        C^{(\TtwothreeB{V},\overbar{\psi}_4)}_2    & =\frac{1}{m_B^2\bar{\sigma}}(m_B(\bar{\sigma}-2(\bar{\sigma}-2)u-2)-6m_1)   \,, \\
        C^{(\TtwothreeB{V},\overbar{\psi}_4)}_3    & =  \frac{2}{m_B^2\bar{\sigma}} (m_B^3\bar{\sigma}^2(\bar{\sigma}-2(\bar{\sigma}-1)u-1)+m_B^2m_1\bar{\sigma}(3\bar{\sigma}-2)\\&+m_B(2u-1)(m_1^2(\bar{\sigma}+1)+q^2(\bar{\sigma}-1))-3m_1^3+m_1q^2(3-4\bar{\sigma}))   \,.
    \end{aligned}
\end{equation}
For $\chi_4$:
\begin{equation}
    \begin{aligned}
        C^{(\TtwothreeB{V},\overbar{\chi}_4)}_2    & = \frac{1}{m_B^2\bar{\sigma}}(m_B(\bar{\sigma}+(4-6\bar{\sigma})u-2)+6m_1)        \,, \\
        C^{(\TtwothreeB{V},\overbar{\chi}_4)}_3    & =  \frac{2}{m_B^2\bar{\sigma}}(m_B^3(\bar{\sigma}-1)\bar{\sigma}^2(2u+1)-m_B^2m_1\bar{\sigma}^2+m_B(m_1^2(-\bar{\sigma}+2u-1)\\&-q^2(\bar{\sigma}-1)((4\bar{\sigma}-2)u+1))+3m_1^3+m_1q^2(4\bar{\sigma}-3))  \,.
    \end{aligned}
\end{equation}

%%%%%%%%%%%%%%%%%%%%%%%%%%%
\newpage
\section{Plots of the Form Factors}
\label{app:plots}

This appendix is dedicated to illustrate our numerical results for the form factors in
relation to previous results obtained from LCSRs with $B$-meson LCDAs \cite{Khodjamirian:2005ea,Khodjamirian:2006st,Faller:2008tr}
and to results obtained from LQCD.

\begin{figure}[hb]
    \centering
    \begin{tabular}{c p{0.05\textwidth} c}
         \includegraphics[width=.39\textwidth]{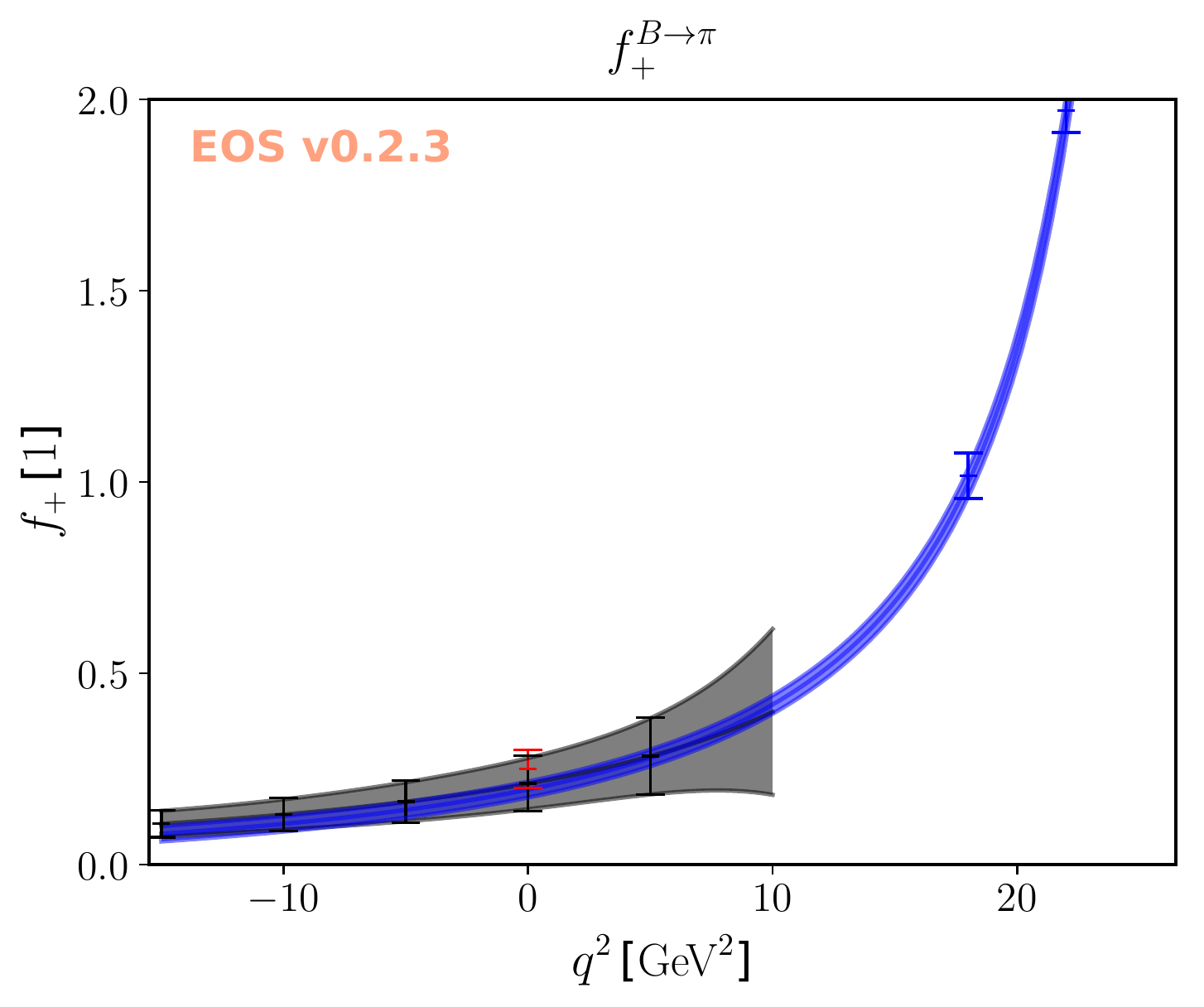}   & &
         \includegraphics[width=.39\textwidth]{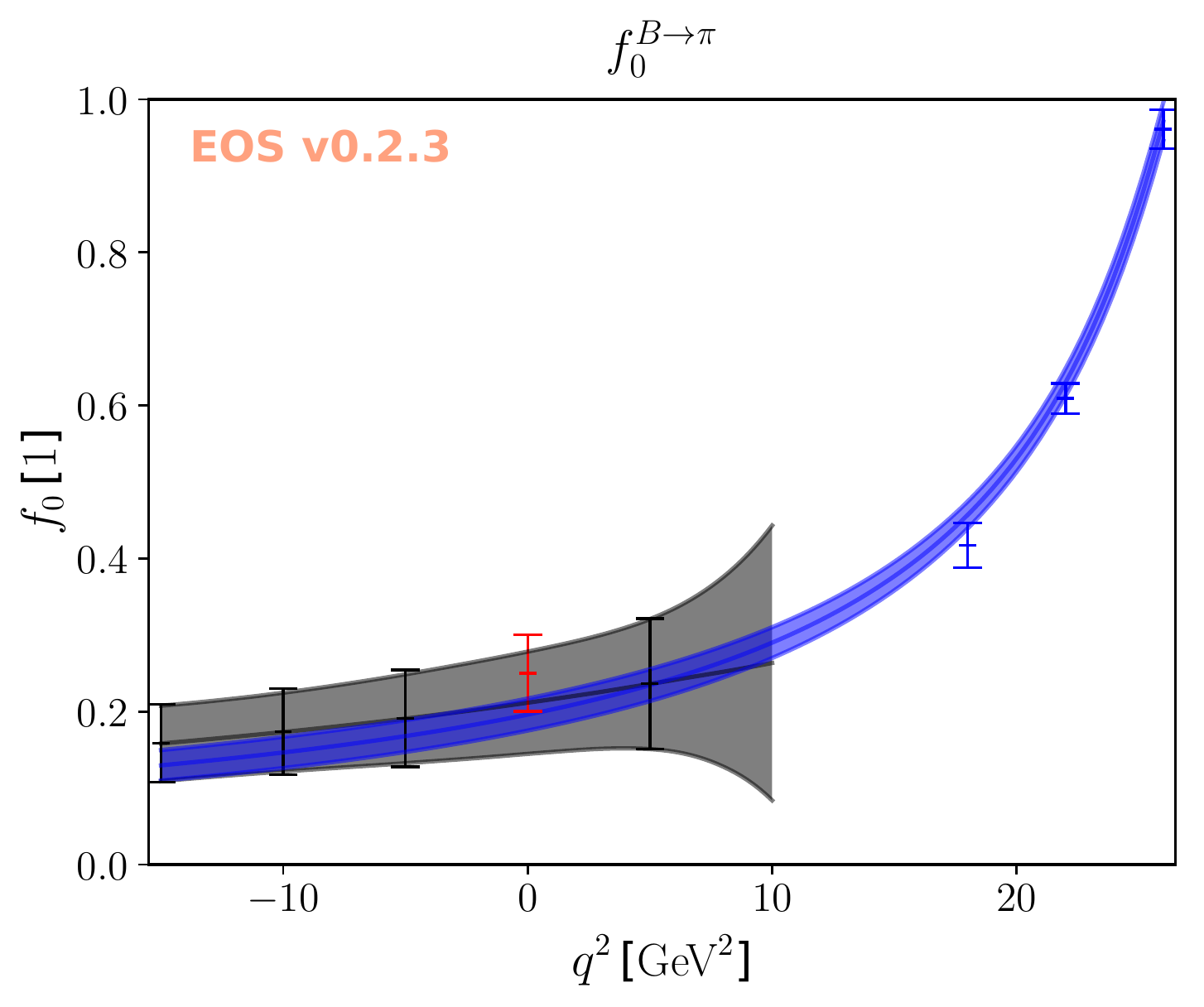}   \\
         \includegraphics[width=.39\textwidth]{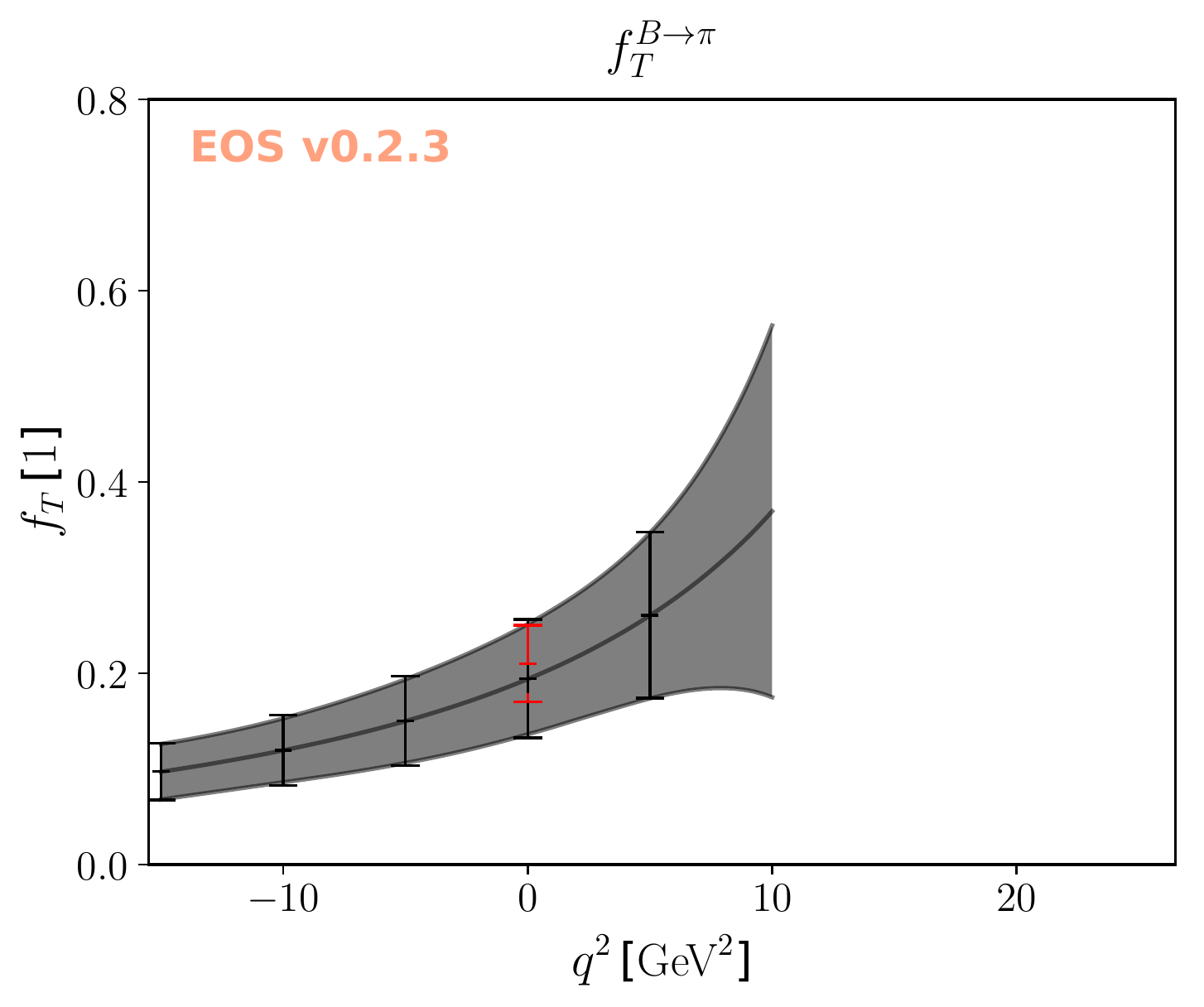}
    \end{tabular}
    \caption{%
    \label{fig:plots-B-to-pi}
    Plots of our results (gray points) and LQCD results from ref.~\cite{Lattice:2015tia}
    (blue points) for the $B\to \pi$
    form factors. Central values and $68\%$ probability envelopes as functions of $q^2$
    from fits to our results only (gray) and a combination of our results and LQCD
    results (blue) are shown as well.
    Previous results from LCSRs using $B$-LCDAs \cite{Khodjamirian:2006st} at $q^2=0$
    are not used in the fits and shown in red for illustrative purpose only.
    Solid lines represent the central values, and shaded areas illustrate the $68\%$ probability envelope.
    }
\end{figure}

\FloatBarrier

\begin{figure}[h!]
    \centering
    \begin{tabular}{c p{0.05\textwidth} c}
         \includegraphics[width=.39\textwidth]{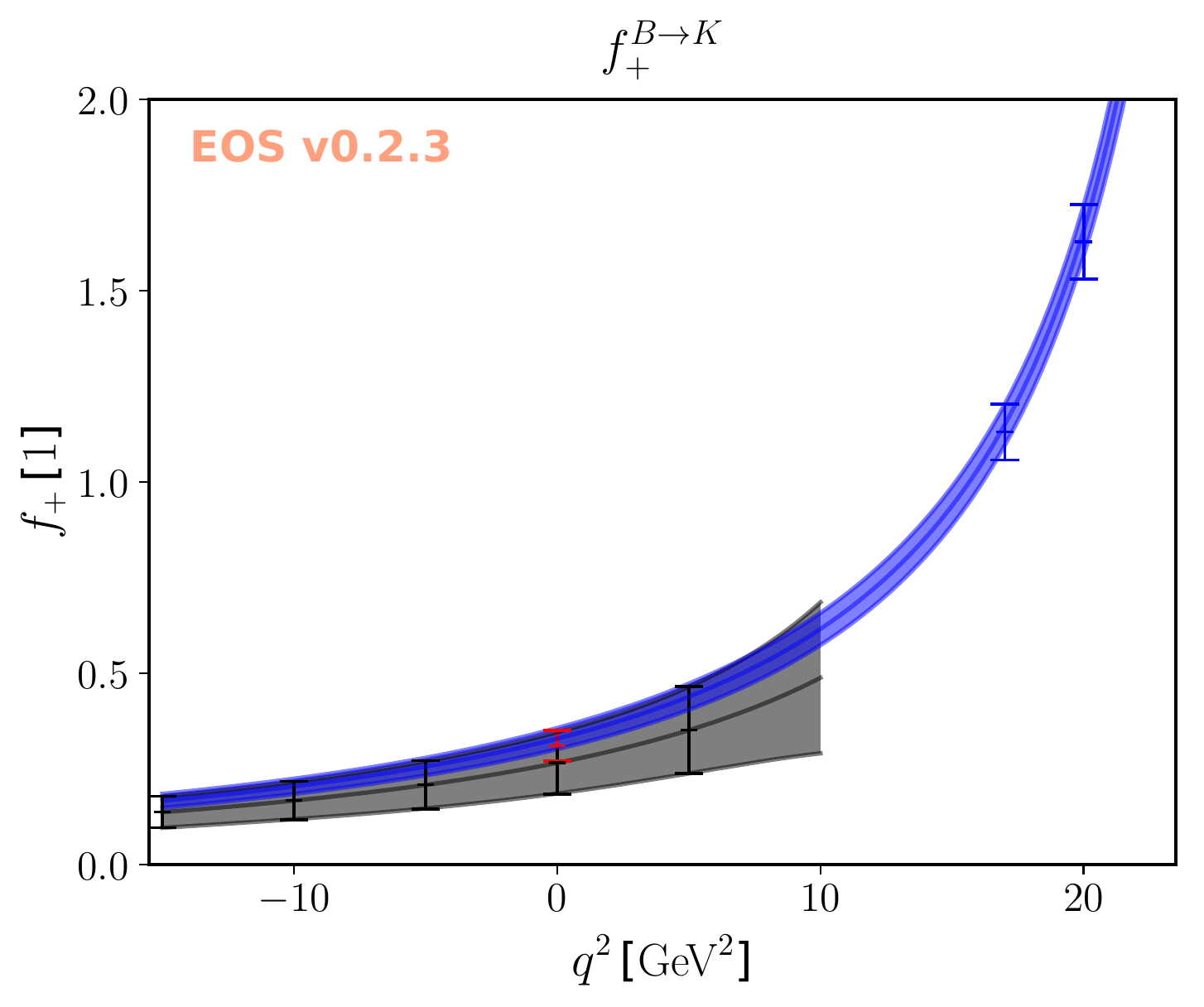}   & &
         \includegraphics[width=.39\textwidth]{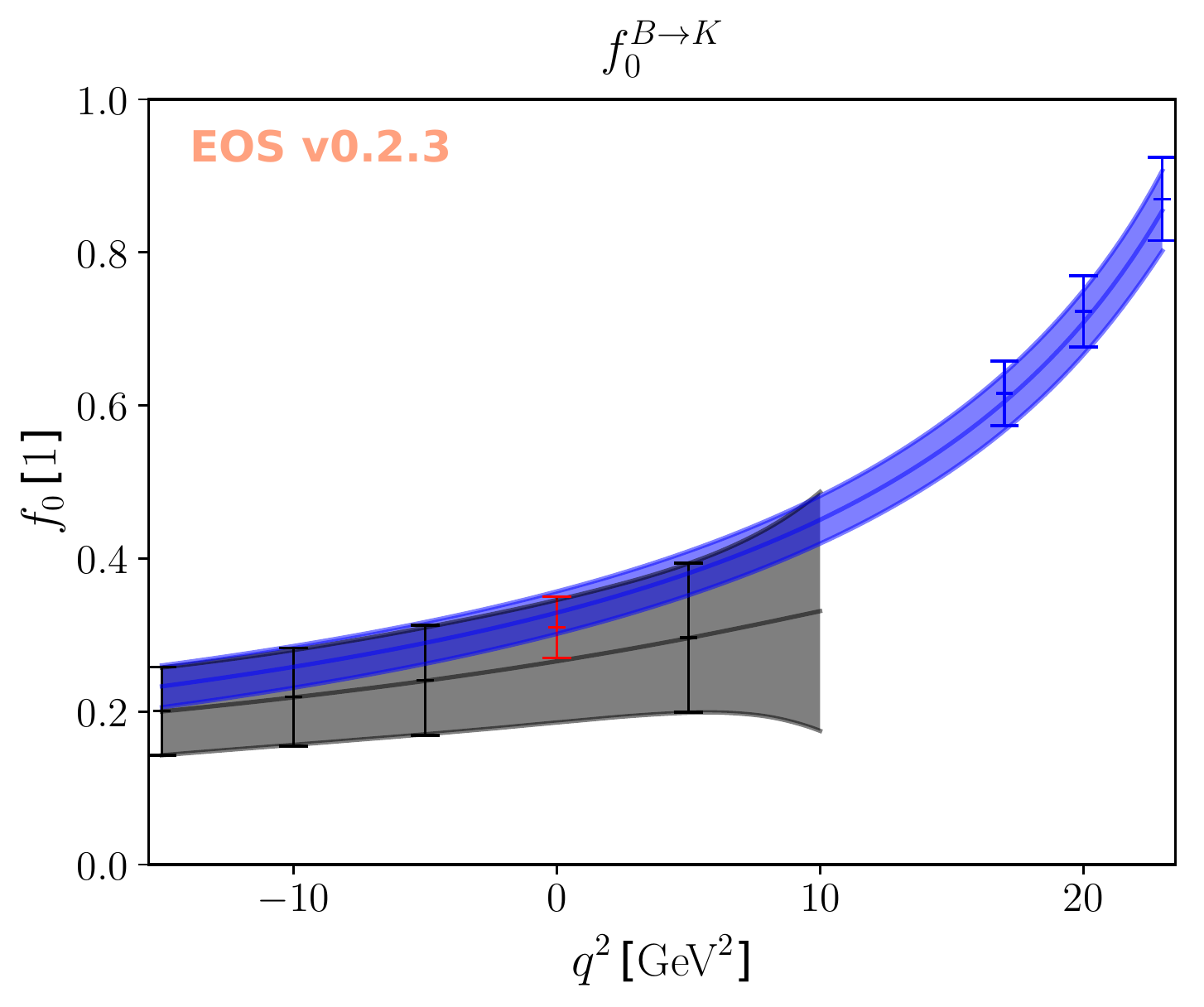}   \\
         \includegraphics[width=.39\textwidth]{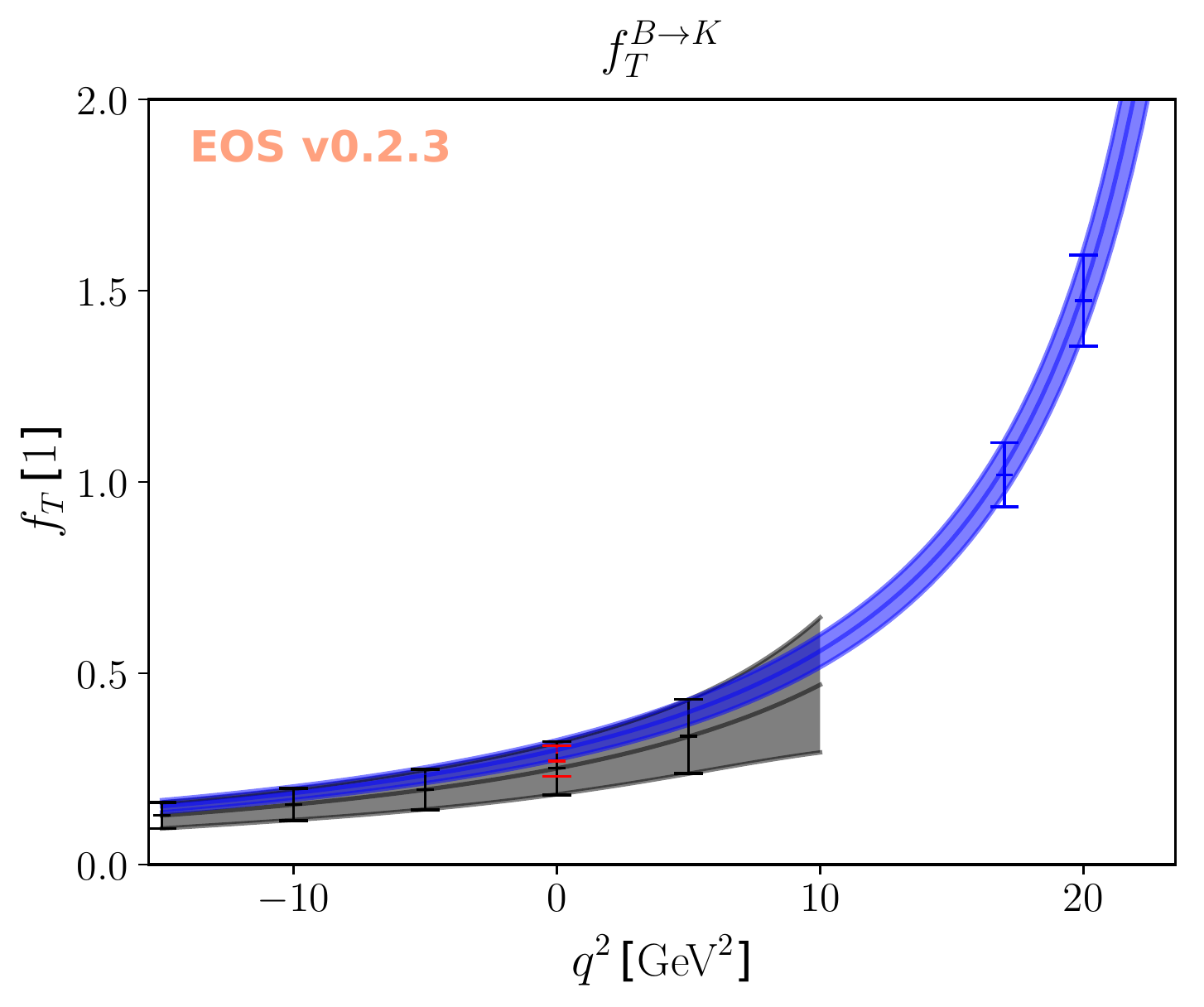}
    \end{tabular}
    \caption{%
    \label{tab:plots-B-to-K}
    Plot of $B\to K$ form factors, LQCD results from ref.~\cite{Bouchard:2013pna}. For a description see \reffig{plots-B-to-pi}.
    }
\end{figure}

\begin{figure}[h!]
    \centering
    \begin{tabular}{c p{0.05\textwidth} c}
         \includegraphics[width=.39\textwidth]{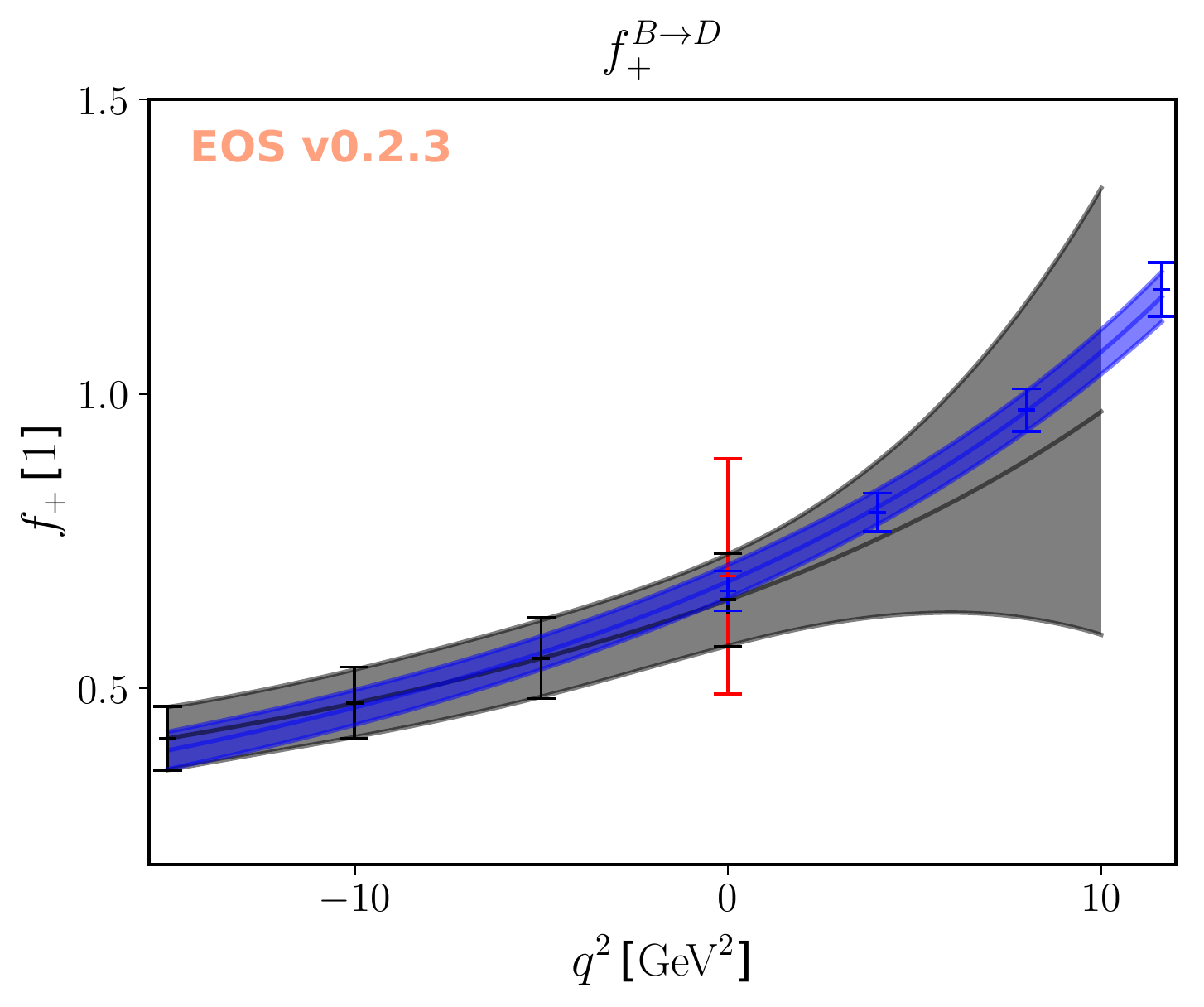}   & &
         \includegraphics[width=.39\textwidth]{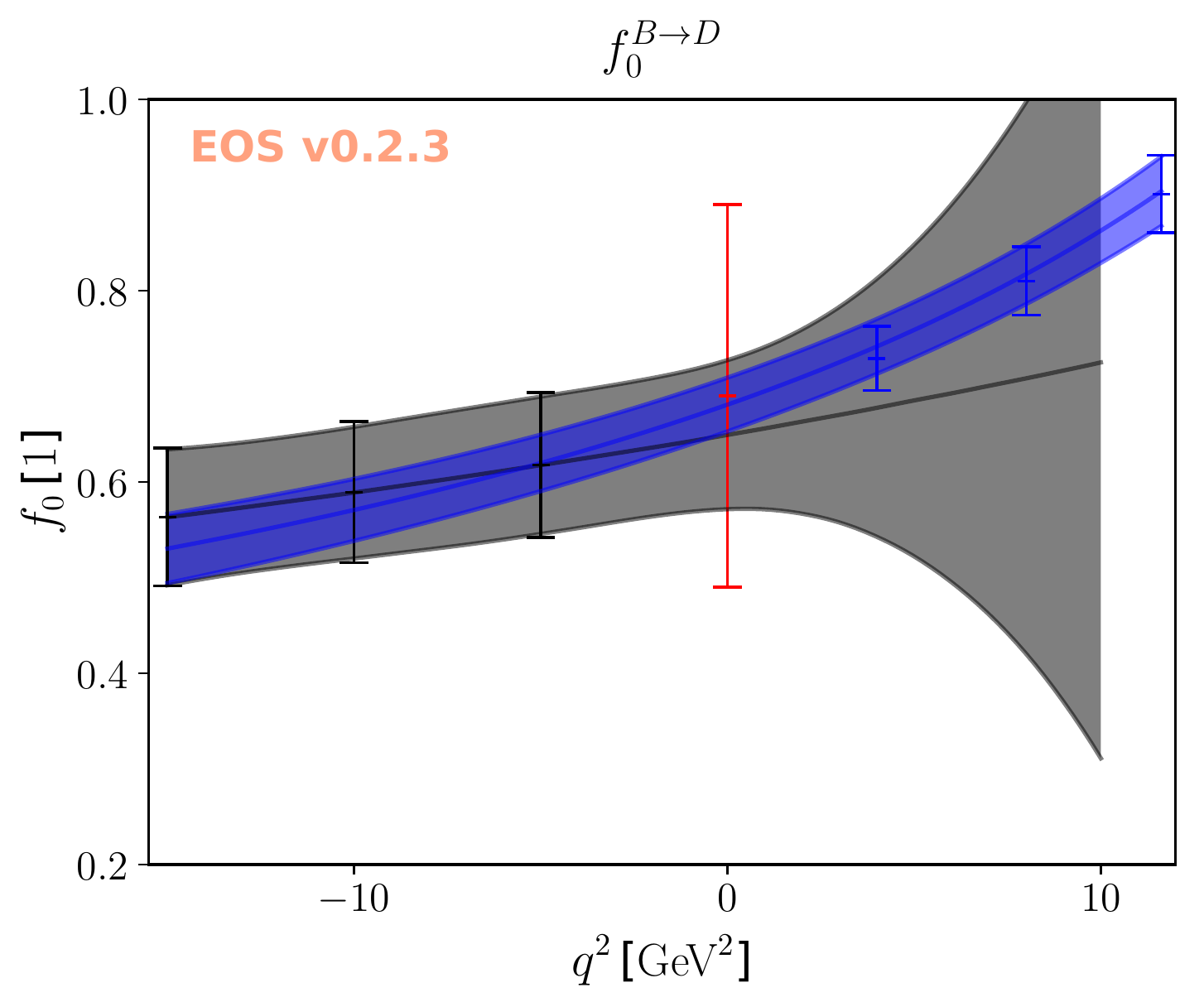}   \\
         \includegraphics[width=.39\textwidth]{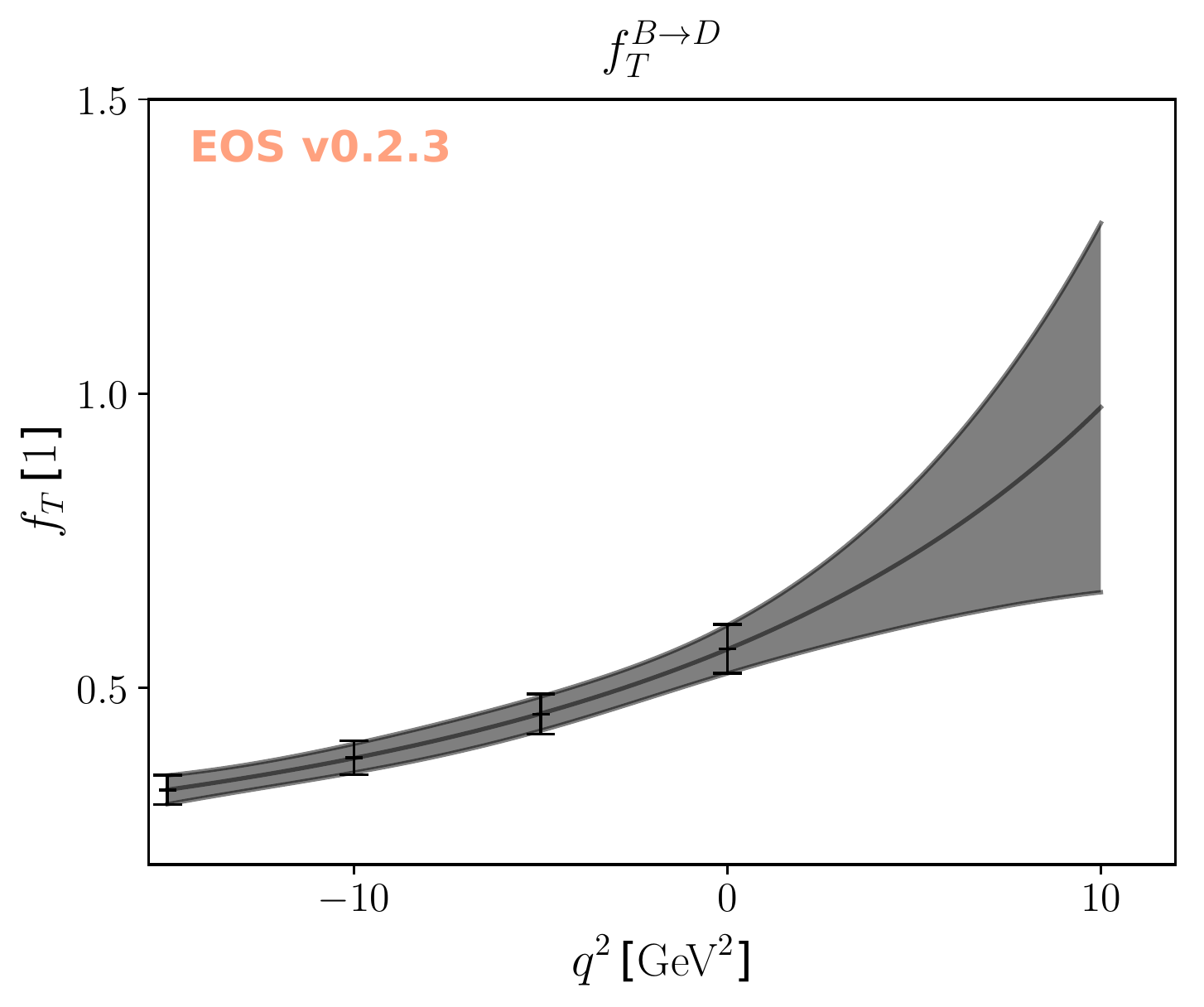}
    \end{tabular}
    \caption{%
    \label{tab:plots-B-to-D}
    Plot of $B\to \bar{D}$ form factors, LQCD results from ref.~\cite{Na:2015kha}. For a description see \reffig{plots-B-to-pi}.
    }
\end{figure}

\FloatBarrier

\begin{figure}[p]
    \centering
    \begin{tabular}{cc}
         \includegraphics[width=.40\textwidth]{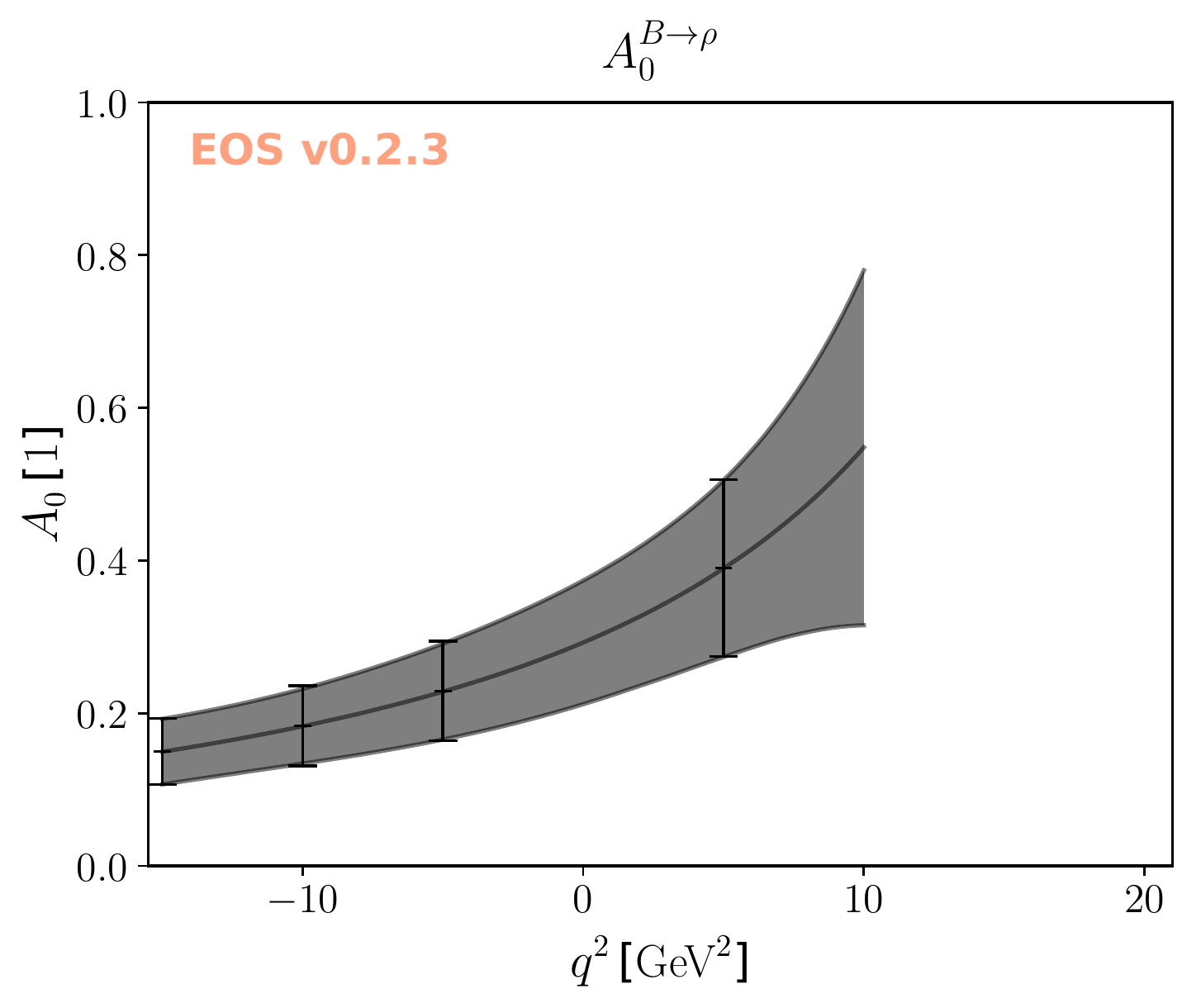}   &
         \includegraphics[width=.40\textwidth]{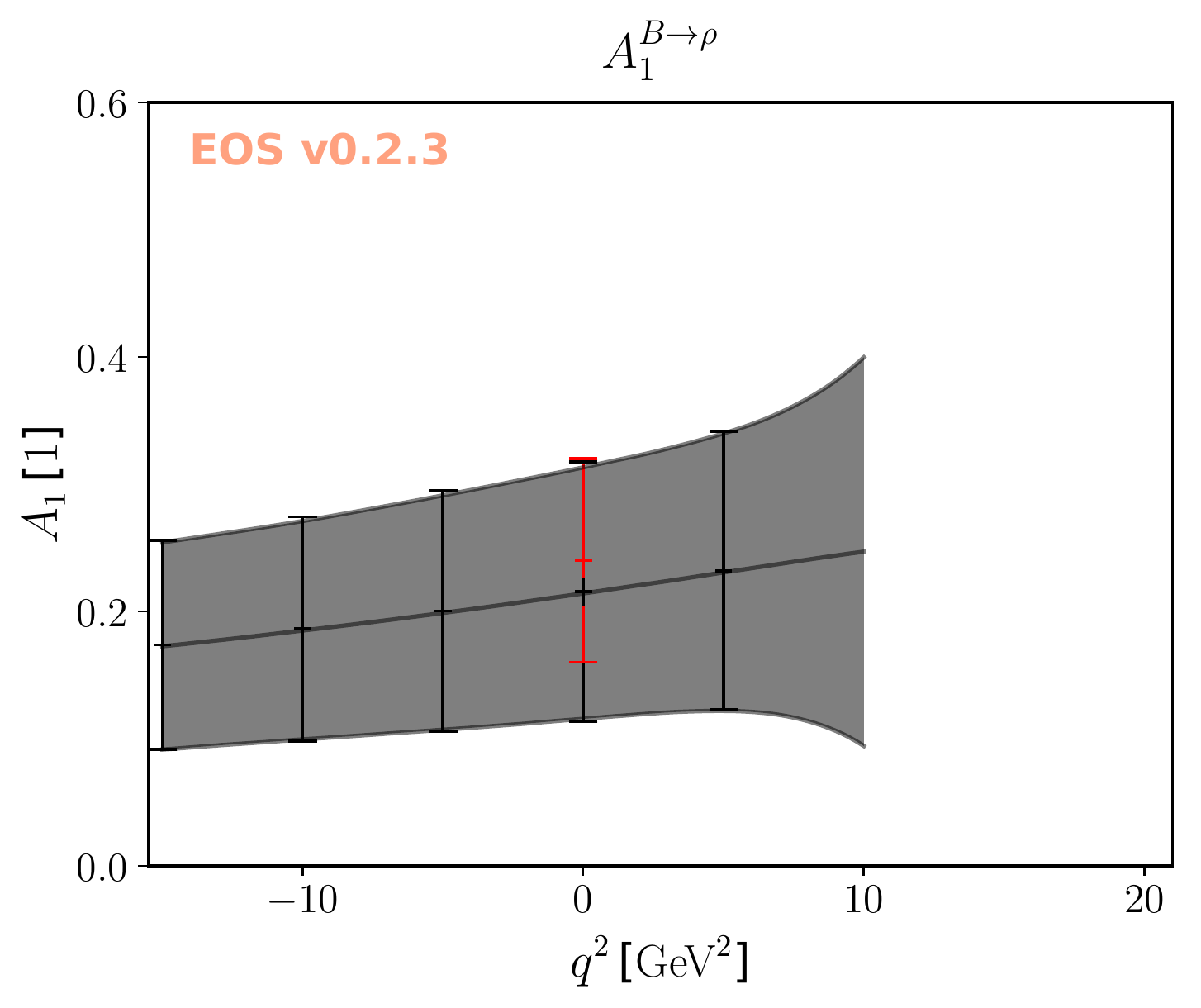}   \\
         \includegraphics[width=.40\textwidth]{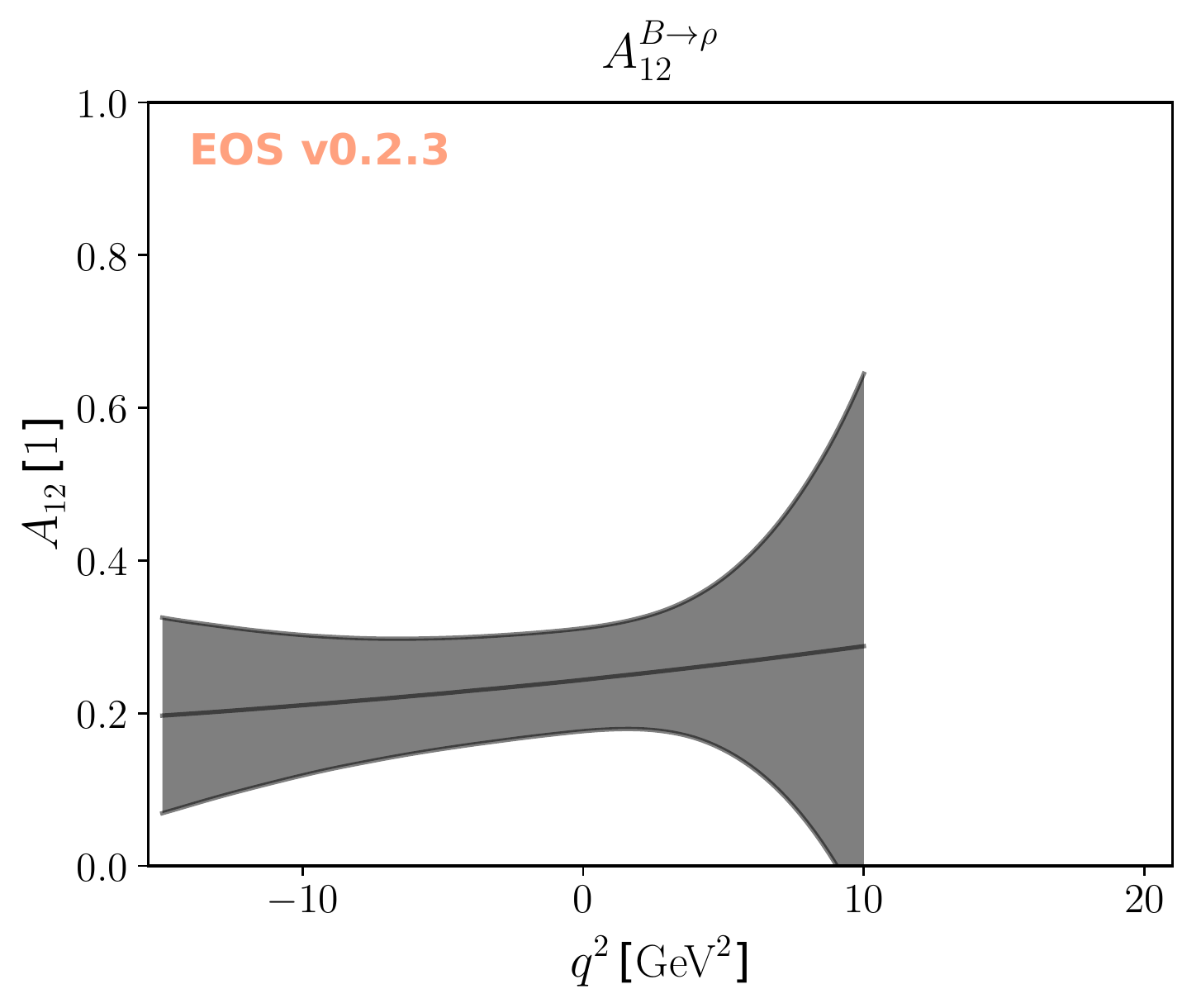}  &
         \includegraphics[width=.40\textwidth]{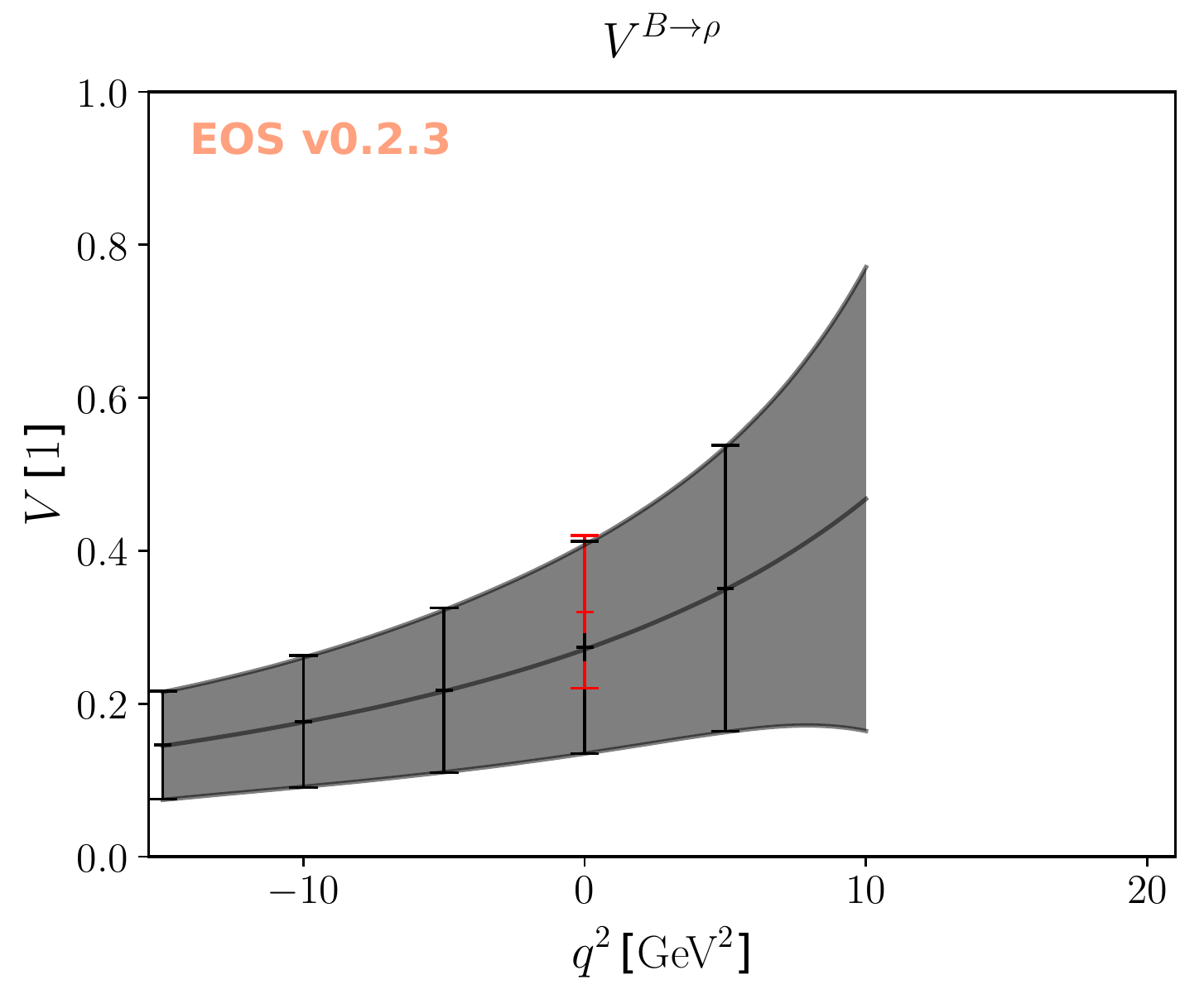}    \\
         \includegraphics[width=.40\textwidth]{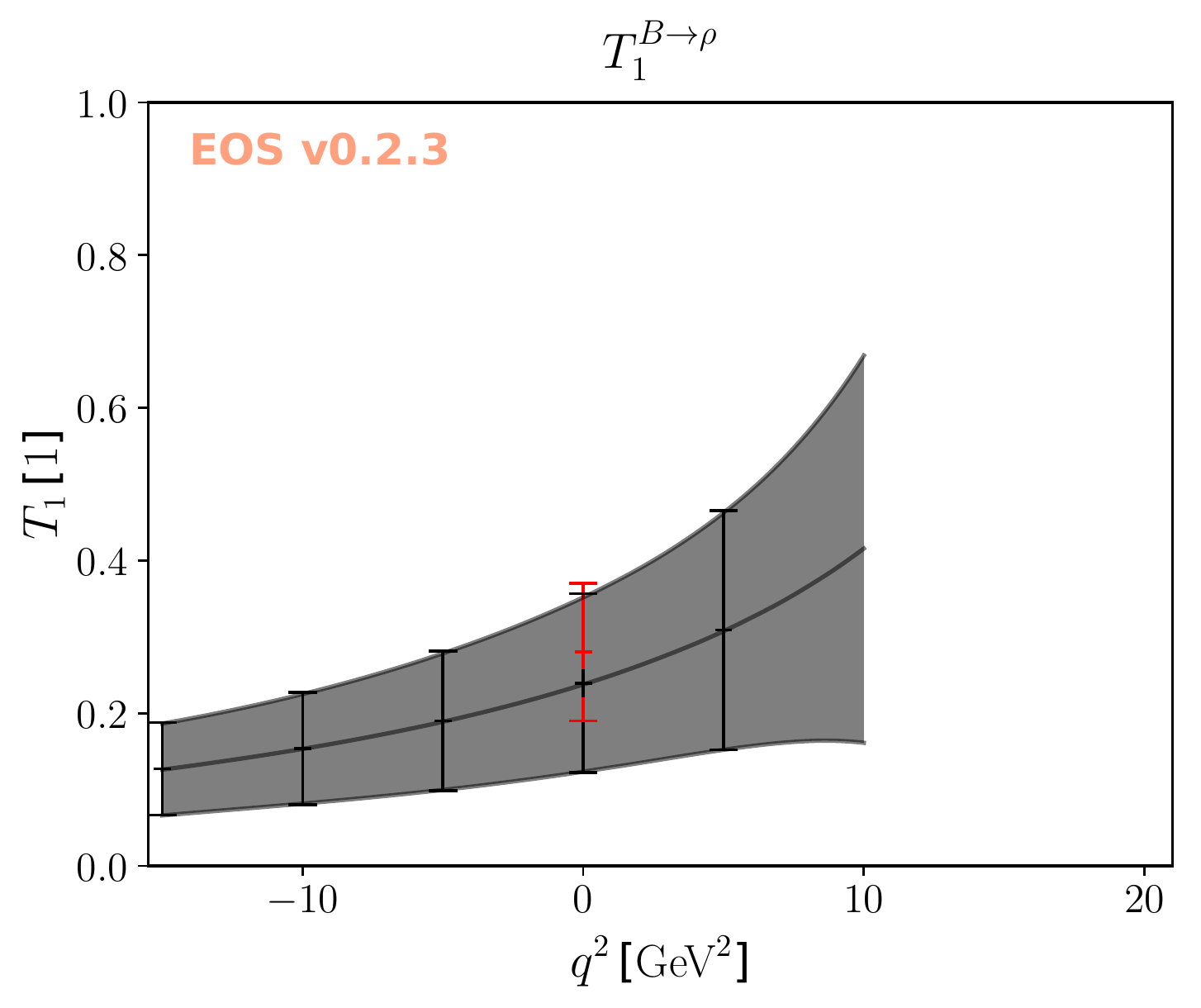}  &
         \includegraphics[width=.40\textwidth]{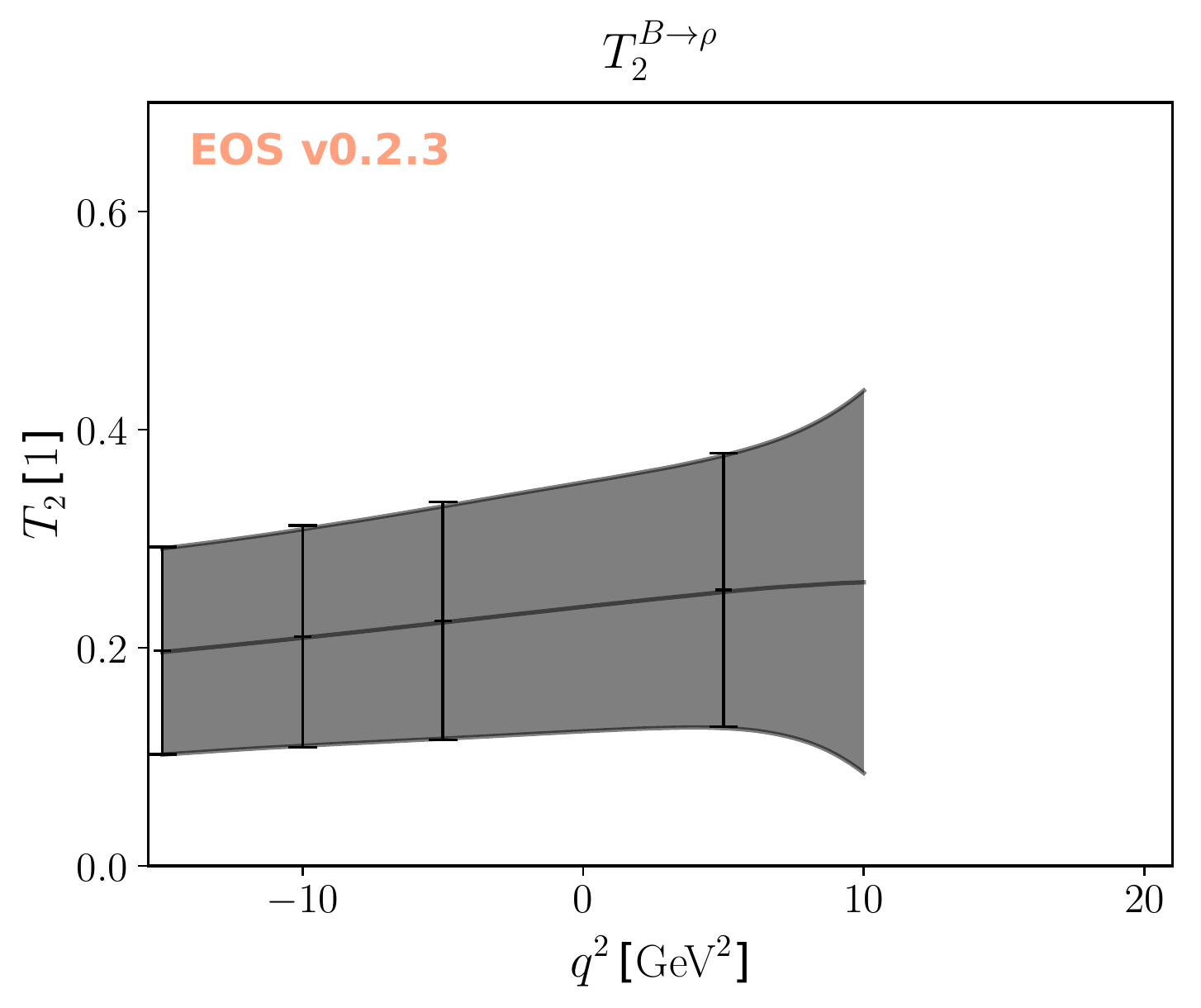}    \\
    \end{tabular}
    \centering
    \begin{tabular}{c}
         \includegraphics[width=.40\textwidth]{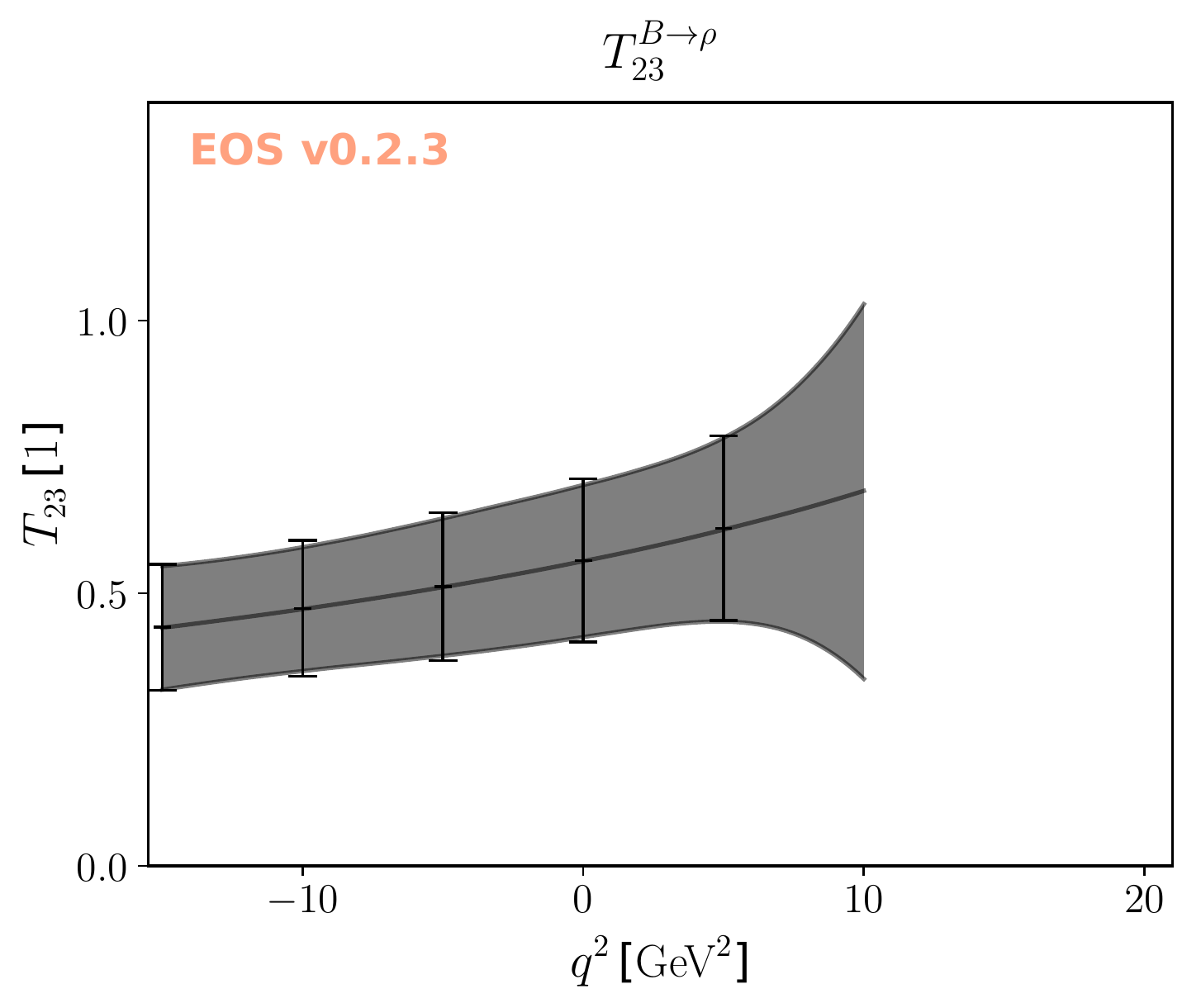}
    \end{tabular}
    \caption{%
    \label{tab:plots-B-to-rho}
    Plot of $B\to \rho$ form factors, no LQCD results available. For a description see \reffig{plots-B-to-pi}.
    }
\end{figure}

\begin{figure}[p]
    \centering
    \begin{tabular}{cc}
         \includegraphics[width=.40\textwidth]{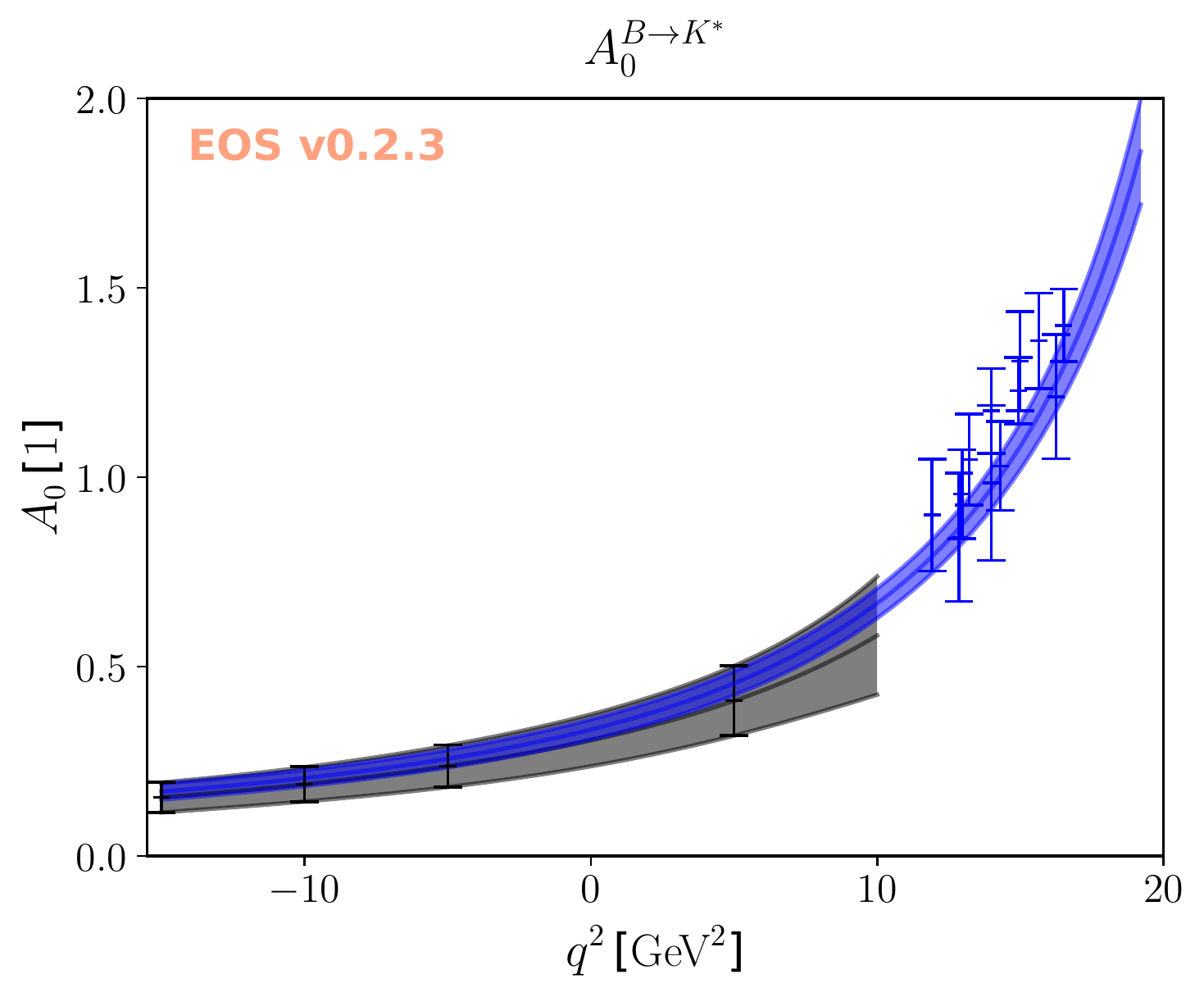}   &
         \includegraphics[width=.40\textwidth]{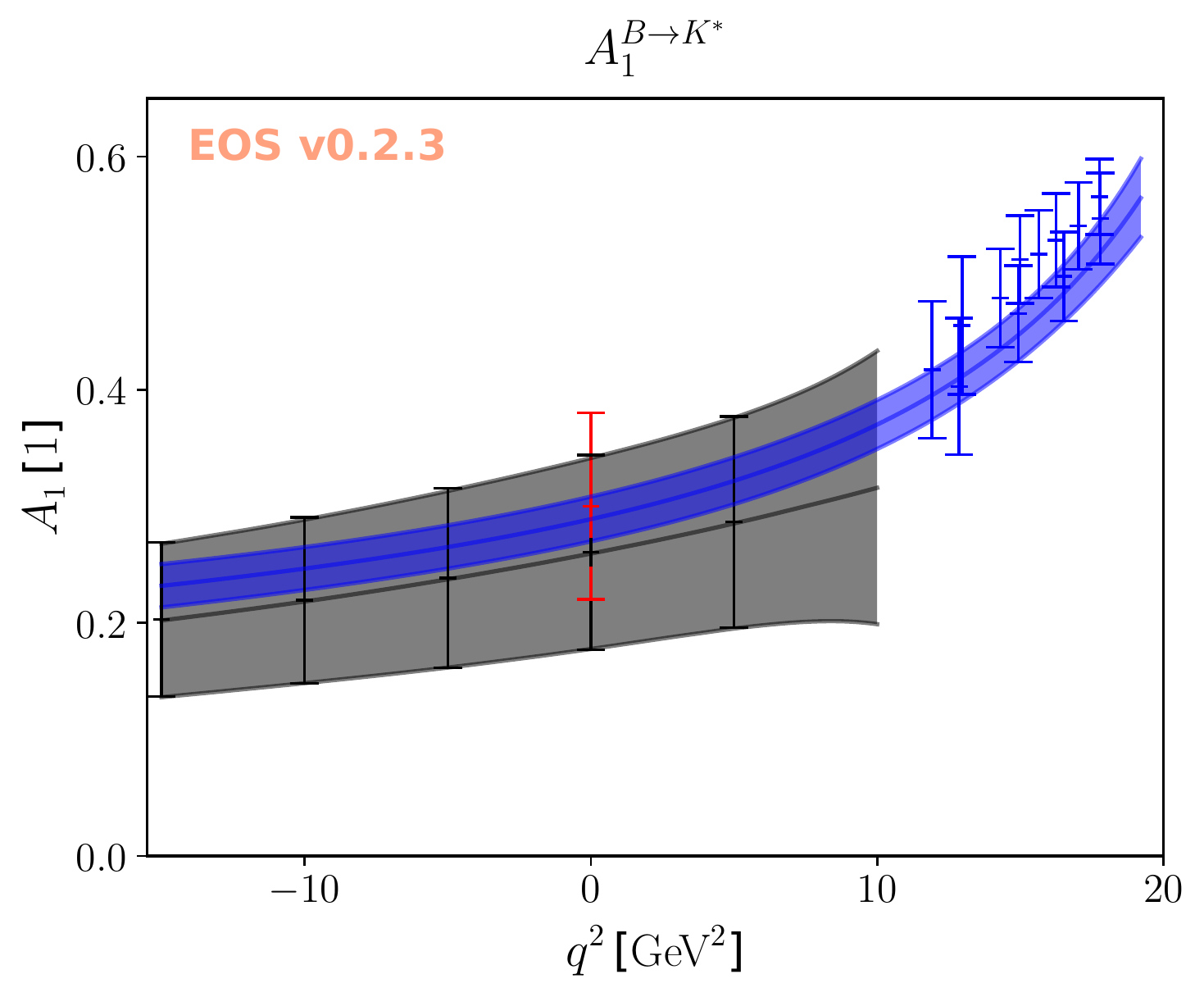}   \\
         \includegraphics[width=.40\textwidth]{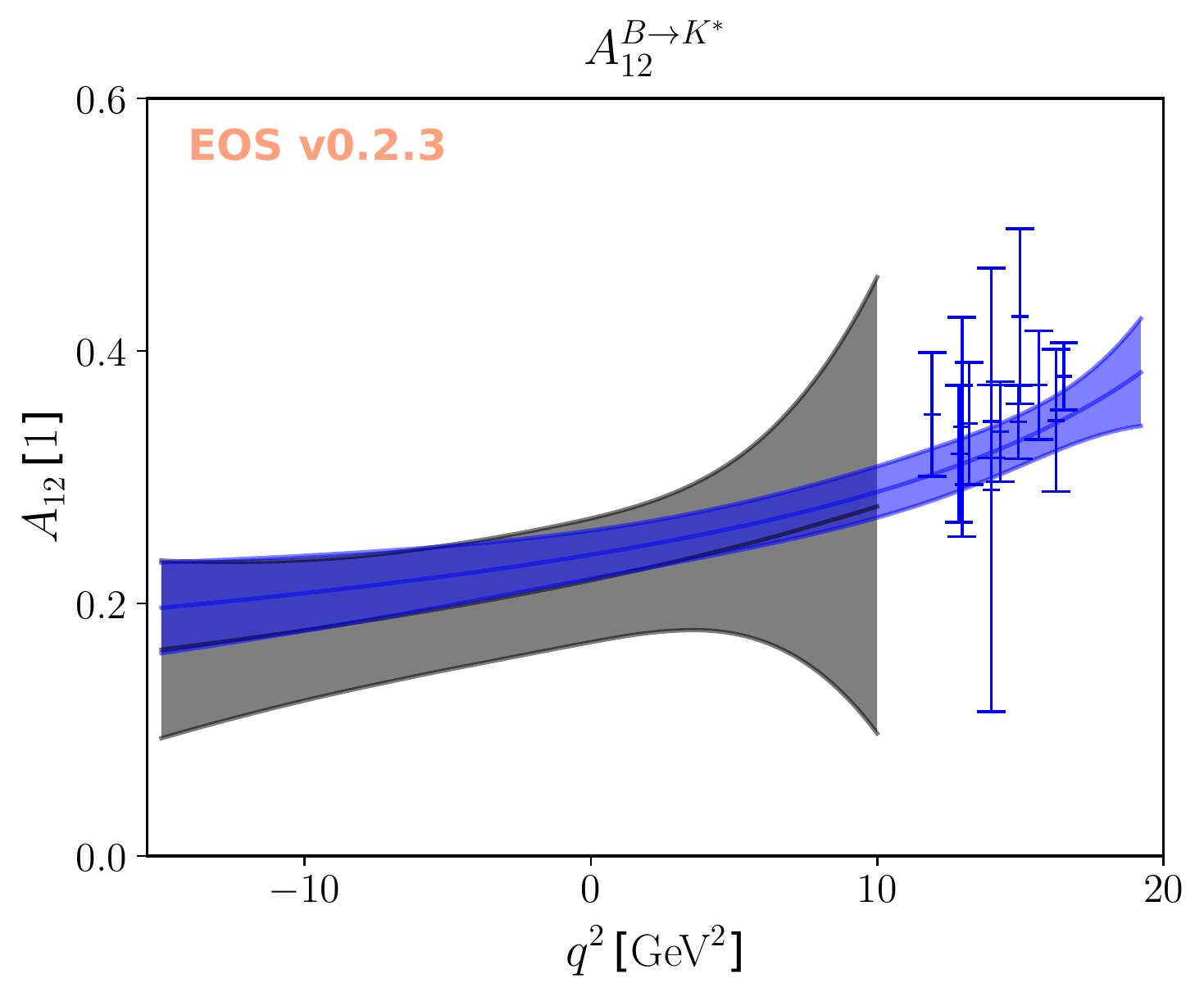}  &
         \includegraphics[width=.40\textwidth]{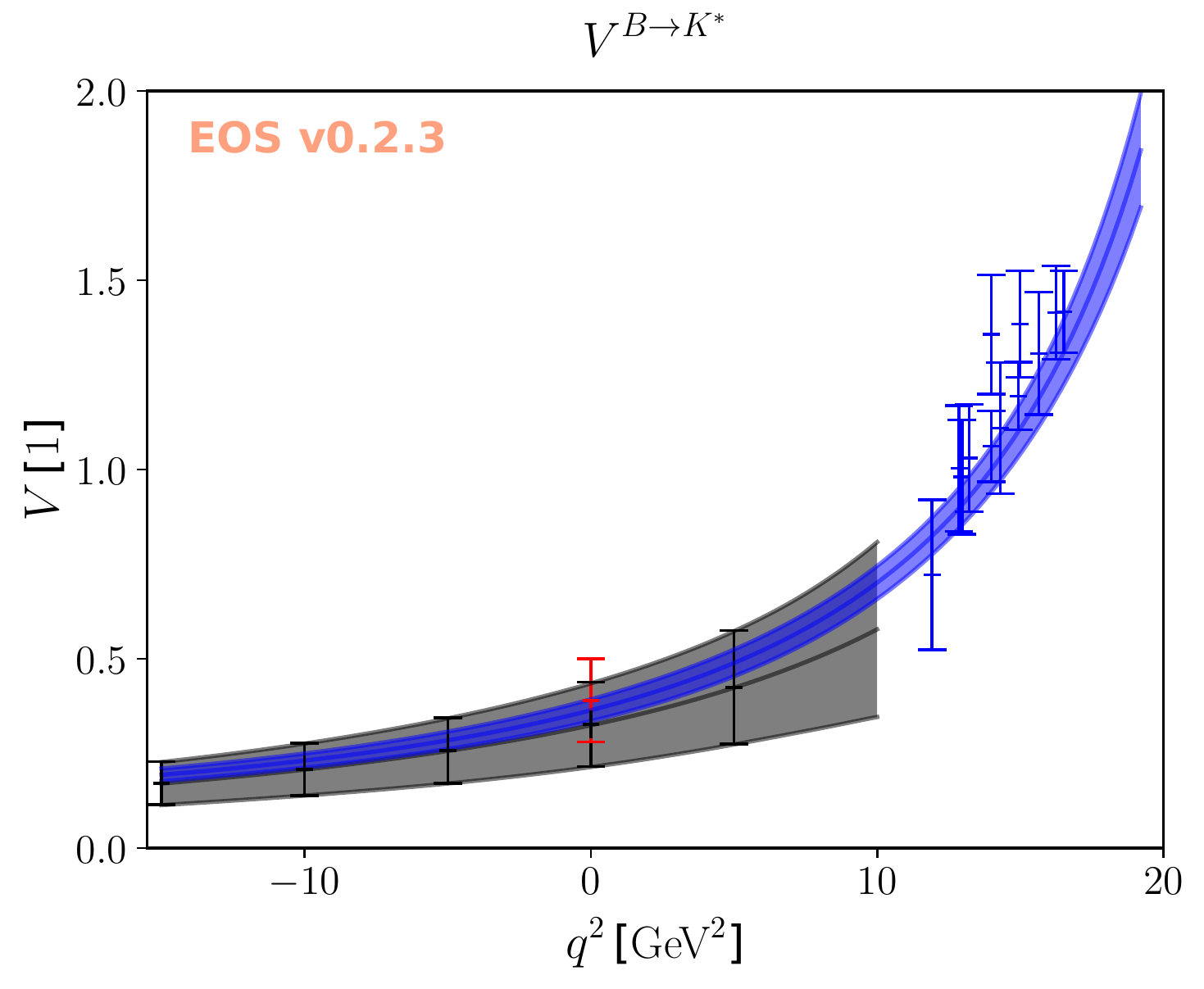}    \\
         \includegraphics[width=.40\textwidth]{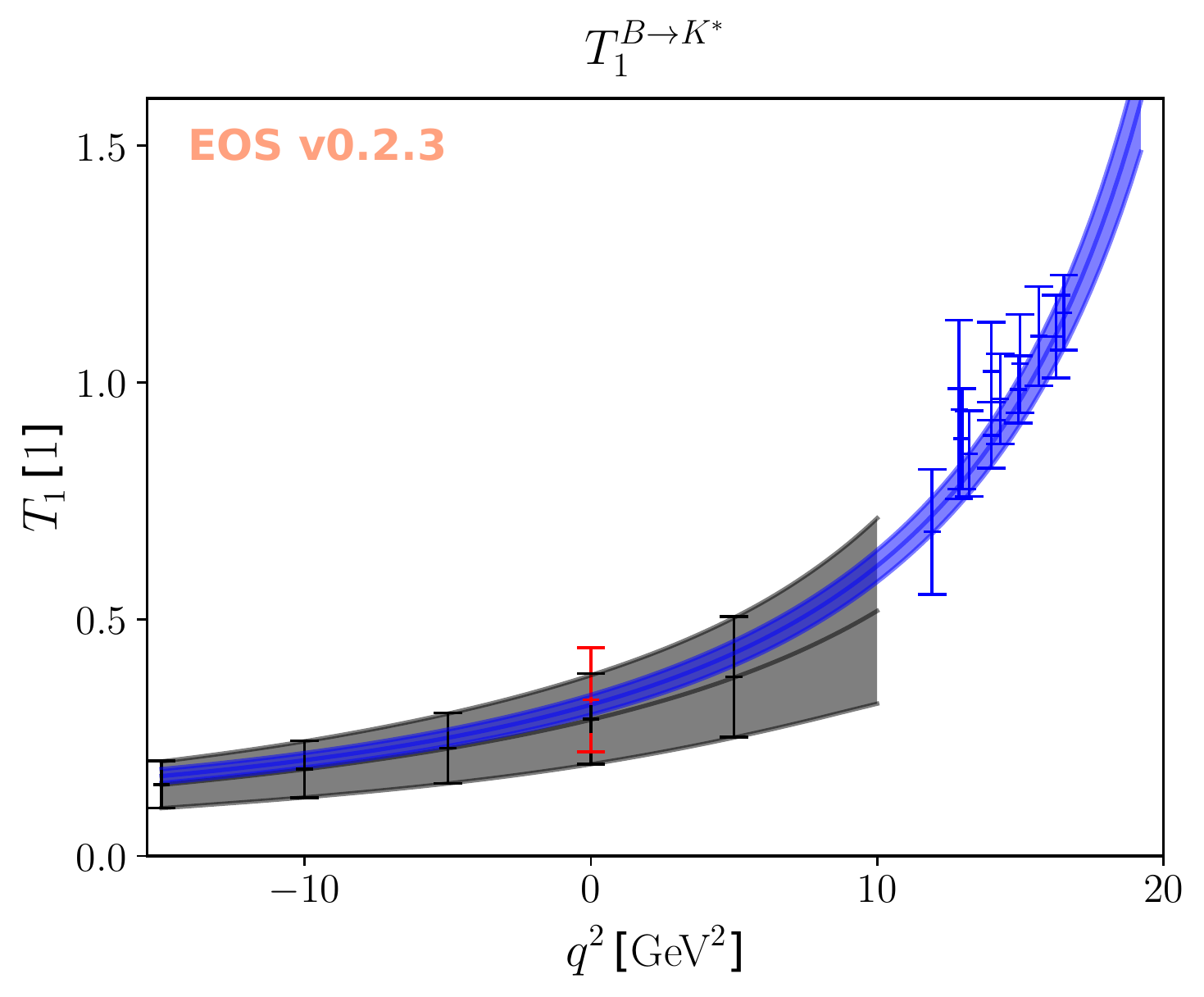}  &
         \includegraphics[width=.40\textwidth]{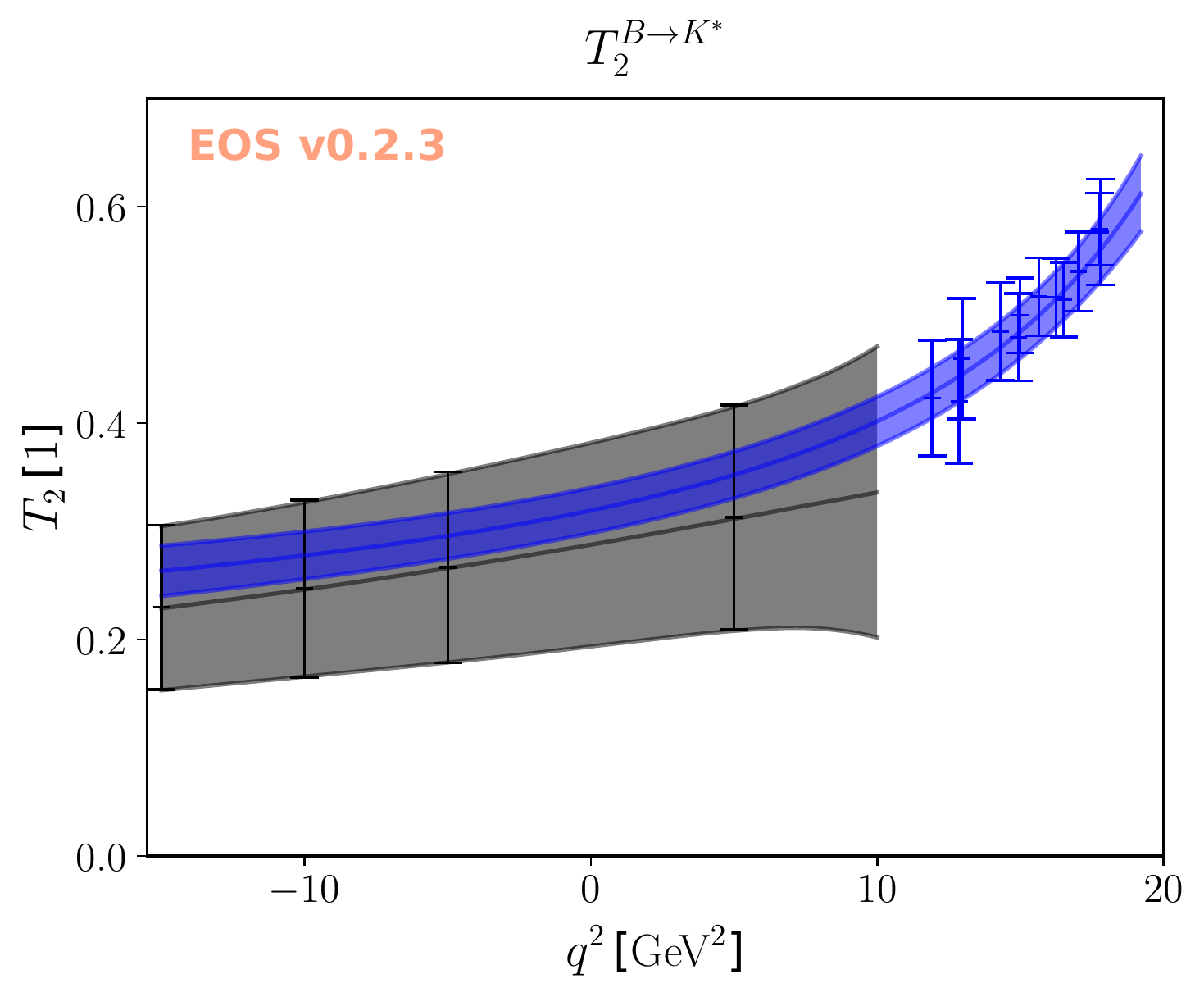}    \\
    \end{tabular}
    \centering
    \begin{tabular}{c}
         \includegraphics[width=.40\textwidth]{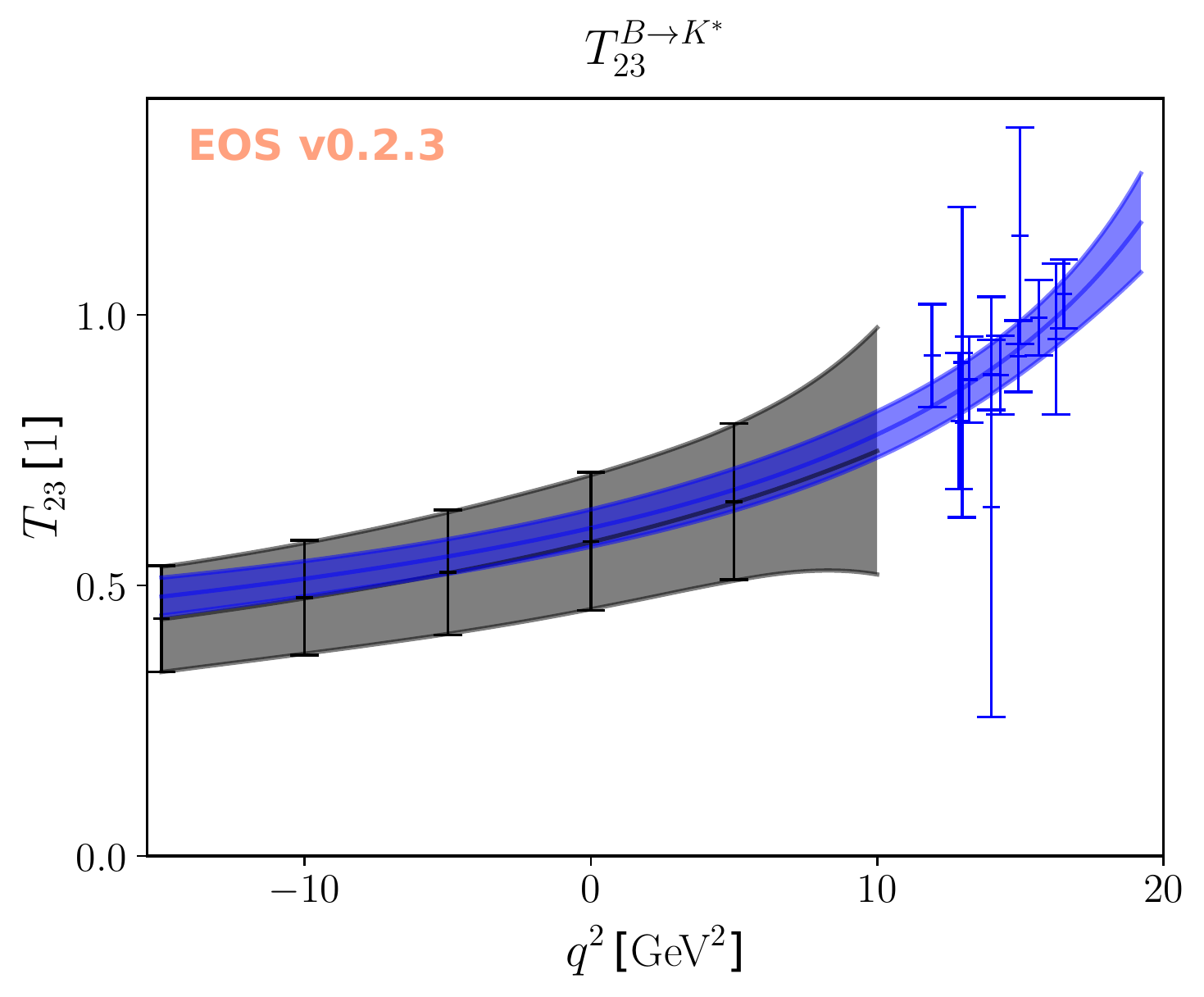}
    \end{tabular}
    \caption{%
    \label{tab:plots-B-to-Kstar}
    Plot of $B\to K^*$ form factors, LQCD results from refs.~\cite{Horgan:2013hoa,Horgan:2015vla}. For a description see \reffig{plots-B-to-pi}.
    }
\end{figure}

\begin{figure}[p]
    \centering
    \begin{tabular}{cc}
         \includegraphics[width=.40\textwidth]{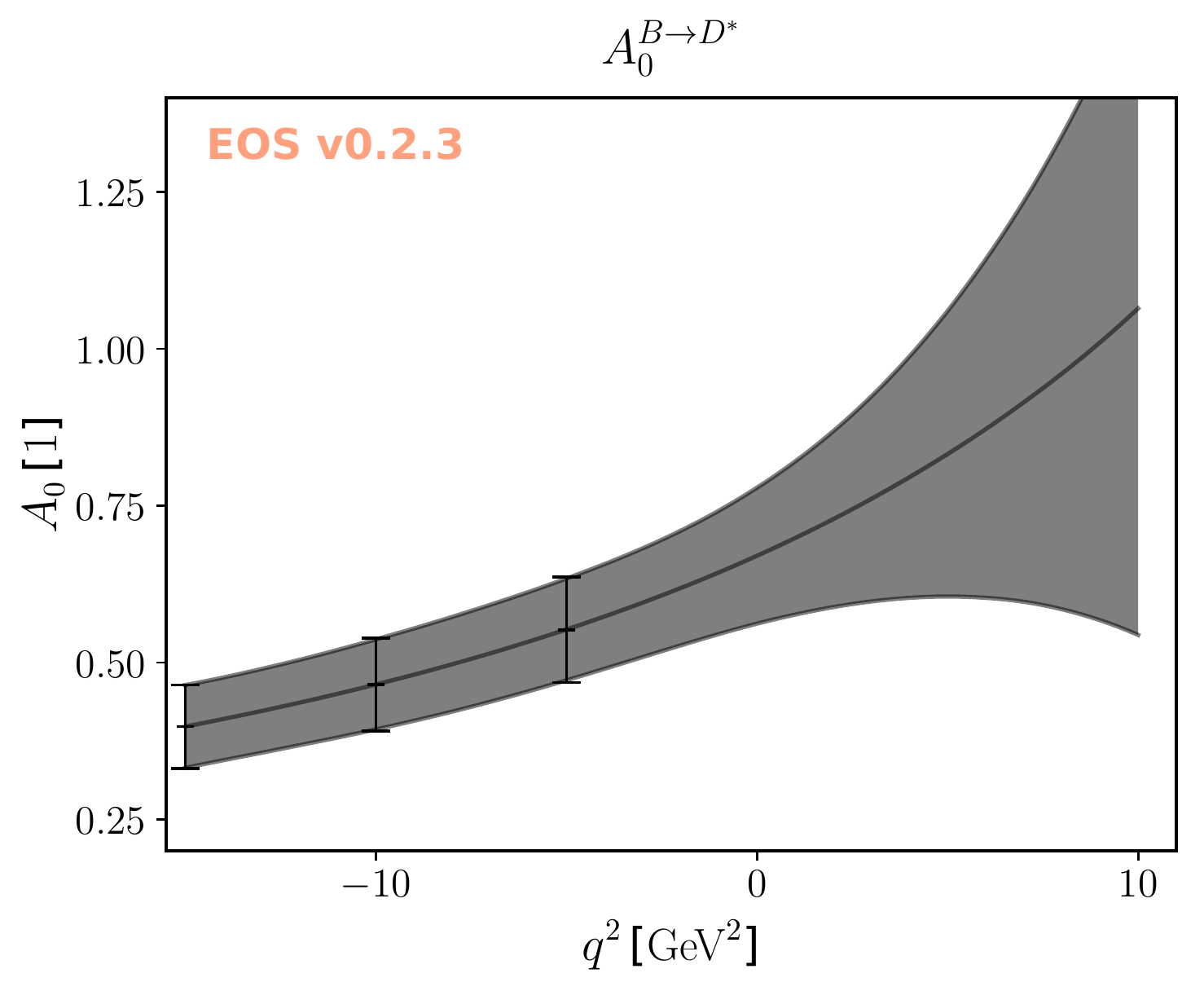}   &
         \includegraphics[width=.40\textwidth]{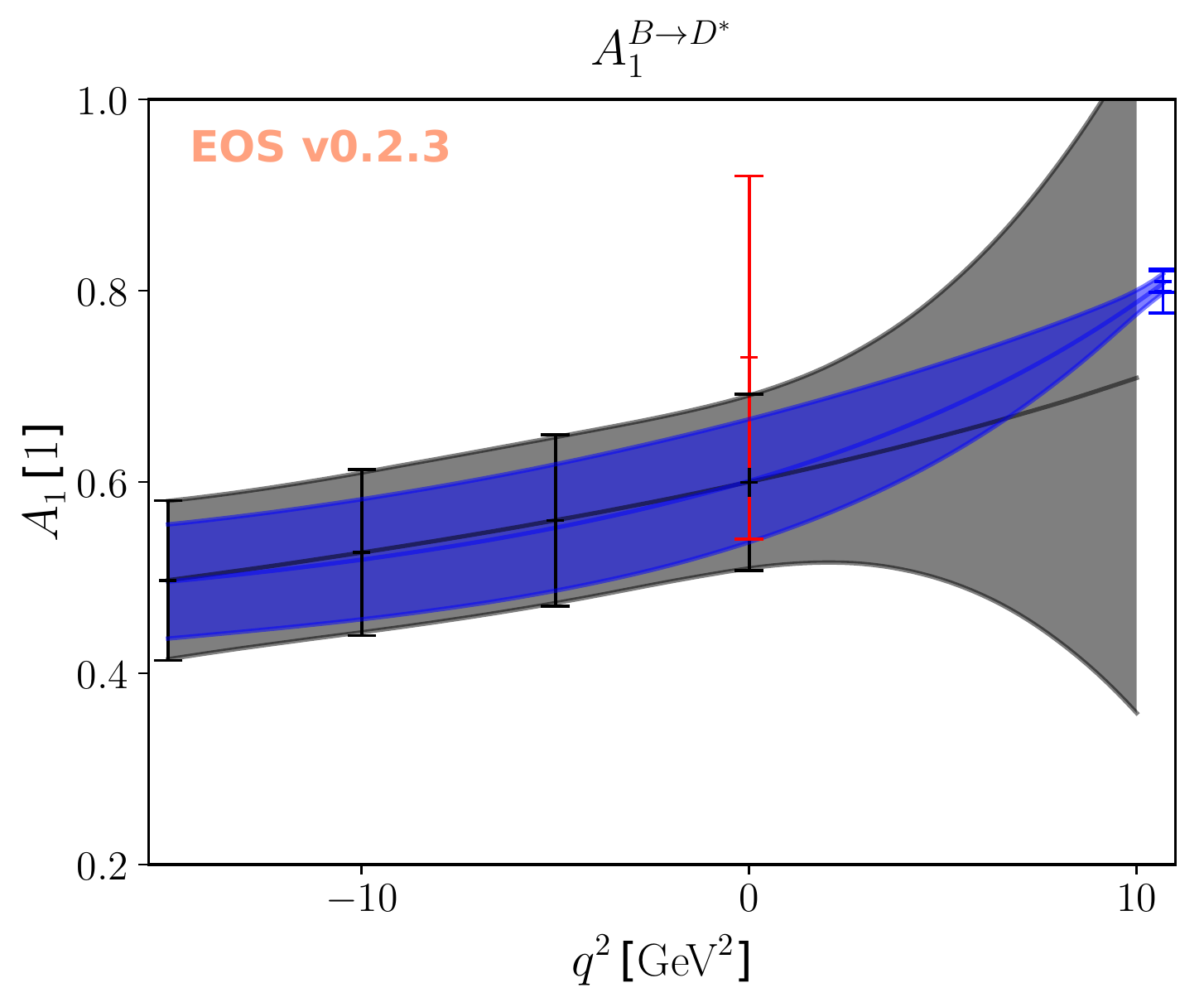}   \\
         \includegraphics[width=.40\textwidth]{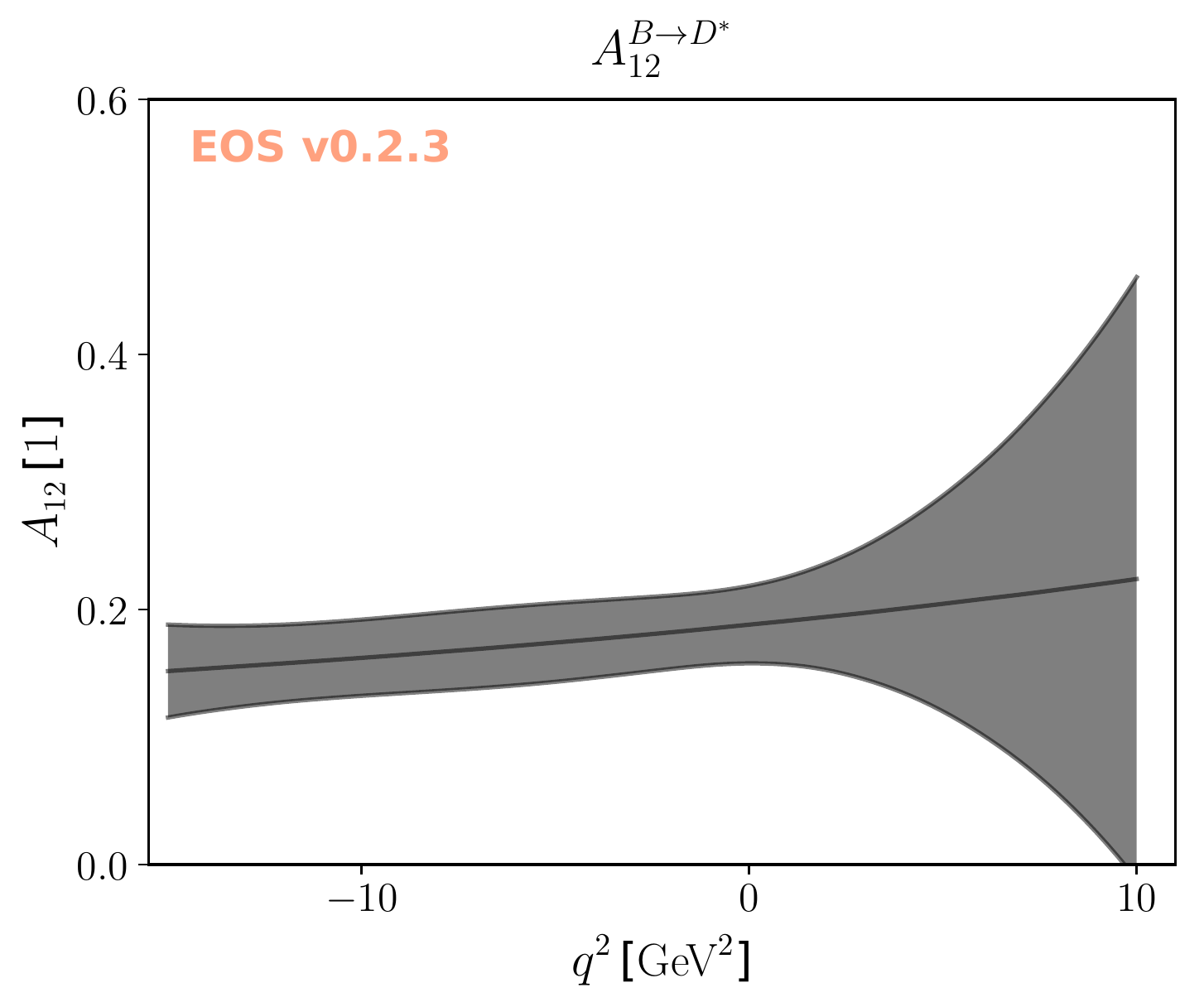}  &
         \includegraphics[width=.40\textwidth]{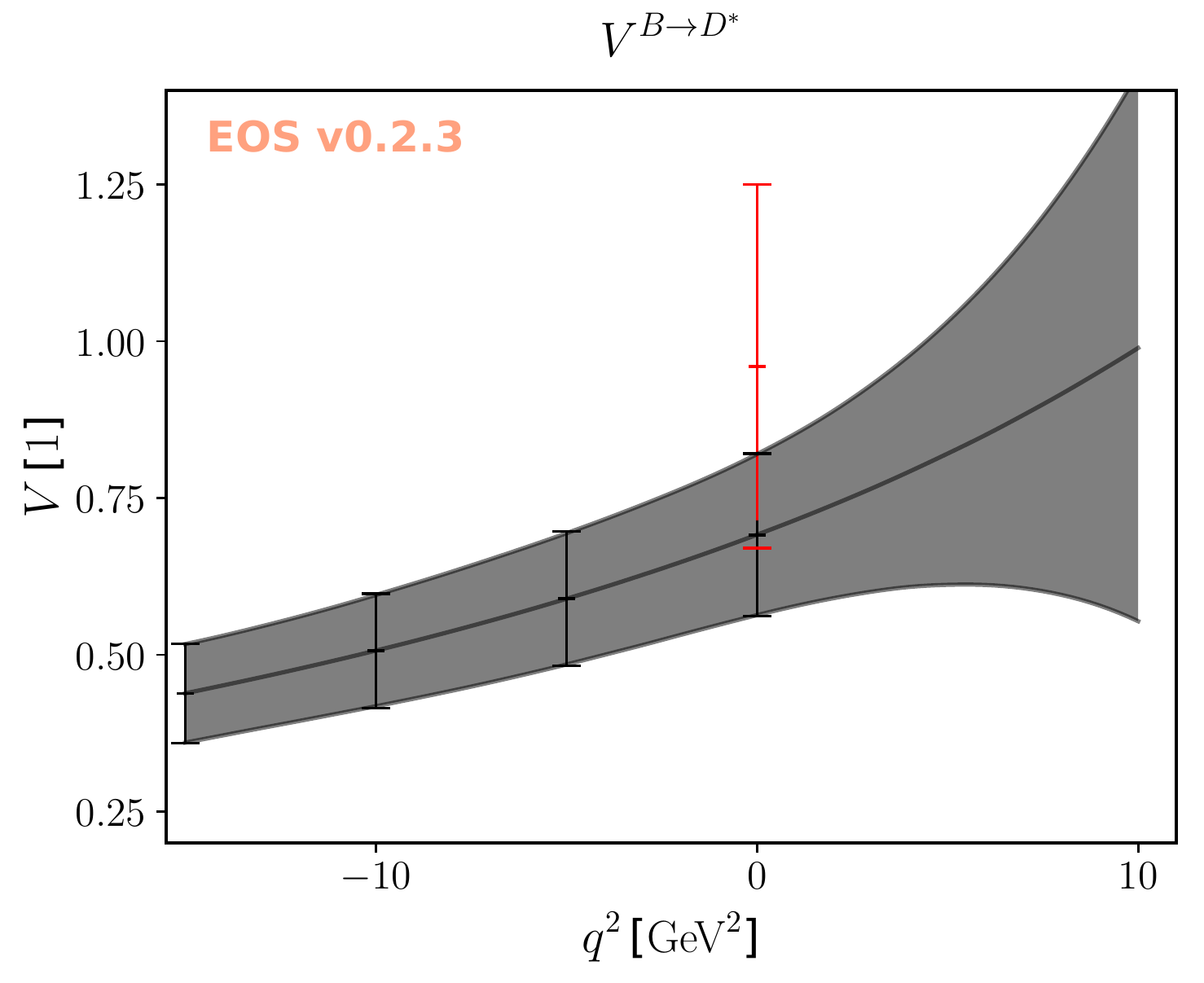}    \\
         \includegraphics[width=.40\textwidth]{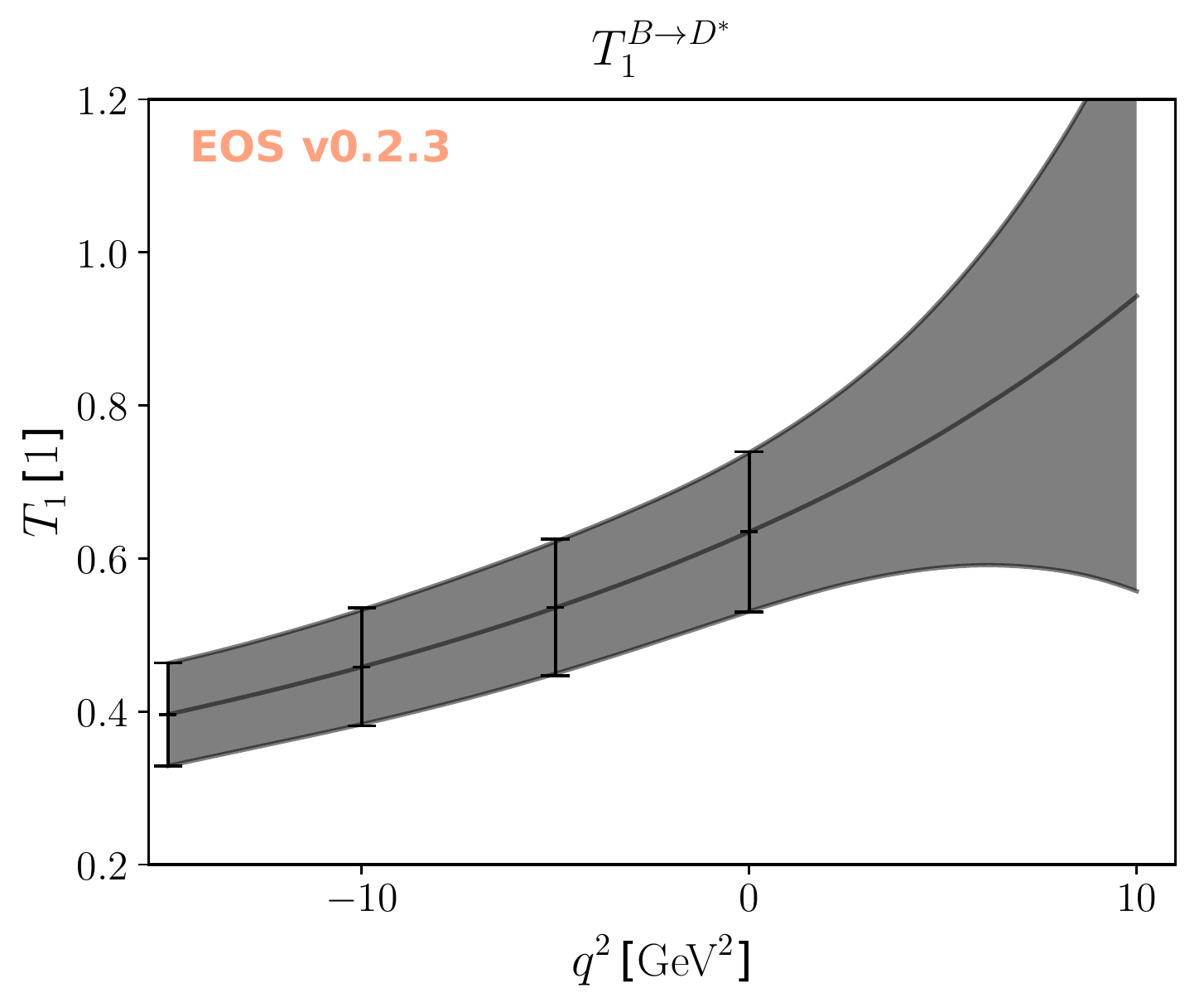}  &
         \includegraphics[width=.40\textwidth]{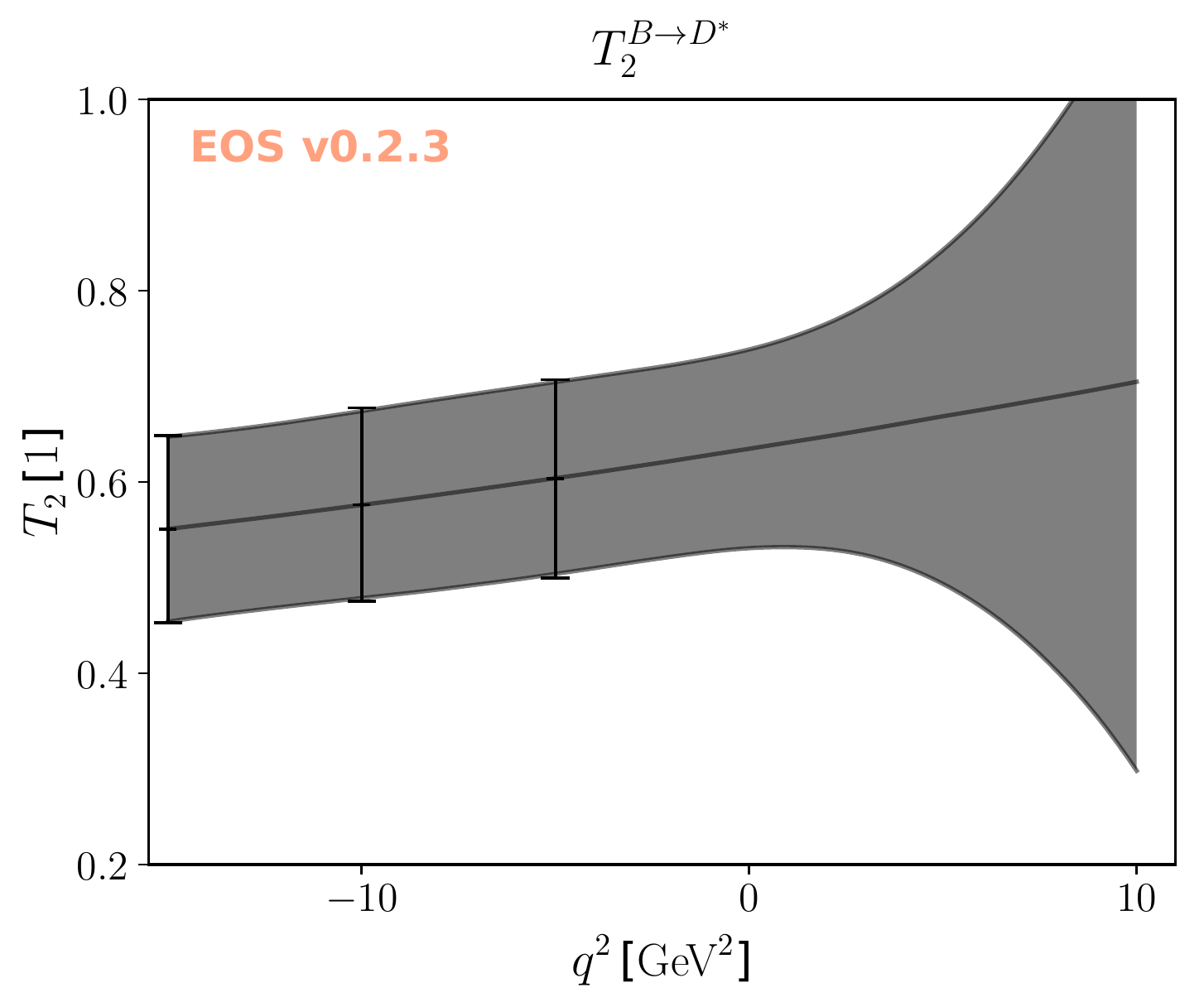}    \\
    \end{tabular}
    \centering
    \begin{tabular}{c}
         \includegraphics[width=.40\textwidth]{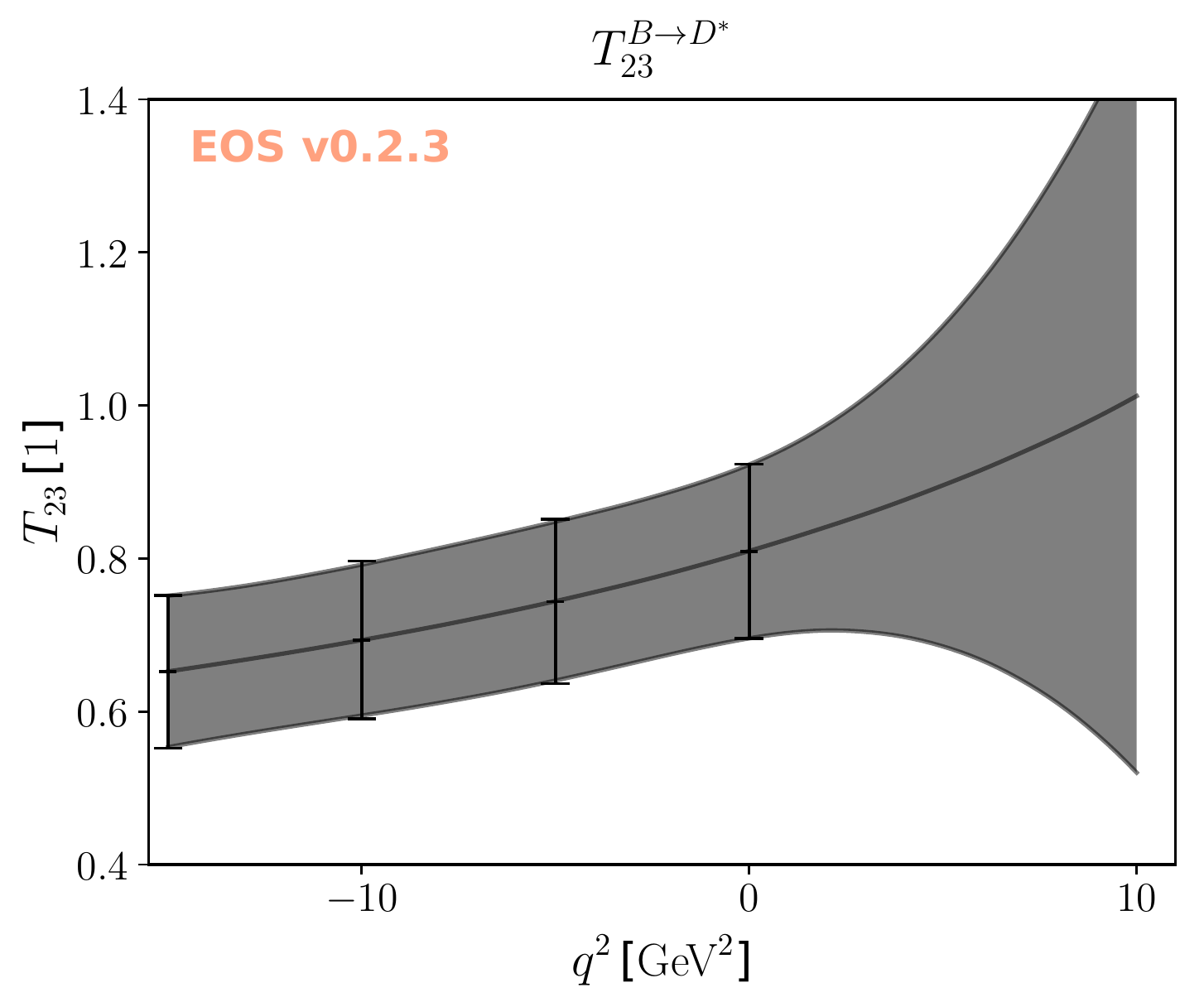}
    \end{tabular}
    \caption{%
    \label{tab:plots-B-to-Dstar}
    Plot of $B\to \bar{D}^*$ form factors, LQCD results from refs.~\cite{Bailey:2014tva,Harrison:2017fmw}. For a description see \reffig{plots-B-to-pi}.
    }
\end{figure}

\FloatBarrier

\bibliographystyle{JHEPmod}
\bibliography{references}

\end{document}

%% file: tables/tab-ff-details.tex
\begin{tabular}{|c|c|c|c|c|}
    \hline
                & \multicolumn{3}{|c|}{$2$ pt.} &\\
    form factor & $\phi_\pm$ & $g_+$ & $g_-^\text{WW}$ & $3$-pt. [$10^{-2}$] \\ \hline
\hline
    $\fp{ \pi  }$ & $0.28$ & $+0.00$ & $-0.06$ & $-0.00$ \\ \hline
    $\fT{ \pi  }$ & $0.25$ & $+0.01$ & $-0.07$ & $-0.29$ \\ \hline
\hline
    $\fp{ K    }$ & $0.35$ & $+0.00$ & $-0.08$ & $-0.01$ \\ \hline
    $\fT{ K    }$ & $0.33$ & $+0.02$ & $-0.09$ & $-0.37$ \\ \hline
\hline
    $\fp{ D    }$ & $0.84$ & $+0.02$ & $-0.21$ & $-0.03$ \\ \hline
    $\fT{ D    }$ & $0.65$ & $+0.33$ & $-0.41$ & $-0.52$ \\ \hline
\hline
    $\Aone{ \rho }$ & $0.28$ & $-0.08$ & $+0.01$ & $-0.19$ \\ \hline
    $\Aonetwo{ \rho }$ & $0.31$ & $+0.01$ & $-0.07$ & $-0.10$ \\ \hline
    $\V{ \rho }$ & $0.37$ & $-0.11$ & $-0.00$ & $-0.34$ \\ \hline
    $\Tone{ \rho }$ & $0.32$ & $-0.09$ & $+0.01$ & $-0.25$ \\ \hline
    $\Ttwothree{ \rho }$ & $0.69$ & $+0.07$ & $-0.18$ & $-0.96$ \\ \hline
\hline
    $\Aone{ K^*  }$ & $0.33$ & $-0.08$ & $+0.01$ & $-0.21$ \\ \hline
    $\Aonetwo{ K^*  }$ & $0.26$ & $+0.01$ & $-0.05$ & $-0.06$ \\ \hline
    $\V{ K^*  }$ & $0.44$ & $-0.12$ & $-0.00$ & $-0.38$ \\ \hline
    $\Tone{ K^*  }$ & $0.37$ & $-0.10$ & $+0.01$ & $-0.28$ \\ \hline
    $\Ttwothree{ K^*  }$ & $0.68$ & $+0.04$ & $-0.14$ & $-0.84$ \\ \hline
\hline
    $\Aone{ D^*  }$ & $0.73$ & $-0.17$ & $+0.04$ & $-0.10$ \\ \hline
    $\Aonetwo{ D^*  }$ & $0.21$ & $+0.01$ & $-0.03$ & $-0.01$ \\ \hline
    $\V{ D^*  }$ & $1.02$ & $-0.29$ & $-0.04$ & $-0.38$ \\ \hline
    $\Tone{ D^*  }$ & $0.83$ & $-0.21$ & $+0.01$ & $-0.19$ \\ \hline
    $\Ttwothree{ D^*  }$ & $0.88$ & $+0.08$ & $-0.15$ & $-0.37$ \\ \hline
\end{tabular}

%% file: tables/tab-ff-comparison.tex
\begin{tabular}{|c|c|c|c|}
    \hline
    form factor & our result & literature & reference \\ \hline
\hline
    \multirow{ 6 }{*}{ $\fp{ \pi  }$ } & \multirow{ 6 }{*}{ $0.21 \pm 0.07$ } & $0.258 \pm 0.031$ & \cite{Ball:2004ye} \\
    & & $0.25 \pm 0.05$ & \cite{Khodjamirian:2006st} \\
    & & $0.28 \pm 0.05$ & \cite{Khodjamirian:2011ub} \\
    & & $0.31 \pm 0.02$ & \cite{Imsong:2014oqa} \\
    & & $0.281 \pm 0.038$ & \cite{Wang:2015vgv} \\
    & & $0.301 \pm 0.023$ & \cite{Khodjamirian:2017fxg} \\
    \hline
    \multirow{ 4 }{*}{ $\fT{ \pi  }$ } & \multirow{ 4 }{*}{ $0.19 \pm 0.06$ } & $0.253 \pm 0.028$ & \cite{Ball:2004ye} \\
    & & $0.21 \pm 0.04$ & \cite{Khodjamirian:2006st} \\
    & & $0.273 \pm 0.021$ & \cite{Khodjamirian:2017fxg} \\
    & & $0.26 \pm 0.06$ & \cite{Lu:2018cfc} \\
    \hline
\hline
    \multirow{ 4 }{*}{ $\fp{ K    }$ } & \multirow{ 4 }{*}{ $0.27 \pm 0.08$ } & $0.331 \pm 0.041$ & \cite{Ball:2004ye} \\
    & & $0.31 \pm 0.04$ & \cite{Khodjamirian:2006st} \\
    & & $0.395 \pm 0.033$ & \cite{Khodjamirian:2017fxg} \\
    & & $0.364 \pm 0.05$ & \cite{Lu:2018cfc} \\
    \hline
    \multirow{ 4 }{*}{ $\fT{ K    }$ } & \multirow{ 4 }{*}{ $0.25 \pm 0.07$ } & $0.358 \pm 0.037$ & \cite{Ball:2004ye} \\
    & & $0.27 \pm 0.04$ & \cite{Khodjamirian:2006st} \\
    & & $0.381 \pm 0.027$ & \cite{Khodjamirian:2017fxg} \\
    & & $0.363 \pm 0.08$ & \cite{Lu:2018cfc} \\
    \hline
\hline
    \multirow{ 2 }{*}{ $\fp{ D    }$ } & \multirow{ 2 }{*}{ $0.65 \pm 0.08$ } & $0.69 \pm 0.2$ & \cite{Faller:2008tr} \\
    & & $0.673 \pm 0.063$ & \cite{Wang:2017jow} \\
    \hline
    \multirow{ 1 }{*}{ $\fT{ D    }$ } & \multirow{ 1 }{*}{ $0.57 \pm 0.05$ } & --- & --- \\
    \hline
\hline
    \multirow{ 2 }{*}{ $\Aone{ \rho }$ } & \multirow{ 2 }{*}{ $0.22 \pm 0.10$ } & $0.24 \pm 0.08$ & \cite{Khodjamirian:2006st} \\
    & & $0.262 \pm 0.026$ & \cite{Straub:2015ica} \\
    \hline
    \multirow{ 1 }{*}{ $\Atwo{ \rho }$ } & \multirow{ 1 }{*}{ $0.19 \pm 0.11$ } & $0.21 \pm 0.09$ & \cite{Khodjamirian:2006st} \\
    \hline
    \multirow{ 2 }{*}{ $\V{ \rho }$ } & \multirow{ 2 }{*}{ $0.27 \pm 0.14$ } & $0.32 \pm 0.10$ & \cite{Khodjamirian:2006st} \\
    & & $0.327 \pm 0.031$ & \cite{Straub:2015ica} \\
    \hline
    \multirow{ 2 }{*}{ $\Tone{ \rho }$ } & \multirow{ 2 }{*}{ $0.24 \pm 0.12$ } & $0.28 \pm 0.09$ & \cite{Khodjamirian:2006st} \\
    & & $0.272 \pm 0.026$ & \cite{Straub:2015ica} \\
    \hline
    \multirow{ 1 }{*}{ $\Ttwothree{ \rho }$ } & \multirow{ 1 }{*}{ $0.56 \pm 0.15$ } & $0.747 \pm 0.076$ & \cite{Straub:2015ica} \\
    \hline
\hline
    \multirow{ 2 }{*}{ $\Aone{ K^*  }$ } & \multirow{ 2 }{*}{ $0.26 \pm 0.08$ } & $0.30 \pm 0.08$ & \cite{Khodjamirian:2006st} \\
    & & $0.269 \pm 0.029$ & \cite{Straub:2015ica} \\
    \hline
    \multirow{ 1 }{*}{ $\Atwo{ K^*  }$ } & \multirow{ 1 }{*}{ $0.24 \pm 0.09$ } & $0.26 \pm 0.08$ & \cite{Khodjamirian:2006st} \\
    \hline
    \multirow{ 2 }{*}{ $\V{ K^*  }$ } & \multirow{ 2 }{*}{ $0.33 \pm 0.11$ } & $0.39 \pm 0.11$ & \cite{Khodjamirian:2006st} \\
    & & $0.341 \pm 0.036$ & \cite{Straub:2015ica} \\
    \hline
    \multirow{ 2 }{*}{ $\Tone{ K^*  }$ } & \multirow{ 2 }{*}{ $0.29 \pm 0.10$ } & $0.33 \pm 0.10$ & \cite{Khodjamirian:2006st} \\
    & & $0.282 \pm 0.031$ & \cite{Straub:2015ica} \\
    \hline
    \multirow{ 1 }{*}{ $\Ttwothree{ K^*  }$ } & \multirow{ 1 }{*}{ $0.58 \pm 0.13$ } & $0.668 \pm 0.083$ & \cite{Straub:2015ica} \\
    \hline
\hline
    \multirow{ 1 }{*}{ $\Aone{ D^*  }$ } & \multirow{ 1 }{*}{ $0.60 \pm 0.09$ } & $0.73 \pm 0.19$ & \cite{Faller:2008tr} \\
    \hline
    \multirow{ 1 }{*}{ $\Atwo{ D^*  }$ } & \multirow{ 1 }{*}{ $0.51 \pm 0.09$ } & $0.66 \pm 0.30$ & \cite{Faller:2008tr} \\
    \hline
    \multirow{ 1 }{*}{ $\V{ D^*  }$ } & \multirow{ 1 }{*}{ $0.69 \pm 0.13$ } & $0.96 \pm 0.29$ & \cite{Faller:2008tr} \\
    \hline
    \multirow{ 1 }{*}{ $\Tone{ D^*  }$ } & \multirow{ 1 }{*}{ $0.63 \pm 0.10$ } & --- & --- \\
    \hline
    \multirow{ 1 }{*}{ $\Ttwothree{ D^*  }$ } & \multirow{ 1 }{*}{ $0.81 \pm 0.11$ } & --- & --- \\
    \hline
\end{tabular}

%% file: draft.bbl
\providecommand{\href}[2]{#2}\begingroup\raggedright\begin{thebibliography}{100}

\bibitem{Albrecht:2017odf}
J.~Albrecht, F.~Bernlochner, M.~Kenzie, S.~Reichert, D.~Straub and A.~Tully,
  \emph{{Future prospects for exploring present day anomalies in flavour
  physics measurements with Belle II and LHCb}},
  \href{https://arxiv.org/abs/1709.10308}{{\ttfamily 1709.10308}}.

\bibitem{Albrecht:2018vsa}
J.~Albrecht, S.~Reichert and D.~van Dyk, \emph{{Status of rare exclusive $B$
  meson decays in 2018}},
  \href{https://doi.org/10.1142/S0217751X18300168}{\emph{Int. J. Mod. Phys.}
  {\bfseries A33} (2018) 1830016}
  [\href{https://arxiv.org/abs/1806.05010}{{\ttfamily 1806.05010}}].

\bibitem{Wilson:1974sk}
K.~G. Wilson, \emph{{Confinement of Quarks}},
  [,319(1974)]\href{https://doi.org/10.1103/PhysRevD.10.2445}{\emph{Phys. Rev.}
  {\bfseries D10} (1974) 2445}.

\bibitem{Bigi:2017jbd}
D.~Bigi, P.~Gambino and S.~Schacht, \emph{{$R(D^*)$, $|V_{cb}|$, and the Heavy
  Quark Symmetry relations between form factors}},
  \href{https://doi.org/10.1007/JHEP11(2017)061}{\emph{JHEP} {\bfseries 11}
  (2017) 061} [\href{https://arxiv.org/abs/1707.09509}{{\ttfamily
  1707.09509}}].

\bibitem{Jung:2018lfu}
M.~Jung and D.~M. Straub, \emph{{Constraining new physics in $b\to c\ell\nu$
  transitions}},  \href{https://arxiv.org/abs/1801.01112}{{\ttfamily
  1801.01112}}.

\bibitem{Khodjamirian:2005ea}
A.~Khodjamirian, T.~Mannel and N.~Offen, \emph{{B-meson distribution amplitude
  from the $B \to \pi$ form-factor}},
  \href{https://doi.org/10.1016/j.physletb.2005.06.021}{\emph{Phys. Lett.}
  {\bfseries B620} (2005) 52}
  [\href{https://arxiv.org/abs/hep-ph/0504091}{{\ttfamily hep-ph/0504091}}].

\bibitem{Khodjamirian:2006st}
A.~Khodjamirian, T.~Mannel and N.~Offen, \emph{{Form-factors from light-cone
  sum rules with B-meson distribution amplitudes}},
  \href{https://doi.org/10.1103/PhysRevD.75.054013}{\emph{Phys. Rev.}
  {\bfseries D75} (2007) 054013}
  [\href{https://arxiv.org/abs/hep-ph/0611193}{{\ttfamily hep-ph/0611193}}].

\bibitem{Faller:2008tr}
S.~Faller, A.~Khodjamirian, C.~Klein and T.~Mannel, \emph{{$B \to D^{(*)}$ Form
  Factors from QCD Light-Cone Sum Rules}},
  \href{https://doi.org/10.1140/epjc/s10052-009-0968-4}{\emph{Eur. Phys. J.}
  {\bfseries C60} (2009) 603}
  [\href{https://arxiv.org/abs/0809.0222}{{\ttfamily 0809.0222}}].

\bibitem{Khodjamirian:2010vf}
A.~Khodjamirian, T.~Mannel, A.~A. Pivovarov and Y.~M. Wang, \emph{{Charm-loop
  effect in $B \to K^{(*)} \ell^{+} \ell^{-}$ and $B\to K^*\gamma$}},
  \href{https://doi.org/10.1007/JHEP09(2010)089}{\emph{JHEP} {\bfseries 09}
  (2010) 089} [\href{https://arxiv.org/abs/1006.4945}{{\ttfamily 1006.4945}}].

\bibitem{Cheng:2017smj}
S.~Cheng, A.~Khodjamirian and J.~Virto, \emph{{$B\to\pi\pi$ Form Factors from
  Light-Cone Sum Rules with $B$-meson Distribution Amplitudes}},
  \href{https://doi.org/10.1007/JHEP05(2017)157}{\emph{JHEP} {\bfseries 05}
  (2017) 157} [\href{https://arxiv.org/abs/1701.01633}{{\ttfamily
  1701.01633}}].

\bibitem{Ball:2004ye}
P.~Ball and R.~Zwicky, \emph{{New results on $B \to \pi, K, \eta$ decay
  formfactors from light-cone sum rules}},
  \href{https://doi.org/10.1103/PhysRevD.71.014015}{\emph{Phys. Rev.}
  {\bfseries D71} (2005) 014015}
  [\href{https://arxiv.org/abs/hep-ph/0406232}{{\ttfamily hep-ph/0406232}}].

\bibitem{Ball:2004rg}
P.~Ball and R.~Zwicky, \emph{{$B_{d,s} \to \rho, \omega, K^*, \phi$ decay
  form-factors from light-cone sum rules revisited}},
  \href{https://doi.org/10.1103/PhysRevD.71.014029}{\emph{Phys. Rev.}
  {\bfseries D71} (2005) 014029}
  [\href{https://arxiv.org/abs/hep-ph/0412079}{{\ttfamily hep-ph/0412079}}].

\bibitem{Khodjamirian:2009ys}
A.~Khodjamirian, C.~Klein, T.~Mannel and N.~Offen, \emph{{Semileptonic charm
  decays $D \to \pi \ell \bar\nu_\ell$ and $D \to K \ell \bar\nu_\ell$ from QCD
  Light-Cone Sum Rules}},
  \href{https://doi.org/10.1103/PhysRevD.80.114005}{\emph{Phys. Rev.}
  {\bfseries D80} (2009) 114005}
  [\href{https://arxiv.org/abs/0907.2842}{{\ttfamily 0907.2842}}].

\bibitem{Khodjamirian:2011ub}
A.~Khodjamirian, T.~Mannel, N.~Offen and Y.~M. Wang, \emph{{$B \to \pi \ell
  \nu_l$ Width and $|V_{ub}|$ from QCD Light-Cone Sum Rules}},
  \href{https://doi.org/10.1103/PhysRevD.83.094031}{\emph{Phys. Rev.}
  {\bfseries D83} (2011) 094031}
  [\href{https://arxiv.org/abs/1103.2655}{{\ttfamily 1103.2655}}].

\bibitem{Bharucha:2012wy}
A.~Bharucha, \emph{{Two-loop Corrections to the $B \to \pi$ Form Factor from
  QCD Sum Rules on the Light-Cone and $|V_{ub}|$}},
  \href{https://doi.org/10.1007/JHEP05(2012)092}{\emph{JHEP} {\bfseries 05}
  (2012) 092} [\href{https://arxiv.org/abs/1203.1359}{{\ttfamily 1203.1359}}].

\bibitem{Imsong:2014oqa}
I.~Sentitemsu~Imsong, A.~Khodjamirian, T.~Mannel and D.~van Dyk,
  \emph{{Extrapolation and unitarity bounds for the $B \to \pi$ form factor}},
  \href{https://doi.org/10.1007/JHEP02(2015)126}{\emph{JHEP} {\bfseries 02}
  (2015) 126} [\href{https://arxiv.org/abs/1409.7816}{{\ttfamily 1409.7816}}].

\bibitem{Straub:2015ica}
A.~Bharucha, D.~M. Straub and R.~Zwicky, \emph{{$B\to V\ell^+\ell^-$ in the
  Standard Model from light-cone sum rules}},
  \href{https://doi.org/10.1007/JHEP08(2016)098}{\emph{JHEP} {\bfseries 08}
  (2016) 098} [\href{https://arxiv.org/abs/1503.05534}{{\ttfamily
  1503.05534}}].

\bibitem{Khodjamirian:2017fxg}
A.~Khodjamirian and A.~V. Rusov, \emph{{$B_{s}\to K \ell \nu_\ell$ and $B_{(s)}
  \to \pi (K) \ell^+\ell^-$ decays at large recoil and CKM matrix elements}},
  \href{https://doi.org/10.1007/JHEP08(2017)112}{\emph{JHEP} {\bfseries 08}
  (2017) 112} [\href{https://arxiv.org/abs/1703.04765}{{\ttfamily
  1703.04765}}].

\bibitem{DeFazio:2005dx}
F.~De~Fazio, T.~Feldmann and T.~Hurth, \emph{{Light-cone sum rules in
  soft-collinear effective theory}},  [Erratum: Nucl.
  Phys.B800,405(2008)]\href{https://doi.org/10.1016/j.nuclphysb.2008.03.022,
  10.1016/j.nuclphysb.2005.09.047}{\emph{Nucl. Phys.} {\bfseries B733} (2006)
  1} [\href{https://arxiv.org/abs/hep-ph/0504088}{{\ttfamily hep-ph/0504088}}].

\bibitem{DeFazio:2007hw}
F.~De~Fazio, T.~Feldmann and T.~Hurth, \emph{{SCET sum rules for $B \to P$ and
  $B \to V$ transition form factors}},
  \href{https://doi.org/10.1088/1126-6708/2008/02/031}{\emph{JHEP} {\bfseries
  02} (2008) 031} [\href{https://arxiv.org/abs/0711.3999}{{\ttfamily
  0711.3999}}].

\bibitem{Wang:2015vgv}
Y.-M. Wang and Y.-L. Shen, \emph{{QCD corrections to $B\to \pi$ form factors
  from light-cone sum rules}},
  \href{https://doi.org/10.1016/j.nuclphysb.2015.07.016}{\emph{Nucl. Phys.}
  {\bfseries B898} (2015) 563}
  [\href{https://arxiv.org/abs/1506.00667}{{\ttfamily 1506.00667}}].

\bibitem{Wang:2017jow}
Y.-M. Wang, Y.-B. Wei, Y.-L. Shen and C.-D. L{\"u}, \emph{{Perturbative
  corrections to $B \to \bar{D}$ form factors in QCD}},
  \href{https://doi.org/10.1007/JHEP06(2017)062}{\emph{JHEP} {\bfseries 06}
  (2017) 062} [\href{https://arxiv.org/abs/1701.06810}{{\ttfamily
  1701.06810}}].

\bibitem{Braun:2017liq}
V.~M. Braun, Y.~Ji and A.~N. Manashov, \emph{{Higher-twist B-meson Distribution
  Amplitudes in HQET}},
  \href{https://doi.org/10.1007/JHEP05(2017)022}{\emph{JHEP} {\bfseries 05}
  (2017) 022} [\href{https://arxiv.org/abs/1703.02446}{{\ttfamily
  1703.02446}}].

\bibitem{Geyer:2005fb}
B.~Geyer and O.~Witzel, \emph{{B-meson distribution amplitudes of geometric
  twist vs. dynamical twist}},
  \href{https://doi.org/10.1103/PhysRevD.72.034023}{\emph{Phys. Rev.}
  {\bfseries D72} (2005) 034023}
  [\href{https://arxiv.org/abs/hep-ph/0502239}{{\ttfamily hep-ph/0502239}}].

\bibitem{Kawamura:2001jv}
H.~Kawamura, J.~Kodaira, C.-F. Qiao and K.~Tanaka, \emph{{Transverse momentum
  distribution in B mesons in the heavy quark limit: The Wandzura-Wilczek
  part}}, \href{https://doi.org/10.1142/S021773230300985X}{\emph{Mod. Phys.
  Lett.} {\bfseries A18} (2003) 799}
  [\href{https://arxiv.org/abs/hep-ph/0112174}{{\ttfamily hep-ph/0112174}}].

\bibitem{Faller:2013dwa}
S.~Faller, T.~Feldmann, A.~Khodjamirian, T.~Mannel and D.~van Dyk,
  \emph{{Disentangling the Decay Observables in $B^- \to
  \pi^+\pi^-\ell^-\bar\nu_\ell$}},
  \href{https://doi.org/10.1103/PhysRevD.89.014015}{\emph{Phys. Rev.}
  {\bfseries D89} (2014) 014015}
  [\href{https://arxiv.org/abs/1310.6660}{{\ttfamily 1310.6660}}].

\bibitem{Kang:2013jaa}
X.-W. Kang, B.~Kubis, C.~Hanhart and U.-G. Meißner, \emph{{$B_{l4}$ decays and
  the extraction of $|V_{ub}|$}},
  \href{https://doi.org/10.1103/PhysRevD.89.053015}{\emph{Phys. Rev.}
  {\bfseries D89} (2014) 053015}
  [\href{https://arxiv.org/abs/1312.1193}{{\ttfamily 1312.1193}}].

\bibitem{Hambrock:2015aor}
C.~Hambrock and A.~Khodjamirian, \emph{{Form factors in $\bar B^0 \to
  \pi\pi\ell\bar\nu_\ell$ from QCD light-cone sum rules}},
  \href{https://doi.org/10.1016/j.nuclphysb.2016.02.035}{\emph{Nucl. Phys.}
  {\bfseries B905} (2016) 373}
  [\href{https://arxiv.org/abs/1511.02509}{{\ttfamily 1511.02509}}].

\bibitem{Boer:2016iez}
P.~Böer, T.~Feldmann and D.~van Dyk, \emph{{QCD Factorization Theorem for $B
  \to \pi\pi\ell\nu$ Decays at Large Dipion Masses}},
  \href{https://doi.org/10.1007/JHEP02(2017)133}{\emph{JHEP} {\bfseries 02}
  (2017) 133} [\href{https://arxiv.org/abs/1608.07127}{{\ttfamily
  1608.07127}}].

\bibitem{Cheng:2017sfk}
S.~Cheng, A.~Khodjamirian and J.~Virto, \emph{{Timelike-helicity $B\to \pi\pi$
  form factor from light-cone sum rules with dipion distribution amplitudes}},
  \href{https://doi.org/10.1103/PhysRevD.96.051901}{\emph{Phys. Rev.}
  {\bfseries D96} (2017) 051901}
  [\href{https://arxiv.org/abs/1709.00173}{{\ttfamily 1709.00173}}].

\bibitem{Feldmann:2018kqr}
T.~Feldmann, D.~van Dyk and K.~K. Vos, \emph{{Revisiting $B \to \pi\pi \ell
  \nu$ at large dipion masses}},
  \href{https://doi.org/10.1007/JHEP10(2018)030}{\emph{JHEP} {\bfseries 10}
  (2018) 030} [\href{https://arxiv.org/abs/1807.01924}{{\ttfamily
  1807.01924}}].

\bibitem{Meissner:2013pba}
U.-G. Meißner and W.~Wang, \emph{{${\bf B_s\to K^{(*)} \ell\bar \nu}$, Angular
  Analysis, S-wave Contributions and ${\bf |V_{ub}|}$}},
  \href{https://doi.org/10.1007/JHEP01(2014)107}{\emph{JHEP} {\bfseries 01}
  (2014) 107} [\href{https://arxiv.org/abs/1311.5420}{{\ttfamily 1311.5420}}].

\bibitem{Meissner:2013hya}
U.-G. Meißner and W.~Wang, \emph{{Generalized Heavy-to-Light Form Factors in
  Light-Cone Sum Rules}},
  \href{https://doi.org/10.1016/j.physletb.2014.02.009}{\emph{Phys. Lett.}
  {\bfseries B730} (2014) 336}
  [\href{https://arxiv.org/abs/1312.3087}{{\ttfamily 1312.3087}}].

\bibitem{Das:2014sra}
D.~Das, G.~Hiller, M.~Jung and A.~Shires, \emph{{The $ \overline{B}\to
  \overline{K}\pi \ell \ell $ and $ {\overline{B}}_s\ \to \overline{K}K\ell
  \ell $ distributions at low hadronic recoil}},
  \href{https://doi.org/10.1007/JHEP09(2014)109}{\emph{JHEP} {\bfseries 09}
  (2014) 109} [\href{https://arxiv.org/abs/1406.6681}{{\ttfamily 1406.6681}}].

\bibitem{Feldmann:2015xsa}
T.~Feldmann, B.~Müller and D.~van Dyk, \emph{{Analyzing $b\to u$ transitions
  in semileptonic $\bar{B}_s \to K^{*+}(\to K \pi)\ell^-\bar\nu_\ell$ decays}},
  \href{https://doi.org/10.1103/PhysRevD.92.034013}{\emph{Phys. Rev.}
  {\bfseries D92} (2015) 034013}
  [\href{https://arxiv.org/abs/1503.09063}{{\ttfamily 1503.09063}}].

\bibitem{DescotesGenon:2018xxx}
S.~Descotes-Genon, A.~Khodjamirian and J.~Virto, \emph{{Light-Cone Sum Rules
  for $B\to K\pi$ Form Factors and Applications to Rare Decays}},  to appear.

\bibitem{DescotesGenon:2018yyy}
S.~Descotes-Genon, A.~Khodjamirian, J.~Virto and K.~K. Vos, \emph{{Light-Cone
  Sum Rules for $S$-wave $B\to M_1 M_2$ Form Factors}},  work in progress.

\bibitem{Wang:2015uea}
W.-F. Wang, H.-n. Li, W.~Wang and C.-D. Lü, \emph{{$S$-wave resonance
  contributions to the $B^0_{(s)}\to J/\psi\pi^+\pi^-$ and
  $B_s\to\pi^+\pi^-\mu^+\mu^-$ decays}},
  \href{https://doi.org/10.1103/PhysRevD.91.094024}{\emph{Phys. Rev.}
  {\bfseries D91} (2015) 094024}
  [\href{https://arxiv.org/abs/1502.05483}{{\ttfamily 1502.05483}}].

\bibitem{Krankl:2015fha}
S.~Kränkl, T.~Mannel and J.~Virto, \emph{{Three-body non-leptonic B decays and
  QCD factorization}},
  \href{https://doi.org/10.1016/j.nuclphysb.2015.08.004}{\emph{Nucl. Phys.}
  {\bfseries B899} (2015) 247}
  [\href{https://arxiv.org/abs/1505.04111}{{\ttfamily 1505.04111}}].

\bibitem{Daub:2015xja}
J.~T. Daub, C.~Hanhart and B.~Kubis, \emph{{A model-independent analysis of
  final-state interactions in $ {\overline{B}}_{d/s}^0\to J/\psi \pi \pi $}},
  \href{https://doi.org/10.1007/JHEP02(2016)009}{\emph{JHEP} {\bfseries 02}
  (2016) 009} [\href{https://arxiv.org/abs/1508.06841}{{\ttfamily
  1508.06841}}].

\bibitem{Cheng:2016shb}
H.-Y. Cheng, C.-K. Chua and Z.-Q. Zhang, \emph{{Direct CP Violation in
  Charmless Three-body Decays of $B$ Mesons}},
  \href{https://doi.org/10.1103/PhysRevD.94.094015}{\emph{Phys. Rev.}
  {\bfseries D94} (2016) 094015}
  [\href{https://arxiv.org/abs/1607.08313}{{\ttfamily 1607.08313}}].

\bibitem{Wang:2016rlo}
W.-F. Wang and H.-n. Li, \emph{{Quasi-two-body decays $B\to K\rho\to K\pi\pi$
  in perturbative QCD approach}},
  \href{https://doi.org/10.1016/j.physletb.2016.10.026}{\emph{Phys. Lett.}
  {\bfseries B763} (2016) 29}
  [\href{https://arxiv.org/abs/1609.04614}{{\ttfamily 1609.04614}}].

\bibitem{Albaladejo:2016mad}
M.~Albaladejo, J.~T. Daub, C.~Hanhart, B.~Kubis and B.~Moussallam, \emph{{How
  to employ $ {\overline{B}}_d^0\to J/\psi \left(\pi \eta, \overline{K}K\right)
  $ decays to extract information on $\pi\eta$ scattering}},
  \href{https://doi.org/10.1007/JHEP04(2017)010}{\emph{JHEP} {\bfseries 04}
  (2017) 010} [\href{https://arxiv.org/abs/1611.03502}{{\ttfamily
  1611.03502}}].

\bibitem{Klein:2017xti}
R.~Klein, T.~Mannel, J.~Virto and K.~K. Vos, \emph{{CP Violation in Multibody
  $B$ Decays from QCD Factorization}},
  \href{https://doi.org/10.1007/JHEP10(2017)117}{\emph{JHEP} {\bfseries 10}
  (2017) 117} [\href{https://arxiv.org/abs/1708.02047}{{\ttfamily
  1708.02047}}].

\bibitem{Balitsky:1988fi}
I.~I. Balitsky and V.~M. Braun, \emph{{Nonlocal Operator Expansion for
  Structure Functions of $e^+ e^-$ Annihilation}},
  \href{https://doi.org/10.1016/0370-2693(89)90733-8}{\emph{Phys. Lett.}
  {\bfseries B222} (1989) 123}.

\bibitem{EOS-v0.2.3}
D.~van Dyk, N.~Gubernari, A.~Kokulu et~al., ``{EOS release v0.2.3}.''
\newblock
  \href{https://github.com/eos/eos/releases/tag/v0.2.3}{https://github.com/eos/eos/releases/tag/v0.2.3}.

\bibitem{EOS}
D.~van Dyk, C.~Bobeth, F.~Beaujean et~al., \emph{{EOS --- A HEP program for
  Flavor Observables}},  2018.

\bibitem{Beneke:2018wjp}
M.~Beneke, V.~M. Braun, Y.~Ji and Y.-B. Wei, \emph{{Radiative leptonic decay
  $B\to \gamma \ell \nu_\ell$ with subleading power corrections}},
  \href{https://arxiv.org/abs/1804.04962}{{\ttfamily 1804.04962}}.

\bibitem{Bazavov:2017lyh}
A.~Bazavov et~al., \emph{{$B$- and $D$-meson leptonic decay constants from
  four-flavor lattice QCD}},
  \href{https://arxiv.org/abs/1712.09262}{{\ttfamily 1712.09262}}.

\bibitem{Nishikawa:2011qk}
T.~Nishikawa and K.~Tanaka, \emph{{QCD Sum Rules for Quark-Gluon Three-Body
  Components in the B Meson}},
  \href{https://doi.org/10.1016/j.nuclphysb.2013.12.007}{\emph{Nucl. Phys.}
  {\bfseries B879} (2014) 110}
  [\href{https://arxiv.org/abs/1109.6786}{{\ttfamily 1109.6786}}].

\bibitem{Shifman:1978bx}
M.~A. Shifman, A.~I. Vainshtein and V.~I. Zakharov, \emph{{QCD and Resonance
  Physics. Theoretical Foundations}},
  \href{https://doi.org/10.1016/0550-3213(79)90022-1}{\emph{Nucl. Phys.}
  {\bfseries B147} (1979) 385}.

\bibitem{Colangelo:2000dp}
P.~Colangelo and A.~Khodjamirian, \emph{{QCD sum rules, a modern perspective}},
   \href{https://arxiv.org/abs/hep-ph/0010175}{{\ttfamily hep-ph/0010175}}.

\bibitem{Khodjamirian:2003xk}
A.~Khodjamirian, T.~Mannel and M.~Melcher, \emph{{Flavor SU(3) symmetry in
  charmless B decays}},
  \href{https://doi.org/10.1103/PhysRevD.68.114007}{\emph{Phys. Rev.}
  {\bfseries D68} (2003) 114007}
  [\href{https://arxiv.org/abs/hep-ph/0308297}{{\ttfamily hep-ph/0308297}}].

\bibitem{Aoki:2016frl}
S.~Aoki et~al., \emph{{Review of lattice results concerning low-energy particle
  physics}}, \href{https://doi.org/10.1140/epjc/s10052-016-4509-7}{\emph{Eur.
  Phys. J.} {\bfseries C77} (2017) 112}
  [\href{https://arxiv.org/abs/1607.00299}{{\ttfamily 1607.00299}}].

\bibitem{Follana:2007uv}
{\scshape HPQCD, UKQCD} collaboration, E.~Follana, C.~T.~H. Davies, G.~P.
  Lepage and J.~Shigemitsu, \emph{{High Precision determination of the $\pi$,
  $K$, $D$ and $D_s$ decay constants from lattice QCD}},
  \href{https://doi.org/10.1103/PhysRevLett.100.062002}{\emph{Phys. Rev. Lett.}
  {\bfseries 100} (2008) 062002}
  [\href{https://arxiv.org/abs/0706.1726}{{\ttfamily 0706.1726}}].

\bibitem{Bazavov:2010hj}
{\scshape MILC} collaboration, A.~Bazavov et~al., \emph{{Results for light
  pseudoscalar mesons}}, {\emph{PoS} {\bfseries LATTICE2010} (2010) 074}
  [\href{https://arxiv.org/abs/1012.0868}{{\ttfamily 1012.0868}}].

\bibitem{Arthur:2012yc}
{\scshape RBC, UKQCD} collaboration, R.~Arthur et~al., \emph{{Domain Wall QCD
  with Near-Physical Pions}},
  \href{https://doi.org/10.1103/PhysRevD.87.094514}{\emph{Phys. Rev.}
  {\bfseries D87} (2013) 094514}
  [\href{https://arxiv.org/abs/1208.4412}{{\ttfamily 1208.4412}}].

\bibitem{Dowdall:2013rya}
R.~J. Dowdall, C.~T.~H. Davies, G.~P. Lepage and C.~McNeile, \emph{{Vus from pi
  and K decay constants in full lattice QCD with physical u, d, s and c
  quarks}}, \href{https://doi.org/10.1103/PhysRevD.88.074504}{\emph{Phys. Rev.}
  {\bfseries D88} (2013) 074504}
  [\href{https://arxiv.org/abs/1303.1670}{{\ttfamily 1303.1670}}].

\bibitem{Carrasco:2014poa}
N.~Carrasco et~al., \emph{{Leptonic decay constants $f_{K},f_{D},$ and
  $f_{{D}_{s}}$ with $N_{f} = 2+1+1$ twisted-mass lattice QCD}},
  \href{https://doi.org/10.1103/PhysRevD.91.054507}{\emph{Phys. Rev.}
  {\bfseries D91} (2015) 054507}
  [\href{https://arxiv.org/abs/1411.7908}{{\ttfamily 1411.7908}}].

\bibitem{Bazavov:2014wgs}
{\scshape Fermilab Lattice, MILC} collaboration, A.~Bazavov et~al.,
  \emph{{Charmed and light pseudoscalar meson decay constants from four-flavor
  lattice QCD with physical light quarks}},
  \href{https://doi.org/10.1103/PhysRevD.90.074509}{\emph{Phys. Rev.}
  {\bfseries D90} (2014) 074509}
  [\href{https://arxiv.org/abs/1407.3772}{{\ttfamily 1407.3772}}].

\bibitem{Gelhausen:2013wia}
P.~Gelhausen, A.~Khodjamirian, A.~A. Pivovarov and D.~Rosenthal, \emph{{Decay
  constants of heavy-light vector mesons from QCD sum rules}},  [Erratum: Phys.
  Rev.D91,099901(2015)]\href{https://doi.org/10.1103/PhysRevD.88.014015,
  10.1103/PhysRevD.91.099901, 10.1103/PhysRevD.89.099901}{\emph{Phys. Rev.}
  {\bfseries D88} (2013) 014015}
  [\href{https://arxiv.org/abs/1305.5432}{{\ttfamily 1305.5432}}].

\bibitem{Braun:PrivateCommunications}
V.~Braun, \emph{private communications.}, .

\bibitem{Lu:2018cfc}
C.-D. Lü, Y.-L. Shen, Y.-M. Wang and Y.-B. Wei, \emph{{QCD calculations of $B
  \to \pi, K$ form factors with higher-twist corrections}},
  \href{https://arxiv.org/abs/1810.00819}{{\ttfamily 1810.00819}}.

\bibitem{Bouchard:2013pna}
{\scshape HPQCD} collaboration, C.~Bouchard, G.~P. Lepage, C.~Monahan, H.~Na
  and J.~Shigemitsu, \emph{{Rare decay $B \to K \ell^+ \ell^-$ form factors
  from lattice QCD}},  [Erratum: Phys.
  Rev.D88,no.7,079901(2013)]\href{https://doi.org/10.1103/PhysRevD.88.079901,
  10.1103/PhysRevD.88.054509}{\emph{Phys. Rev.} {\bfseries D88} (2013) 054509}
  [\href{https://arxiv.org/abs/1306.2384}{{\ttfamily 1306.2384}}].

\bibitem{Horgan:2013hoa}
R.~R. Horgan, Z.~Liu, S.~Meinel and M.~Wingate, \emph{{Lattice QCD calculation
  of form factors describing the rare decays $B \to K^* \ell^+ \ell^-$ and $B_s
  \to \phi \ell^+ \ell^-$}},
  \href{https://doi.org/10.1103/PhysRevD.89.094501}{\emph{Phys. Rev.}
  {\bfseries D89} (2014) 094501}
  [\href{https://arxiv.org/abs/1310.3722}{{\ttfamily 1310.3722}}].

\bibitem{Bailey:2014tva}
{\scshape Fermilab Lattice, MILC} collaboration, J.~A. Bailey et~al.,
  \emph{{Update of $|V_{cb}|$ from the $\bar{B}\to D^*\ell\bar{\nu}$ form
  factor at zero recoil with three-flavor lattice QCD}},
  \href{https://doi.org/10.1103/PhysRevD.89.114504}{\emph{Phys. Rev.}
  {\bfseries D89} (2014) 114504}
  [\href{https://arxiv.org/abs/1403.0635}{{\ttfamily 1403.0635}}].

\bibitem{Horgan:2015vla}
R.~R. Horgan, Z.~Liu, S.~Meinel and M.~Wingate, \emph{{Rare $B$ decays using
  lattice QCD form factors}},
  \href{https://doi.org/10.22323/1.214.0372}{\emph{PoS} {\bfseries LATTICE2014}
  (2015) 372} [\href{https://arxiv.org/abs/1501.00367}{{\ttfamily
  1501.00367}}].

\bibitem{Na:2015kha}
{\scshape HPQCD} collaboration, H.~Na, C.~M. Bouchard, G.~P. Lepage, C.~Monahan
  and J.~Shigemitsu, \emph{{$B \rightarrow D l \nu$ form factors at nonzero
  recoil and extraction of $|V_{cb}|$}},  [Erratum: Phys.
  Rev.D93,no.11,119906(2016)]\href{https://doi.org/10.1103/PhysRevD.93.119906,
  10.1103/PhysRevD.92.054510}{\emph{Phys. Rev.} {\bfseries D92} (2015) 054510}
  [\href{https://arxiv.org/abs/1505.03925}{{\ttfamily 1505.03925}}].

\bibitem{Lattice:2015tia}
{\scshape Fermilab Lattice, MILC} collaboration, J.~A. Bailey et~al.,
  \emph{{$|V_{ub}|$ from $B\to\pi\ell\nu$ decays and (2+1)-flavor lattice
  QCD}}, \href{https://doi.org/10.1103/PhysRevD.92.014024}{\emph{Phys. Rev.}
  {\bfseries D92} (2015) 014024}
  [\href{https://arxiv.org/abs/1503.07839}{{\ttfamily 1503.07839}}].

\bibitem{Harrison:2017fmw}
{\scshape HPQCD} collaboration, J.~Harrison, C.~Davies and M.~Wingate,
  \emph{{Lattice QCD calculation of the ${{B}_{(s)}\to D_{(s)}^{*}\ell{\nu}}$
  form factors at zero recoil and implications for ${|V_{cb}|}$}},
  \href{https://doi.org/10.1103/PhysRevD.97.054502}{\emph{Phys. Rev.}
  {\bfseries D97} (2018) 054502}
  [\href{https://arxiv.org/abs/1711.11013}{{\ttfamily 1711.11013}}].

\bibitem{Straub:2018kue}
D.~M. Straub, \emph{{flavio: a Python package for flavour and precision
  phenomenology in the Standard Model and beyond}},
  \href{https://arxiv.org/abs/1810.08132}{{\ttfamily 1810.08132}}.

\bibitem{Detmold:2015aaa}
W.~Detmold, C.~Lehner and S.~Meinel, \emph{{$\Lambda_b \to p \ell^-
  \bar{\nu}_\ell$ and $\Lambda_b \to \Lambda_c \ell^- \bar{\nu}_\ell$ form
  factors from lattice QCD with relativistic heavy quarks}},
  \href{https://doi.org/10.1103/PhysRevD.92.034503}{\emph{Phys. Rev.}
  {\bfseries D92} (2015) 034503}
  [\href{https://arxiv.org/abs/1503.01421}{{\ttfamily 1503.01421}}].

\bibitem{Aaij:2014pli}
{\scshape LHCb} collaboration, R.~Aaij et~al., \emph{{Differential branching
  fractions and isospin asymmetries of $B \to K^{(*)} \mu^+ \mu^-$ decays}},
  \href{https://doi.org/10.1007/JHEP06(2014)133}{\emph{JHEP} {\bfseries 06}
  (2014) 133} [\href{https://arxiv.org/abs/1403.8044}{{\ttfamily 1403.8044}}].

\bibitem{Aaij:2015esa}
{\scshape LHCb} collaboration, R.~Aaij et~al., \emph{{Angular analysis and
  differential branching fraction of the decay $B^0_s\to\phi\mu^+\mu^-$}},
  \href{https://doi.org/10.1007/JHEP09(2015)179}{\emph{JHEP} {\bfseries 09}
  (2015) 179} [\href{https://arxiv.org/abs/1506.08777}{{\ttfamily
  1506.08777}}].

\bibitem{Aaij:2016flj}
{\scshape LHCb} collaboration, R.~Aaij et~al., \emph{{Measurements of the
  S-wave fraction in $B^{0}\rightarrow K^{+}\pi^{-}\mu^{+}\mu^{-}$ decays and
  the $B^{0}\rightarrow K^{\ast}(892)^{0}\mu^{+}\mu^{-}$ differential branching
  fraction}},  [Erratum:
  JHEP04,142(2017)]\href{https://doi.org/10.1007/JHEP11(2016)047,
  10.1007/JHEP04(2017)142}{\emph{JHEP} {\bfseries 11} (2016) 047}
  [\href{https://arxiv.org/abs/1606.04731}{{\ttfamily 1606.04731}}].

\bibitem{Aaij:2015xza}
{\scshape LHCb} collaboration, R.~Aaij et~al., \emph{{Differential branching
  fraction and angular analysis of $\Lambda^{0}_{b} \rightarrow \Lambda
  \mu^+\mu^-$ decays}},  [Erratum:
  JHEP09,145(2018)]\href{https://doi.org/10.1007/JHEP09(2018)145,
  10.1007/JHEP06(2015)115}{\emph{JHEP} {\bfseries 06} (2015) 115}
  [\href{https://arxiv.org/abs/1503.07138}{{\ttfamily 1503.07138}}].

\bibitem{Aaij:2015oid}
{\scshape LHCb} collaboration, R.~Aaij et~al., \emph{{Angular analysis of the
  $B^{0} \to K^{*0} \mu^{+} \mu^{-}$ decay using 3 fb$^{-1}$ of integrated
  luminosity}}, \href{https://doi.org/10.1007/JHEP02(2016)104}{\emph{JHEP}
  {\bfseries 02} (2016) 104}
  [\href{https://arxiv.org/abs/1512.04442}{{\ttfamily 1512.04442}}].

\bibitem{Khachatryan:2015isa}
{\scshape CMS} collaboration, V.~Khachatryan et~al., \emph{{Angular analysis of
  the decay $B^0 \to K^{*0} \mu^+ \mu^-$ from pp collisions at $\sqrt s = 8$
  TeV}}, \href{https://doi.org/10.1016/j.physletb.2015.12.020}{\emph{Phys.
  Lett.} {\bfseries B753} (2016) 424}
  [\href{https://arxiv.org/abs/1507.08126}{{\ttfamily 1507.08126}}].

\bibitem{Aaboud:2018krd}
{\scshape ATLAS} collaboration, M.~Aaboud et~al., \emph{{Angular analysis of
  $B^0_d \rightarrow K^{*}\mu^+\mu^-$ decays in $pp$ collisions at $\sqrt{s}=
  8$ TeV with the ATLAS detector}},
  \href{https://doi.org/10.1007/JHEP10(2018)047}{\emph{JHEP} {\bfseries 10}
  (2018) 047} [\href{https://arxiv.org/abs/1805.04000}{{\ttfamily
  1805.04000}}].

\bibitem{Aaij:2014ora}
{\scshape LHCb} collaboration, R.~Aaij et~al., \emph{{Test of lepton
  universality using $B^{+}\rightarrow K^{+}\ell^{+}\ell^{-}$ decays}},
  \href{https://doi.org/10.1103/PhysRevLett.113.151601}{\emph{Phys. Rev. Lett.}
  {\bfseries 113} (2014) 151601}
  [\href{https://arxiv.org/abs/1406.6482}{{\ttfamily 1406.6482}}].

\bibitem{Aaij:2017vbb}
{\scshape LHCb} collaboration, R.~Aaij et~al., \emph{{Test of lepton
  universality with $B^{0} \rightarrow K^{*0}\ell^{+}\ell^{-}$ decays}},
  \href{https://doi.org/10.1007/JHEP08(2017)055}{\emph{JHEP} {\bfseries 08}
  (2017) 055} [\href{https://arxiv.org/abs/1705.05802}{{\ttfamily
  1705.05802}}].

\bibitem{Descotes-Genon:2013wba}
S.~Descotes-Genon, J.~Matias and J.~Virto, \emph{{Understanding the $B\to
  K^*\mu^+\mu^-$ Anomaly}},
  \href{https://doi.org/10.1103/PhysRevD.88.074002}{\emph{Phys. Rev.}
  {\bfseries D88} (2013) 074002}
  [\href{https://arxiv.org/abs/1307.5683}{{\ttfamily 1307.5683}}].

\bibitem{Altmannshofer:2013foa}
W.~Altmannshofer and D.~M. Straub, \emph{{New Physics in $B \to K^*\mu\mu$?}},
  \href{https://doi.org/10.1140/epjc/s10052-013-2646-9}{\emph{Eur. Phys. J.}
  {\bfseries C73} (2013) 2646}
  [\href{https://arxiv.org/abs/1308.1501}{{\ttfamily 1308.1501}}].

\bibitem{Beaujean:2013soa}
F.~Beaujean, C.~Bobeth and D.~van Dyk, \emph{{Comprehensive Bayesian analysis
  of rare (semi)leptonic and radiative $B$ decays}},  [Erratum: Eur. Phys.
  J.C74,3179(2014)]\href{https://doi.org/10.1140/epjc/s10052-014-2897-0,
  10.1140/epjc/s10052-014-3179-6}{\emph{Eur. Phys. J.} {\bfseries C74} (2014)
  2897} [\href{https://arxiv.org/abs/1310.2478}{{\ttfamily 1310.2478}}].

\bibitem{Hurth:2013ssa}
T.~Hurth and F.~Mahmoudi, \emph{{On the LHCb anomaly in B $\to
  K^*\ell^+\ell^-$}},
  \href{https://doi.org/10.1007/JHEP04(2014)097}{\emph{JHEP} {\bfseries 04}
  (2014) 097} [\href{https://arxiv.org/abs/1312.5267}{{\ttfamily 1312.5267}}].

\bibitem{Descotes-Genon:2015uva}
S.~Descotes-Genon, L.~Hofer, J.~Matias and J.~Virto, \emph{{Global analysis of
  $b\to s\ell\ell$ anomalies}},
  \href{https://doi.org/10.1007/JHEP06(2016)092}{\emph{JHEP} {\bfseries 06}
  (2016) 092} [\href{https://arxiv.org/abs/1510.04239}{{\ttfamily
  1510.04239}}].

\bibitem{Altmannshofer:2017fio}
W.~Altmannshofer, C.~Niehoff, P.~Stangl and D.~M. Straub, \emph{{Status of the
  $B\rightarrow K^*\mu ^+\mu ^-$ anomaly after Moriond 2017}},
  \href{https://doi.org/10.1140/epjc/s10052-017-4952-0}{\emph{Eur. Phys. J.}
  {\bfseries C77} (2017) 377}
  [\href{https://arxiv.org/abs/1703.09189}{{\ttfamily 1703.09189}}].

\bibitem{Altmannshofer:2017yso}
W.~Altmannshofer, P.~Stangl and D.~M. Straub, \emph{{Interpreting Hints for
  Lepton Flavor Universality Violation}},
  \href{https://doi.org/10.1103/PhysRevD.96.055008}{\emph{Phys. Rev.}
  {\bfseries D96} (2017) 055008}
  [\href{https://arxiv.org/abs/1704.05435}{{\ttfamily 1704.05435}}].

\bibitem{Capdevila:2017bsm}
B.~Capdevila, A.~Crivellin, S.~Descotes-Genon, J.~Matias and J.~Virto,
  \emph{{Patterns of New Physics in $b\to s\ell^+\ell^-$ transitions in the
  light of recent data}},
  \href{https://doi.org/10.1007/JHEP01(2018)093}{\emph{JHEP} {\bfseries 01}
  (2018) 093} [\href{https://arxiv.org/abs/1704.05340}{{\ttfamily
  1704.05340}}].

\bibitem{Geng:2017svp}
L.-S. Geng, B.~Grinstein, S.~Jäger, J.~Martin~Camalich, X.-L. Ren and R.-X.
  Shi, \emph{{Towards the discovery of new physics with lepton-universality
  ratios of $b\to s\ell\ell$ decays}},
  \href{https://doi.org/10.1103/PhysRevD.96.093006}{\emph{Phys. Rev.}
  {\bfseries D96} (2017) 093006}
  [\href{https://arxiv.org/abs/1704.05446}{{\ttfamily 1704.05446}}].

\bibitem{Ciuchini:2017mik}
M.~Ciuchini, A.~M. Coutinho, M.~Fedele, E.~Franco, A.~Paul, L.~Silvestrini
  et~al., \emph{{On Flavourful Easter eggs for New Physics hunger and Lepton
  Flavour Universality violation}},
  \href{https://doi.org/10.1140/epjc/s10052-017-5270-2}{\emph{Eur. Phys. J.}
  {\bfseries C77} (2017) 688}
  [\href{https://arxiv.org/abs/1704.05447}{{\ttfamily 1704.05447}}].

\bibitem{Mahmoudi:2018qsk}
F.~Mahmoudi, T.~Hurth and S.~Neshatpour, \emph{{Updated fits to the present $b
  \to s \ell ^+\ell ^-$ data}},
  \href{https://doi.org/10.5506/APhysPolB.49.1267}{\emph{Acta Phys. Polon.}
  {\bfseries B49} (2018) 1267}.

\bibitem{Hiller:2003js}
G.~Hiller and F.~Kruger, \emph{{More model-independent analysis of $b \to s$
  processes}}, \href{https://doi.org/10.1103/PhysRevD.69.074020}{\emph{Phys.
  Rev.} {\bfseries D69} (2004) 074020}
  [\href{https://arxiv.org/abs/hep-ph/0310219}{{\ttfamily hep-ph/0310219}}].

\bibitem{Bobeth:2007dw}
C.~Bobeth, G.~Hiller and G.~Piranishvili, \emph{{Angular distributions of
  $\bar{B} \to \bar{K} \ell^+\ell^-$ decays}},
  \href{https://doi.org/10.1088/1126-6708/2007/12/040}{\emph{JHEP} {\bfseries
  12} (2007) 040} [\href{https://arxiv.org/abs/0709.4174}{{\ttfamily
  0709.4174}}].

\bibitem{Bordone:2016gaq}
M.~Bordone, G.~Isidori and A.~Pattori, \emph{{On the Standard Model predictions
  for $R_K$ and $R_{K^*}$}},
  \href{https://doi.org/10.1140/epjc/s10052-016-4274-7}{\emph{Eur. Phys. J.}
  {\bfseries C76} (2016) 440}
  [\href{https://arxiv.org/abs/1605.07633}{{\ttfamily 1605.07633}}].

\bibitem{Charles:1998dr}
J.~Charles, A.~Le~Yaouanc, L.~Oliver, O.~Pene and J.~C. Raynal, \emph{{Heavy to
  light form-factors in the heavy mass to large energy limit of QCD}},
  \href{https://doi.org/10.1103/PhysRevD.60.014001}{\emph{Phys. Rev.}
  {\bfseries D60} (1999) 014001}
  [\href{https://arxiv.org/abs/hep-ph/9812358}{{\ttfamily hep-ph/9812358}}].

\bibitem{Beneke:2000wa}
M.~Beneke and T.~Feldmann, \emph{{Symmetry breaking corrections to heavy to
  light B meson form-factors at large recoil}},
  \href{https://doi.org/10.1016/S0550-3213(00)00585-X}{\emph{Nucl. Phys.}
  {\bfseries B592} (2001) 3}
  [\href{https://arxiv.org/abs/hep-ph/0008255}{{\ttfamily hep-ph/0008255}}].

\bibitem{Bobeth:2017vxj}
C.~Bobeth, M.~Chrzaszcz, D.~van Dyk and J.~Virto, \emph{{Long-distance effects
  in $B\rightarrow K^*\ell \ell $ from analyticity}},
  \href{https://doi.org/10.1140/epjc/s10052-018-5918-6}{\emph{Eur. Phys. J.}
  {\bfseries C78} (2018) 451}
  [\href{https://arxiv.org/abs/1707.07305}{{\ttfamily 1707.07305}}].

\bibitem{Horgan:2013pva}
R.~R. Horgan, Z.~Liu, S.~Meinel and M.~Wingate, \emph{{Calculation of $B^0 \to
  K^{*0} \mu^+ \mu^-$ and $B_s^0 \to \phi \mu^+ \mu^-$ observables using form
  factors from lattice QCD}},
  \href{https://doi.org/10.1103/PhysRevLett.112.212003}{\emph{Phys. Rev. Lett.}
  {\bfseries 112} (2014) 212003}
  [\href{https://arxiv.org/abs/1310.3887}{{\ttfamily 1310.3887}}].

\bibitem{Boyd:1995cf}
C.~G. Boyd, B.~Grinstein and R.~F. Lebed, \emph{{Model independent extraction
  of $|V_{cb}|$ using dispersion relations}},
  \href{https://doi.org/10.1016/0370-2693(95)00480-9}{\emph{Phys. Lett.}
  {\bfseries B353} (1995) 306}
  [\href{https://arxiv.org/abs/hep-ph/9504235}{{\ttfamily hep-ph/9504235}}].

\bibitem{Caprini:1997mu}
I.~Caprini, L.~Lellouch and M.~Neubert, \emph{{Dispersive bounds on the shape
  of $\bar{B}\to D^{(*)} \ell\bar\nu$ form-factors}},
  \href{https://doi.org/10.1016/S0550-3213(98)00350-2}{\emph{Nucl. Phys.}
  {\bfseries B530} (1998) 153}
  [\href{https://arxiv.org/abs/hep-ph/9712417}{{\ttfamily hep-ph/9712417}}].

\bibitem{Fajfer:2012vx}
S.~Fajfer, J.~F. Kamenik and I.~Nisandzic, \emph{{On the $B \to D^* \tau \bar
  \nu_{\tau}$ Sensitivity to New Physics}},
  \href{https://doi.org/10.1103/PhysRevD.85.094025}{\emph{Phys. Rev.}
  {\bfseries D85} (2012) 094025}
  [\href{https://arxiv.org/abs/1203.2654}{{\ttfamily 1203.2654}}].

\bibitem{Bernlochner:2017jka}
F.~U. Bernlochner, Z.~Ligeti, M.~Papucci and D.~J. Robinson, \emph{{Combined
  analysis of semileptonic $B$ decays to $D$ and $D^*$: $R(D^{(*)})$,
  $|V_{cb}|$, and new physics}},  [Erratum: Phys.
  Rev.D97,no.5,059902(2018)]\href{https://doi.org/10.1103/PhysRevD.95.115008,
  10.1103/PhysRevD.97.059902}{\emph{Phys. Rev.} {\bfseries D95} (2017) 115008}
  [\href{https://arxiv.org/abs/1703.05330}{{\ttfamily 1703.05330}}].

\bibitem{Grozin:1996pq}
A.~G. Grozin and M.~Neubert, \emph{{Asymptotics of heavy meson form-factors}},
  \href{https://doi.org/10.1103/PhysRevD.55.272}{\emph{Phys. Rev.} {\bfseries
  D55} (1997) 272} [\href{https://arxiv.org/abs/hep-ph/9607366}{{\ttfamily
  hep-ph/9607366}}].

\end{thebibliography}\endgroup
